 \documentclass[a4paper]{article}
 \usepackage{etoolbox}
 \newbool{PREPRINT}
 \booltrue{PREPRINT}

\ifbool{PREPRINT}{ 
\usepackage[colorlinks,bookmarksopen,bookmarksnumbered,citecolor=red,urlcolor=red]{hyperref}
\usepackage[affil-it]{authblk}

\usepackage{fullpage}
}{
\usepackage{fullpage}
} 

\usepackage{enumitem}
\usepackage{soul}
\usepackage{amsmath,amssymb}
\usepackage{stmaryrd}
\usepackage{graphicx}
\usepackage{hhline}
\usepackage{algorithm,algorithmic}
\usepackage[usenames]{color}
\usepackage[colorlinks,bookmarksopen,bookmarksnumbered,citecolor=red,urlcolor=red]{hyperref}
\usepackage{graphicx}
\usepackage{multirow}
\usepackage[dvipsnames]{xcolor}
\newcommand{\figdir}{./}

\let\oldvec=\vec
\renewcommand{\vec}[1]{\boldsymbol{#1}}

\newcommand{\favg}[1]{\{\!\!\{#1\}\!\!\}}
\newcommand{\fjump}[1]{[\![#1]\!]}
\newcommand{\fdiff}[1]{\langle{\!\langle{#1}\rangle}\!\rangle}
\newcommand{\what}[1]{\widehat{#1}}

\newcommand{\refr}[1]{\overline{#1}}
\newcommand{\unit}[1]{\textsf{#1}}
\newcommand{\uperp}{\vec{u}^\perp} 
\newcommand{\secapp}{\ifbool{PREPRINT}{Appendix~}{}}
\newcommand{\pparagraph}[1]{\ifbool{PREPRINT}{\paragraph{#1.}}{\paragraph{#1}}}
\usepackage{arydshln}
\usepackage{listings}
\lstdefinelanguage[firedrake]{python}[]{python}{%
  emph={[2]Function,MeshHierarchy,FunctionSpaceHierarchy,restrict,prolong,
    assemble,solve,Kernel,par_loop,Tensor,LinearVariationalProblem,LinearVariationalSolver},
  emph={grad,div,dx,dS,dot,action,op2,inner}
}
\definecolor{DarkBlue}{rgb}{0.00,0.00,0.55}
\definecolor{DarkRed}{rgb}{0.55,0.00,0.00}
\definecolor{LightGray}{rgb}{0.8,0.8,0.8}
\definecolor{DarkGreen}{rgb}{0.00,0.55,0.00}
\definecolor{Purple}{rgb}{0.5,0.0,0.5}
\definecolor{Bittersweet}{rgb}{1.0,0.44,0.37}
\title{Multigrid preconditioners for the hybridised Discontinuous Galerkin discretisation of the shallow water equations}
\ifbool{PREPRINT}{
\author[1,2]{Jack~Betteridge}
\author[2,3]{Thomas~H.~Gibson}
\author[1]{Ivan~G.~Graham}
\author[$\dagger$,*,1]{Eike~H.~M\"uller}
\affil[1]{Department of Mathematical Sciences, University of Bath, Bath BA2 7AY, United Kingdom}
\affil[2]{Department of Computing and Department of Mathematics, Imperial College, South Kensington Campus, London SW7 2AZ, United Kingdom}
\affil[3]{Naval Postgraduate School, 1 University Circle, Monterey, CA 93943, USA}
\affil[$\dagger$]{Email: \texttt{e.mueller@bath.ac.uk}}
\affil[*]{Corresponding author}

}{ 
\author[bath,imperial]{Jack~Betteridge}
\author[imperial,nps]{Thomas~H.~Gibson}
\author[bath]{Ivan~G.~Graham}
\author[bath]{Eike~H.~M\"{u}ller\corref{cor1}\fnref{fn1}}
\ead{e.mueller@bath.ac.uk}
\cortext[cor1]{Corresponding author}
\fntext[fn1]{email: \texttt{e.mueller@bath.ac.uk}}
\address[bath]{Department of Mathematical Sciences, University of Bath, Claverton Down, Bath BA2 7AY, United Kingdom}
\address[imperial]{Department of Computing and Department of Mathematics, Imperial College, South Kensington Campus, London SW7 2AZ, United Kingdom}
\address[nps]{Naval Postgraduate School, 1 University Circle, Monterey, CA 93943, USA} 
}
\date{\today}
\begin{document}
\ifbool{PREPRINT}{ 
\maketitle
}{} 
\begin{abstract}
  Numerical climate- and weather-prediction models require the fast solution of the equations of fluid dynamics. Discontinuous Galerkin (DG) discretisations have several advantageous properties. They can be used for arbitrary domains and support a structured data layout, which is particularly important on modern chip architectures. For smooth solutions, higher order approximations can be particularly efficient since errors decrease exponentially in the polynomial degree. Due to the wide separation of timescales in atmospheric dynamics, semi-implicit time integrators are highly efficient, since the implicit treatment of fast waves avoids tight constraints on the time step size, and can therefore improve overall efficiency. However, if implicit-explicit (IMEX) integrators are used, a large linear system of equations has to be solved in every time step. A particular problem for DG discretisations of velocity-pressure systems is that the normal Schur-complement reduction to an elliptic system for the pressure is not possible since the numerical fluxes introduce artificial diffusion terms. For the shallow water equations, which form an important model system, hybridised DG methods have been shown to overcome this issue. However, no attention has been paid to the efficient solution of the resulting linear system of equations. In this paper we address this issue and show that the elliptic system for the flux unknowns can be solved efficiently by using a non-nested multigrid algorithm. The method is implemented in the Firedrake library and we demonstrate the excellent performance of the algorithm both for an idealised stationary flow problem in a flat domain and for non-stationary setups in spherical geometry from the well-known testsuite in [Williamson et al. (1992) JCP, 102(1), pp.211-224]. In the latter case the performance of our bespoke multigrid preconditioner (although itself not highly optimised) is comparable to that of a highly optimised direct solver.
\end{abstract}
\ifbool{PREPRINT}{ 
%
\textbf{keywords}:
\newcommand{\sep}{,}
}{ 
\begin{keyword}
} 
multigrid\sep\ elliptic PDE\sep\ Hybridised Discontinuous Galerkin \sep\ preconditioners\sep\ atmospheric modelling
\ifbool{PREPRINT}{}{ 
\end{keyword}
} 
\maketitle
\section{Introduction}
The dynamics of geophysical fluids is characterised by phenomena which occur at very different time scales. While large-scale flow in the atmosphere is described by the slow evolution of quasi-geostrophic modes and non-linearities are weak, fast acoustic waves play an important r\^{o}le in coupling phenomena at different scales and rapid adjustment to geostrophic balance. However, acoustic waves carry very little energy and hence do not have to be represented to high accuracy in computer simulations. Efficient numerical models for weather- and climate-prediction often exploit this scale-separation by using semi-implicit time integrators. Treating the fast waves implicitly (and thereby removing any numerical instabilities due to acoustic modes) avoids tight restrictions on the time step size. It is sufficient to chose the timestep such that it allows the stable and accurate representation of the slow modes, which may be treated explicitly. Examples of popular time-integrators are semi-implicit semi-Lagrangian methods used by the UK Met Office \cite{Wood2014} and implicit-explicit (IMEX) methods \cite{Ascher1995,Ascher1997,Pareschi2000,Kennedy2003} (see also \cite{Weller2013} for applications in atmospheric modelling).

In a computer model the equations of fluid-dynamics have to be represented on a grid with finite resolution. While simple finite-difference- or finite-volumes schemes on structured meshes (such as latitude-longitude grids) work well for local area models, they suffer from the convergence of grid lines at the poles on global grids \cite{Staniforth2012}. For a fully explicit time discretisation, this leads to artificially tight constraints on the timestep size due to the CFL condition because of the small grid cells at the poles. Although semi-implicit discretisations are not restricted by this stability constraint and can be run with larger timestep sizes, the linear system for the pressure correction, which has to be solved in every time step, becomes increasingly ill-conditioned and expensive to solve due to large Courant number at the poles. Furthermore, all-to-all communications limit scalability in massively parallel implementations. Finite element methods on quasi-uniform unstructured global grids overcome this problem, but care has to be taken when solving the resulting linear systems which arise in semi-implicit timestepping. The standard approach uses a Schur-complement reduction to a pressure system. Naively, this results in a dense Schur-complement operator in pressure space. For conforming finite element methods this can be traced back to the fact that the mass-matrices are non-diagonal. In this case the problem is easily overcome by under-integrating and co-locating the quadrature points with nodal basis points, or forming an approximate Schur-complement \cite{Mitchell2016}.

Discontinuous Galerkin (DG) discretisations have recently been explored as a promising alternative in atmospheric modelling; the review in \cite{Marras2016} highlights their excellent parallel scalability and discusses current issues of the approach. DG methods have several advantages, especially on modern chip architectures: since the unknowns are associated with cells of the mesh, data access can be coalesced, which is particularly important for chips with wide vector units. For sufficiently smooth solutions, higher order methods are numerically efficient since errors decrease exponentially as the polynomial degree of the DG approximation increases. Matrix-free implementations are particularly efficient on chips with limited memory bandwidth since they have a very high computational intensity; sum factorisation techniques can further reduce the computational complexity \cite{Muething2017}. Unfortunately, for DG methods the numerical flux introduces off-diagonal matrix elements which cause issues for semi-implicit discretisations. The flux generates artificial diffusion terms, which are numerical artifacts and can not be removed in an obvious way. This rules out the naive Schur-complement approach for solving the implicit linear system.

In \cite{Kang2020} a solution to this problem is presented for the shallow water equations (SWEs), a popular two-dimensional model system for atmospheric dynamics. The authors use the hybridised Discontinuous Galerkin (HDG) method (see e.g. \cite{Cockburn2009a,Cockburn2009b,Egger2009,Nguyen2009}), which is adapted for the SWEs in \cite{BuiThanh2015,BuiThanh2016}. Introducing a set of flux-unknowns on the facets of the mesh allows the reduction to a sparse Schur-complement system in the flux space. In \cite{Kang2020} this problem is assembled and solved with a direct method via LU factorisation. While this works for small problems, it is expected that the cost of this direct solver grows rapidly as the problem size increases, even if regularities in the sparsity pattern are exploited (see \cite{Duff2017} for a general overview of direct methods). This approach is therefore clearly unfeasible for large scale applications, especially if they are eventually to be used in time critical operational forecasting systems.

The central aim of this paper is to present an alternative solution which overcomes those issues and to demonstrate the computational performance of our approach. We introduce a new non-nested multigrid method for solving the linear systems arising in IMEX time integrators for the SWEs, when discretised using a hybridised (high order) DG method. The method is strongly influenced by the work of Gopalakrishnan and Tan \cite{Gopalakrishnan2009} and Cockburn, Dubois, Gopalakrishnan and Tan \cite{Cockburn2014}. In particular \cite{Gopalakrishnan2009} introduces a non-nested multigrid V-cycle for the hybridised mixed method for standard second order elliptic PDEs discretised using Raviart Thomas elements for the velocity approximation. This has been extended in \cite{Cockburn2014} to the multigrid solution of the HDG discretisation for the same elliptic PDE. In this paper we adapt this approach to solve the linear and non-linear shallow water equations. The non-linear SWEs present additional challenges since they require the construction of a Lax-Friedrich flux, which results in a vector-valued flux space \cite{BuiThanh2016} for which the method in \cite{Cockburn2014} is not directly applicable.

More specifically, the new contributions of this paper are:
\begin{enumerate}
\item The non-nested multigrid method of \cite{Cockburn2014} is extended to the HDG solution of the linearised SWEs with IMEX methods.
\item To solve the non-linear SWEs by this approach we construct novel intergrid- and coarse-level operators for vector-valued flux spaces.
\item Our HDG solvers and multigrid algorithms are implemented in the Slate language \cite{Gibson2019}, which forms part of the open source Firedrake library \cite{Rathgeber2017} for the solution of finite element problems based on composable abstractions.
\item Based on this, the superior performance of the non-nested multigrid approach (compared to a naive Schur-complement preconditioner) for both the linear- and non-linear SWEs is demonstrated for a simplified stationary test flow in a flat domain.
\item For more complex non-stationary flow problems from the standard Williamson et al. test suite \cite{Williamson1992} the non-nested multigrid preconditioner is compared to a highly-optimised direct solver from the MUMPS package \cite{MUMPS1,MUMPS2}.
\end{enumerate}
Our approach differs from the more algebraic method presented in \cite{Wildey2018,Muralikrishnan2019}, which constructs a multigrid hierarchy directly on the trace-space defined on the skeleton of the mesh; this is achieved by using $hp$-refinement and statically condensing out unknowns on fine edges which are not represented on the coarser mesh.

We stress that the aim of this work is to quantify the advantages of replacing the direct solver (such as the one used in \cite{Kang2020}) with our new multigrid-method. Our experiments on the stationary flow problem indicate substantial advantages of our new multigrid method. While for flows in a spherical geometry both considered solvers show similar performance on a single compute node, it is highly likely that an iterative solver with a multigrid preconditioner will be superior to a direct method when solving very large problems on massively parallel machines.

In the long run it would be important to properly compare the HDG method to other discretisation schemes such as finite volume. However a proper treatment of this would need a detailed assessment of both accuracy and efficiency of implementations and solvers. While we feel this is worth doing, it is well beyond the scope of the present paper.
\pparagraph{Structure}
This paper is organised as follows: In Section \ref{sec:SWEs} we briefly review the physics of the shallow water equations. The hybridised DG discretisation is described in Section \ref{sec:Methods} and the non-nested multigrid preconditioner for the resulting linear systems in IMEX timestepping is presented in Section \ref{sec:IMEXHDG}. The specific integrators considered in this paper are introduced in Section \ref{sec:timesteppers} with focus on the linear solvers used for implicit methods. The implementation in the Firedrake library is described in Section \ref{sec:Implementation} and results are presented in Section \ref{sec:Results}. We finally conclude and outline directions for future work in Section \ref{sec:Conclusion}. Some more technical issues such as the explicit expressions for key bilinear forms are relegated to the appendices.
\section{Shallow water equations}\label{sec:SWEs}
The shallow water equations (SWEs) describe the evolution of a layer of fluid under gravitational and inertial forces. In a rotating frame a Coriolis term has to be included as well. The SWEs can be written in conservative form for the geopotential height $\phi(\vec{x},t)$ and the column-integrated momentum $\vec{u}(\vec{x},t)$, which are defined for all points $\vec{x}\in\Omega$ in a bounded domain $\Omega\subseteq\mathbb{R}^2$. The time evolution of $\phi$ and $\vec{u}=(u,v)$ is described by the following two equations:
\begin{equation}
\begin{aligned}
\frac{\partial\phi}{\partial t} + c_g \nabla \cdot \vec{u} &= 0 \qquad\text{(mass conservation)}\\
\frac{\partial\vec{u}}{\partial t} + c_g\nabla \cdot \left(\frac{\vec{u}\otimes\vec{u}}{\phi+\phi_B}+\frac{1}{2}\left(\phi_B+\phi\right)^2\right) &= \vec{R} := c_g(\phi_B+\phi)\nabla\phi_B - f\uperp
\qquad\text{(momentum conservation)}
\end{aligned}
\label{eqn:SWE_continuum}
\end{equation}
\begin{figure}
  \begin{center}
    \includegraphics[width=0.4\linewidth]{\figdir/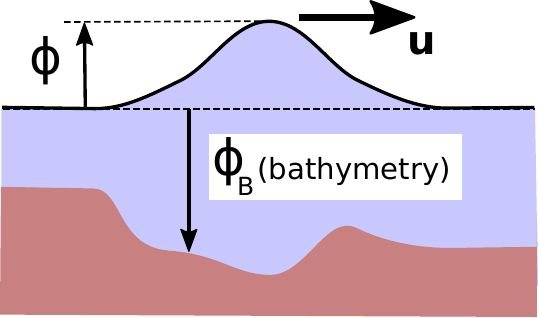}
    \caption{Fields used in the formulation of the shallow water equations}
    \label{fig:SWEschematics}
  \end{center}
\end{figure}
Suitable boundary conditions have to be included at the boundary of $\Omega$.
The geopotential height is defined as $\phi=gH$ where $H$ denotes the perturbation of the fluid relative to its state at rest and $g=9.81\operatorname{m}\operatorname{s}^{-2}$ is the gravitational acceleration. Furthermore $\phi_B=gH_B$ where $H_B$ is the bathymetry of the sea-floor, measured from the surface at rest (see Fig. \ref{fig:SWEschematics}). 
The momentum $\vec{u}(\vec{x},t)$ is related to the column-averaged velocity of the fluid $\vec{U}(\vec{x},t)$ at each point $\vec{x}\in\Omega$ by $\vec{u}=(\phi_B+\phi)\vec{U}$. $f=f(\vec{x})\in\mathbb{R}$ is the latitude-dependent Coriolis parameter, and the vector $\uperp=(-v,u)$ is obtained by rotating $\vec{u}$ by 90 degrees in the counter-clockwise direction.

Eq. \eqref{eqn:SWE_continuum} has been non-dimensionalised as follows: let $\refr{H}_B$ be a reference height, and define $\refr{\phi}_B=g\refr{H}_B$. Further assume that any horizontal distances are measured in units of some reference length scale $\refr{L}$ and times in units of a reference time scale $\refr{T}$. $\phi$, $\phi_B$ are given in units of $\refr{\phi}_B$ and the momentum $\vec{u}$ in units of $\refr{\phi}_B^{3/2}$. The dimensionless reference gravity wave speed which multiplies the flux terms in Eq. \eqref{eqn:SWE_continuum} is then given by
\begin{equation*}
  c_g = \frac{\refr{T}}{\refr{L}}\refr{\phi}_B^{1/2}.
\end{equation*}

For convenience, we introduce the variable $q=(\phi,\vec{u})=(\phi,u,v)$, which allows us to write Eq. \eqref{eqn:SWE_continuum} in compact form as
\begin{equation}
  \frac{\partial q}{\partial t} + \nabla\cdot\mathcal{F}(q) = s(q),\qquad
\text{or equivalently}\qquad
\frac{\partial q_i}{\partial t} + \partial_j \mathcal{F}_{ij}(q) = s_i(q).
\label{eqn:SWE_continuum_compact}
\end{equation}
Here and in the following summation over repeated indices is implicitly assumed. The exact form of $\mathcal{F}:\mathbb{R}^{3}\rightarrow \mathbb{R}^{3\times2}$ and $s:\mathbb{R}^3\rightarrow \mathbb{R}^3$ can be deduced from Eq. \eqref{eqn:SWE_continuum} and is given by (see \cite{Kang2020})
\begin{xalignat*}{2}
  \mathcal{F}(q) &=
  \begin{pmatrix}
    u & v\\
    \frac{u^2}{\phi+\phi_B}+\phi_B\phi + \frac{1}{2}\phi^2 & \frac{uv}{\phi+\phi_B}\\
    \frac{uv}{\phi+\phi_B} & \frac{v^2}{\phi+\phi_B}+\phi_B\phi + \frac{1}{2}\phi^2
  \end{pmatrix}, &
  s(q) &= 
\begin{pmatrix}
0 \\
c_g \phi\partial_1\phi_B + fv \\
c_g \phi\partial_2\phi_B - fu
\end{pmatrix}.
\end{xalignat*}
The SWEs can be linearised about $0$ (this is a good approximation if $\phi,|\vec{u}|\ll \phi_B$ at every point in the domain) to obtain
\begin{xalignat}{2}
\frac{\partial\phi}{\partial t} + c_g\nabla \cdot \vec{u} &= 0, &
\frac{\partial\vec{u}}{\partial t} + c_g\phi_B \nabla \phi
 &= - f\uperp.
\label{eqn:SWE_continuum_linear}
\end{xalignat}
It is important to note that in a non-rotating frame ($f=0$) fast gravity waves are supported by Eq. \eqref{eqn:SWE_continuum_linear}. This can be seen by taking the divergence of the second equation and inserting it into the time derivative of the first, to obtain the wave equation
\begin{equation*}
\frac{\partial^2\phi}{\partial t^2} - c_g^2\nabla \cdot\left(\phi_B\nabla \phi \right) = 0.\end{equation*}
The (local) speed of the gravity waves is given by $c=c_g\sqrt{\phi_B}$. More generally, in a rotating frame the fundamental solutions of the linear SWEs in Eq. \eqref{eqn:SWE_continuum_linear} are inertia-gravity (Poincar\'{e}) waves. In this case the frequency $\omega$ is related to the wave vector $\vec{k}$ by the dispersion relation $\omega=\pm\sqrt{f^2+c_g^2\phi_B\vec{k}^2}$. In analogy to Eq. \eqref{eqn:SWE_continuum_compact}, the linearised SWEs in Eq. \eqref{eqn:SWE_continuum_linear} can be written as 
\begin{equation}
  \frac{\partial q}{\partial t} + \nabla\cdot\mathcal{F}_L(q) = s(q)
\qquad
\Leftrightarrow
\qquad
\frac{\partial q_i}{\partial t} + \partial_j \left(\mathcal{F}_{L}\right)_{ij}(q) = s_{i}(q)
\label{eqn:SWE_continuum_linear_compact}
\end{equation}
with $q=(\phi,u,v)$ and the linear operator $\mathcal{F}_L:\mathbb{R}^3\rightarrow \mathbb{R}^{3\times 2}$ given by
\begin{equation*}
\mathcal{F}_L(q) = c_g
\begin{pmatrix}
u & v\\
\phi_B\phi & 0 \\
0 & \phi_B \phi
\end{pmatrix}.
\end{equation*}
\section{IMEX HDG methods for the shallow water equations}\label{sec:Methods}
\subsection{Spatial discretisation}
To discretise the SWEs with the discontinuous Galerkin (DG) method, we first construct a mesh $\Omega_h$ of non-overlapping cells $K$ that partition the domain such that $\Omega=\cup_{K\in\Omega_h} K$. The mesh spacing is defined as $h=\max_{K\in\Omega_h}\operatorname{diam}(K)$. Let $\mathcal{E}_h=\cup_{K,K'\in\Omega_h} (K\cap K')$ be the set of facets of the mesh; $\mathcal{E}_h$ is often called the ``\textit{skeleton of $\Omega_h$}''. We further assume that the mesh consists of triangular or quadrilateral cells and has no hanging nodes. More specifically, each facet $e\in\mathcal{E}_h$ is a straight line which forms the side of exactly two triangles or quadrilaterals. For simplicity we use periodic boundary conditions in this paper, which implies that all facets are internal.

\begin{figure}
\begin{center}
\begin{minipage}{0.4\linewidth}
  \includegraphics[width=\linewidth]{\figdir/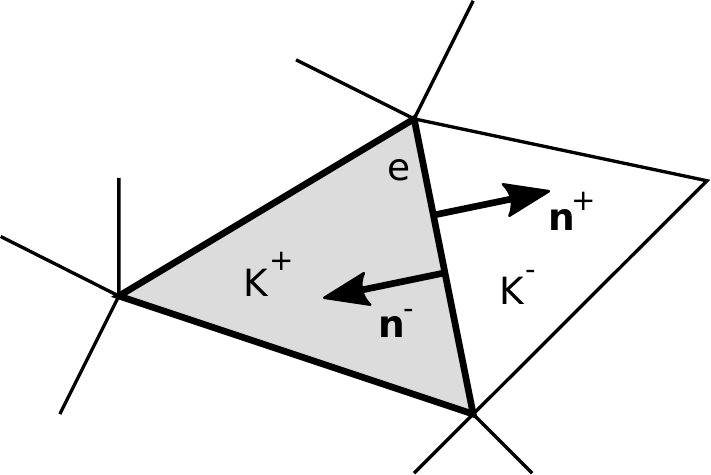}
\end{minipage}
\hspace{8ex}
\begin{minipage}{0.4\linewidth}
\includegraphics[width=\linewidth]{\figdir/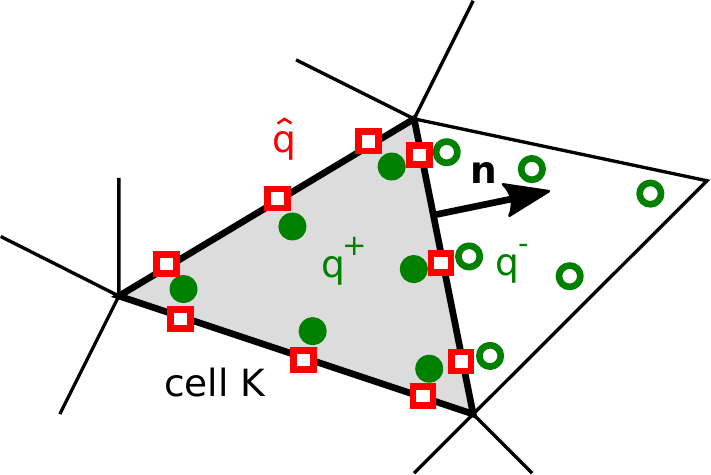}
\end{minipage}
  \caption{Two cells $K^+$, $K^-$ and their respective outer normal vectors $\vec{n}^+$, $\vec{n}^-$ on the joint edge $e=\partial K^+\cap \partial K^-$ (left). Degrees of freedom of the hybridised DG discretisation (right).}
\label{fig:cells}
\end{center}
\end{figure}

By construction, functions defined in the DG space introduced below are not continuous across cell-boundaries. As a result, special care has to be taken when defining quantities on the mesh skeleton $\mathcal{E}_h$. For two neighbouring cells $K^+$ and $K^-$ that share a facet $e=K^+\cap K^-$, let $\vec{n}^-$ be the outward unit normal on the boundary $\partial K^-$ of element $K^-$, and $\vec{n}^+$ the corresponding outward normal of cell $K^+$ (see Fig. \ref{fig:cells}), such that $\vec{n}^+=-\vec{n}^-$ on the shared facet $e$. On each facet $e\in\mathcal{E}_h$, the following average $\favg{\cdot}$, difference $\fdiff{\cdot}$ and jump $\fjump{\cdot}$ operators can be defined for scalar ($s$) and vector valued ($\vec{a}$) quantities:
\begin{xalignat*}{2}
\favg{s} &= \frac{1}{2}\left(s^++s^-\right), &
\favg{\vec{a}} &= \frac{1}{2}\left(\vec{a}^++\vec{a}^-\right),\\
\fdiff{s} &= s^+-s^-, &
\fdiff{\vec{a}} &= \vec{a}^+-\vec{a}^-,\\
\fjump{s} &= s^+\vec{n}^++s^-\vec{n}^-, &
\fjump{\vec{a}} &= \vec{a}^+\cdot\vec{n}^++\vec{a}^-\cdot\vec{n}^-.
\end{xalignat*}
At any point $\vec{x}\in e$ on the facet $e$ and for any scalar- or vector-valued piecewise polynomial functions $s:\Omega\rightarrow \mathbb{R}$, $\vec{a}:\Omega\rightarrow \mathbb{R}^2$ we define $s^{\pm}(\vec{x})=\lim_{\epsilon\rightarrow 0^+} s(\vec{x}\mp\epsilon\vec{n}^{\pm})$ and $\vec{a}^{\pm}(\vec{x})=\lim_{\epsilon\rightarrow 0^+} \vec{a}(\vec{x}\mp\epsilon\vec{n}^\pm)$. Note that the jump $\fjump{s}$ of a scalar quantity $s$ is vector-valued and vice versa.

We write $\mathcal{P}^p(D)$ for the space of polynomials of degree at most $p$ on 
an arbitrary domain $D$. The (scalar-valued) DG space of order $p$ on the mesh $\Omega_h$ is defined as
\begin{equation}
  V_h := \left\{\psi\in L^2(\Omega_h): \psi|_K \in \mathcal{P}^p(K),\forall K\in\Omega_h\right\}.\label{eqn:DG_space_def}
\end{equation}
\pparagraph{Case 1: Linear SWEs}
It is instructive to first discuss the DG discretisation of the linear SWEs in Eq. \eqref{eqn:SWE_continuum_linear_compact}. This can be found by assuming that\footnote{In a slight abuse of notation, we use the same variable names for the continuous fields $q$, $u$, $v$, $\phi$ etc. in Section \ref{sec:SWEs} and for the DG fields discussed from this section onwards.} $q=(\phi,u,v)\in V_h\times V_h\times V_h :=  W_h$, then multiplying by a test function $v\in W_h$ and integrating by parts in each cell $K$ to obtain
\begin{equation}
  \left(\frac{\partial q_i}{\partial t},v_i\right)_K - \left((\mathcal{F}_L)_{ij}(q),\partial_j v_i\right)_K + \langle (\mathcal{F}^*_L)_{ij}(q^+,q^-)n_j,v_i\rangle_{\partial K} = (s_i(q),v_i)_K\qquad\text{for all $v\in W_h$}.
  \label{eqn:SWEweak_form_linear}
\end{equation}
Here $\vec{n}$ is the outward normal on each facet of cell $K$, with $q^{\pm}(\vec{x})=\lim_{\epsilon\rightarrow 0^+} q(\vec{x}\mp\epsilon\vec{n})$. Round and angular brackets denote $L_2$ scalar products in function spaces defined on the cell $K$ and its boundary $\partial K$ respectively:
\begin{xalignat*}{2}
\left(f,g\right)_K &= \left(f,g\right)_{L_2(K)} = \int_K f(\vec{x})g(\vec{x})\; dx, &
\langle f,g\rangle_{\partial K} &= \left( f,g\right)_{L_2(\partial K)} =\int_{\partial K} f(\vec{x})g(\vec{x})\; ds.
\end{xalignat*}
The numerical flux $\mathcal{F}_L^*(q^+,q^-)$ appearing in Eq. \eqref{eqn:SWEweak_form_linear} (still to be defined) will depend on the values $q^+$, $q^-$ of the field on both sides of the facets comprising $\partial K$ and will couple the unknowns between neighbouring cells. For the linear problem both the local Lax-Friedrichs (Rusanov) flux $\mathcal{F}_L^{*\text{(LF)}}$ \cite{Rusanov1961} and the upwind flux $\mathcal{F}_L^{*\text{(up)}}$ (see e.g. \cite{Leveque2002}) are candidates for the numerical flux $\mathcal{F}_L^*$. As shown in \cite{BuiThanh2015,BuiThanh2016}, those are defined as
\begin{subequations}
\begin{align}
  (\mathcal{F}_L^{*\text{(LF)}})_{ij}(q^+,q^-)n_j &= \favg{(\mathcal{F}_L)_{ij}(q)}n_j + \frac{c_g}{2}\sqrt{\phi_B} \fdiff{q_i}\label{eqn:flux_linear_LaxFriedrichs}\\
  (\mathcal{F}_L^{*\text{(up)}})_{ij}(q^+,q^-)n_j &=
\favg{(\mathcal{F}_L)_{ij}(q)}n_j + \frac{c_g}{2}\sqrt{\phi_B} B_{ik}\fdiff{q_k}\label{eqn:flux_linear_upwind} 
\end{align}
\end{subequations}
where the $3\times 3$ matrix $B$ defined on $\partial K$ is given by
\begin{equation*}
  B = \begin{pmatrix}
    1 & 0 & 0\\
    0 & n_x^2 & n_xn_y\\
    0 & n_xn_y & n_y^2
    \end{pmatrix}.
\end{equation*}
If Eq. \eqref{eqn:SWEweak_form_linear} is integrated in time with a semi-implicit method, a coupled sparse linear system for the fields $(\phi,\vec{u})$ has to be solved in each timestep. The standard approach in many atmospheric models is to eliminate the momentum unknowns $\vec{u}$ from the \textit{discrete} equations and solve the resulting elliptic Schur-complement system for the height perturbation $\phi$, followed by reconstruction of the momentum field (see e.g. \cite{Wood2014} for the equivalent procedure applied to the full Navier Stokes equations). Unfortunately, for the DG discretisation considered here, the numerical flux in Eq. \eqref{eqn:SWEweak_form_linear} introduces artificial diffusion terms in the momentum equation. Although those terms disappear in the limit $h\rightarrow 0$, they result in a dense Schur-complement operator which renders this naive Schur-complement approach unfeasible here. It is possible to address this issue by forming an approximate Schur-complement \label{page:approximate_schur} and using the resulting approximate solver as a preconditioner for a Krylov subspace method. A similar approach is pursued to deal with the non-diagonal velocity mass matrix in conforming mixed finite element discretisations in \cite{Mitchell2016,Melvin2019}. However, this requires additional iterations over the large $(\phi,\vec{u})$-system. As our numerical results in Section \ref{sec:Results} demonstrate, this approximate Schur-complement method (which essentially drops the artificial diffusion terms in the preconditioner) is not competitive with the hybridised DG method pursued in this paper since the latter allows the formulation of an exact Schur-complement in a suitably defined finite element space on the facets.
\subsection{Hybridised DG discretisation}
As described in \cite[section 2]{Kang2020}, IMEX time integrators separate the operator which describes the time evolution of the shallow water equations into a linear part, which is treated implicitly, and an explicit non-linear part. By using a \textit{hybridised} DG discretisation for the linear part it is possible to use the efficient non-nested multigrid techniques which are discussed in this paper to solve the implicit linear problem. The explicit part can be discretised with a standard DG method.

The key idea of the hybridised DG method described in \cite{BuiThanh2015,BuiThanh2016} is to introduce a new variable $\what{q}$, which is associated with the skeleton $\mathcal{E}_h$ of the mesh, see Fig. \ref{fig:cells} (right). As we will show below, it is possible to form an exact Schur-complement in $\what{q}$. For this, recall that the facets $e\in\mathcal{E}_h$ are straight lines bounding the cells. On these we define the piecewise polynomial space
\begin{xalignat}{2}
\Lambda_h &= \{\lambda\in L^2(\mathcal{E}_h): \lambda|_e \in \mathcal{P}^p(e), \forall e\in\mathcal{E}_h\} &\text{and set}\qquad
\what{\Lambda}_h &= \Lambda_h\times \Lambda_h\times \Lambda_h.
\label{eqn:DG_trace_def}
\end{xalignat}
This allows the introduction of a flux $(\what{\mathcal{F}}_L)_{ij}(q,\what{q})$ which only depends on the value $q$ inside the cell $K$ and the field $\what{q}$ on its boundary. To see this, recall that for hyperbolic conservation laws the numerical upwind flux can be constructed by solving a Riemann problem on the interface between two cells (see e.g. section 2.4 of \cite{Hesthaven2007}). The solution contains both the field $q$ inside the cell and an intermediate state $q^*$, which is usually eliminated with a jump condition. Since this jump condition involves the value of fields on both sides of the interface, the resulting flux in Eq. \eqref{eqn:flux_linear_upwind} depends on $q^+$ and $q^-$. As argued in \cite{BuiThanh2015,BuiThanh2016}, the HDG method represents the intermediate state as $\what{q}=q^*$ in the space $\what{\Lambda}_h$ instead of eliminating $q^*$; the jump condition is enforced weakly.

$\what{\mathcal{F}}_L$ replaces the numerical flux $\mathcal{F}^*_L$ in Eq. \eqref{eqn:SWEweak_form_linear} and is given by either $\what{\mathcal{F}}_L^{(\text{LF})}$ (Lax-Friedrichs) or $\what{\mathcal{F}}_L^{(\text{up})}$ (upwind) defined on the boundary $\partial K$ of each cell as 
\begin{subequations}
\begin{align}
  (\what{\mathcal{F}}_L^{(\text{LF})})_{ij}(q,\what{q})n_j &= (\mathcal{F}_L)_{ij}(q)n_j + c_g\sqrt{\phi_B}(q_i-\what{q}_i)\label{eqn:flux_hybrid_linear_LaxFriedrichs}\\
(\what{\mathcal{F}}_L^{(\text{up})})_{ij}(q,\what{q})n_j &= (\mathcal{F}_L)_{ij}(q)n_j + c_g\sqrt{\phi_B}B_{ik}(q_k-\what{q}_k).
\label{eqn:flux_hybrid_linear_upwind} 
\end{align}
\end{subequations}
To close the system of equations, continuity of the flux is enforced weakly by requiring that on each facet $e\in \mathcal{E}_h$
\begin{equation}
  \langle \favg{(\what{\mathcal{F}}_L)_{ij}(q,\what{q})n_j},\what{v}_i\rangle_e = 0\qquad\text{for all test functions $\what{v}\in\what{\Lambda}_h$}.
  \label{eqn:flux_condition}
\end{equation}
To simplify notation, introduce the following abbreviations:
\begin{equation}
\begin{aligned}
  \mathcal{L}(q,v) &=\sum_{K\in\Omega_h}\left\{\left((\mathcal{F}_L)_{ij}(q),\partial_jv_i\right)_K - \langle (\mathcal{F}^*_L)_{ij}(q^+,q^-)n_j,v_i\rangle_{\partial K}\right\}\\
\what{\mathcal{L}}(q,\what{q},v) &= \sum_{K\in\Omega_h}\left\{\left((\mathcal{F}_L)_{ij}(q), \partial_jv_i\right)_K - \langle (\what{\mathcal{F}}_L)_{ij}(q,\what{q})n_j,v_i\rangle_{\partial K}\right\}\\
\mathcal{N}_0(q,v) &= \sum_{K\in\Omega_h}(s_{i}(q),v_i)_K\quad\text{(see Eq. \eqref{eqn:SWEweak_form_linear})}\\
\mathcal{M}(q,v) &= \sum_{K\in\Omega_h}(q_i,v_i)_K\\
\Xi(q,\what{q},\what{v}) &= \sum_{e\in\mathcal{E}_h}\langle \favg{(\what{\mathcal{F}}_L)_{ij}(q,\what{q})n_j},\what{v}_i\rangle_{e}
\end{aligned}
\label{eqn:L_definitions}
\end{equation}
With the definitions in Eq. \eqref{eqn:L_definitions}, after summing over all cells $K\in\Omega_h$ the weak form of the linear SWEs discretised with DG in Eq. \eqref{eqn:SWEweak_form_linear} can be written in condensed form as
\begin{equation}
  \frac{\partial \mathcal{M}(q,v)}{\partial t} = \mathcal{N}_0(q,v)+\mathcal{L}(q,v)\qquad\text{for all $v\in W_h$}.\label{eqn:linear_non_hybridized}
\end{equation}
The equivalent HDG discretisation of Eq. \eqref{eqn:SWEweak_form_linear} is obtained by replacing $\mathcal{F}^*_L$ by $\what{\mathcal{F}}_L$ in Eq. \eqref{eqn:linear_non_hybridized} and enforcing the continuity condition in Eq. \eqref{eqn:flux_condition}. This results in
\begin{xalignat}{2}
\frac{\partial \mathcal{M}(q,v)}{\partial t} &= \mathcal{N}_0(q,v)
+ \what{\mathcal{L}}(q,\what{q},v), &\text{subject to}\qquad
\Xi(q,\what{q},\what{v}) &= 0\qquad\text{for all $v\in W_h$, $\what{v}\in\what{\Lambda}_h$}.
\label{eqn:linear_hybridized}
\end{xalignat}
As has been shown in \cite{BuiThanh2016}, the solution $q$ of the original, non-hybridized system in Eq. \eqref{eqn:linear_non_hybridized} is identical to the solution $q$ of the HDG problem given in Eq. \eqref{eqn:linear_hybridized}. In general, the additional field $\what{q}=(\what{\phi},\what{\vec{u}})=(\what{\phi},\what{u},\what{v})\in\what{\Lambda}_h$ introduced in Eq. \eqref{eqn:linear_hybridized} on the facets has three components. As will be discussed in Section \ref{sec:Schur_MG_linear_eqn} below, $\what{\vec{u}}$ can be eliminated for the upwind flux, whereas $\what{\phi}$ does not enter the equations if the Lax-Friedrichs method is used.
\pparagraph{Case 2: non-linear SWEs}
Similarly, the weak form of the non-linear SWEs in Eq. \eqref{eqn:SWE_continuum_compact} can be written for each cell $K\in\Omega_h$ as
\begin{equation}
  \left(\frac{\partial q_i}{\partial t},v_i\right)_K - \left(\mathcal{F}_{ij}(q),\partial_j v_i\right)_K + \langle \mathcal{F}^*_{ij}(q^+,q^-)n_j,v_i\rangle_{\partial K} = (s_i(q),v_i)_K\qquad\text{for all $v\in W_h$}.
  \label{eqn:SWEnonlinear_weak_form}
\end{equation}
where $\mathcal{F}^*=\mathcal{F}^{(\text{LF})}$ is the non-linear Lax-Friedrichs flux
\begin{equation}
  \mathcal{F}^{\text{(LF)}}_{ij}(q^+,q^-)n_j = \favg{\mathcal{F}_{ij}(q)}n_j + \frac{c_g}{2}\tau^*\fdiff{q_i}\label{eqn:flux_LaxFriedrichs}
\end{equation}
with $\tau^* = \max\left\{|\vec{n}\cdot\vec{u}^+|+\sqrt{\phi_B+\phi^+},|\vec{n}\cdot\vec{u}^-|+\sqrt{\phi_B+\phi^-}\right\}$. Further define
\begin{equation}
\mathcal{N}(q,v) = \sum_{K\in\Omega_h}\left\{
\left((\mathcal{F}-\mathcal{F}_L)_{ij}(q),\partial_j v_i\right)_K - \langle (\mathcal{F}^*-\mathcal{F}^*_L)_{ij}(q^+,q^-)n_j,v_i\rangle_{\partial K} + (s_i(q),v_i)_K\right\}
\label{eqn:NL_definitions}
\end{equation}
With the definitions in Eqs. \eqref{eqn:L_definitions} and \eqref{eqn:NL_definitions}, after summing over all cells $K\in\Omega_h$ the weak form of the DG discretisation in Eq. \eqref{eqn:SWEnonlinear_weak_form} can be written in condensed form as
\begin{equation}
  \frac{\partial \mathcal{M}(q,v)}{\partial t} = \mathcal{N}(q,v)+\mathcal{L}(q,v)\qquad\text{for all $v\in W_h$}.\label{eqn:non_linear_non_hybridized}
\end{equation}
As above, the equivalent HDG discretisation of Eq. \eqref{eqn:SWEnonlinear_weak_form} is obtained by replacing $\mathcal{F}^*_L$ by $\what{\mathcal{F}}_L$ in the linear term $\mathcal{L}$ in Eq. \eqref{eqn:non_linear_non_hybridized} and enforcing the continuity condition in Eq. \eqref{eqn:flux_condition}. This results in
\begin{xalignat}{2}
  \frac{\partial \mathcal{M}(q,v)}{\partial t} &= \mathcal{N}(q,v) + \what{\mathcal{L}}(q,\what{q},v), &\text{subject to}\quad
  \Xi(q,\what{q},\what{v}) &= 0\qquad\text{for all $v\in W_h$, $\what{v}\in\what{\Lambda}_h$}.  \label{eqn:nonlinear_hybridized}
\end{xalignat}
By splitting the right hand side into two terms, we have isolated the fast dynamics due to gravity waves in $\what{\mathcal{L}}(q,\what{q},v)$, whereas any slower modes are described by $\mathcal{N}(q,v)$. This observation is crucial for the construction of suitable semi-implicit time integrators, which avoid numerical instabilities due to the fast gravity waves.
\subsection{Time discretisation}\label{sec:IMEXmethods}
To integrate Eq. \eqref{eqn:nonlinear_hybridized} (Eq. \eqref{eqn:linear_hybridized} can be dealt with in exactly the same way) in time, the term $\mathcal{N}(q,v)$ is integrated explicitly, whereas $\what{\mathcal{L}}(q,\what{q},v)$ is treated implicitly. Denoting the solution at time $t=n\Delta t$ by $q^{(n)}$, a simple scheme would for example be the following ``\textit{$\theta$-method}'':
\begin{equation}
  \frac{\mathcal{M}(q^{(n+1)},v)-\mathcal{M}(q^{(n)},v)}{\Delta t}
  = \mathcal{N}(q^{(n)},v) + \theta \what{\mathcal{L}}(q^{(n+1)},\what{q}^{(n+1)},v)
  + (1-\theta) \mathcal{L}(q^{(n)},v)\qquad\text{for all $v\in W_h$}
\label{eqn:theta_method}
\end{equation}
with $\theta\in[0,1]$ and subject to the flux condition Eq. \eqref{eqn:flux_condition}. The equivalent expression for the linear SWEs in Eq. \eqref{eqn:linear_hybridized} can be obtained by replacing $\mathcal{N}\rightarrow\mathcal{N}_0$ in Eq. \eqref{eqn:theta_method}. At each time step, calculating the field $q^{(n+1)}$ requires the solution of the \textit{linear} system
\begin{equation}
\begin{aligned}
  \mathcal{M} (q^{(n+1)},v) - \theta\Delta t \what{\mathcal{L}}(q^{(n+1)},\what{q}^{(n+1)},v)
  &= \mathcal{R}(q^{(n)},\what{q}^{(n)},v) \\
&:= \mathcal{M}(q^{(n)},v) + \Delta t\left(\mathcal{N}(q^{(n)},v)+(1-\theta)\mathcal{L}(q^{(n)},v)\right)\\
\text{subject to}\qquad\Xi(q^{(n+1)},\what{q}^{(n+1)},\what{v}) &= 0\qquad\text{for all $v\in W_h$, $\what{v}\in\what{\Lambda}_h$.}
\end{aligned}
\label{eqn:theta_linear_system}
\end{equation}
Note that for a purely explicit method ($\theta=0$), the system matrix arising from the left hand side of the system in Eq. \eqref{eqn:theta_linear_system} is simply the mass matrix (obviously, in this case Eq. \eqref{eqn:theta_method} reduces to the explicit Euler integrator). On the other hand, $s$-stage IMEX methods \cite{Ascher1995,Ascher1997,Pareschi2000,Kennedy2003, Weller2013} are a generalisation of the method in Eq. \eqref{eqn:theta_method}.
They require the construction of $s$ intermediate states $\{(Q^{(i)},\what{Q}^{(i)})\}_{i=1}^s$ which are obtained by solving the system of linear equations
\begin{equation}
  \begin{aligned}
    \mathcal{M}(Q^{(i)},v) &= \mathcal{M}(q^{(n)},v) + \Delta t\sum_{j=1}^{i-1}
    a_{ij} \mathcal{N}(Q^{(j)},v) + \Delta t \sum_{j=1}^{i}\widetilde{a}_{ij}\what{\mathcal{L}}(Q^{(j)},\what{Q}^{(j)},v),\\
    \text{with}\qquad\Xi(Q^{(i)},\what{Q}^{(i)},\what{v})&=0\qquad\text{for each stage $i=1,\dots,s$ and all $v\in W_h$, $\what{v}\in\what{\Lambda}_h$}.\label{eqn:IMEX_HDG_I}
  \end{aligned}
\end{equation}
The state $q^{(n+1)}$ at the next timestep satisfies
\begin{equation}
  \mathcal{M}(q^{(n+1)},v) = \mathcal{M}(q^{(n)},v) + \Delta t\sum_{i=1}^s b_i \mathcal{N}(Q^{(i)},v) + \Delta t\sum_{i=1}^s \widetilde{b}_i \what{\mathcal{L}}(Q^{(i)},\what{Q}^{(i)},v).\label{eqn:IMEX_HDG_II}
\end{equation}
and is obtained by an additional mass-solve. Each particular IMEX method is determined by a choice of coefficients $a_{ij}$, $\widetilde{a}_{ij}$, $b_i$, $\widetilde{b}_i$ which are commonly known as the Butcher tableau coefficients for the explicit and implicit terms.
The equations in Eq. \eqref{eqn:IMEX_HDG_I} can be solved stage-by-stage. At every stage a \textit{linear} system of the abstract general form
\begin{equation}
  \what{\mathcal{A}}(q,\what{q},v,\what{v})=\mathcal{M}(q,v) - \alpha\Delta t \what{\mathcal{L}}(q,\what{q},v) + \Xi(q,\what{q},\what{v}) = \mathcal{R}(v)
  \qquad\text{for all $v\in W_h$, $\what{v}\in\what{\Lambda}_h$}.
  \label{eqn:general_linear_system}
\end{equation}
with some positive constant $\alpha\in \mathbb{R}^+$ has to be solved\footnote{
Note that the \textit{``$\theta$-method''} in Eq. \eqref{eqn:theta_method} could be written as a 2-stage IMEX method with
\begin{xalignat*}{4}
  a &= \begin{pmatrix}
    0 & 0 \\
    1 & 0
  \end{pmatrix}
  &
  \widetilde{a} &= \begin{pmatrix}
    0 & 0 \\
    1-\theta & \theta
  \end{pmatrix}
  &
  b &= \begin{pmatrix}
    1 \\ 0
  \end{pmatrix}
  &
  \widetilde{b} &= \begin{pmatrix}
    1-\theta\\\theta
  \end{pmatrix}
\end{xalignat*}
but this would be inefficient since it requires an additional mass-solve in the first stage.}. Typically the value of $\alpha$ is different in each stage of the IMEX method and the right hand side $\mathcal{R}$ depends on previously calculated fields. To see that $\what{\mathcal{A}}$ in Eq. \eqref{eqn:general_linear_system} is indeed linear in the combined field $(q,\what{q})\in W_h\times \what{\Lambda}_h$, use the definitions of $\mathcal{M}$, $\what{\mathcal{L}}$ and $\Xi$ in Eqs. \eqref{eqn:flux_hybrid_linear_LaxFriedrichs}, \eqref{eqn:flux_hybrid_linear_upwind}, \eqref{eqn:L_definitions} and observe that
  \begin{equation*}
      \what{\mathcal{A}}(c_1 q_1 + c_2 q_2,c_1 \what{q}_1+c_2\what{q}_2,v,\what{v}) = c_1 \what{\mathcal{A}}(q_1,\what{q}_1,v,\what{v})+c_2\what{\mathcal{A}}(q_2,\what{q}_2,v,\what{v})\qquad\text{for all $v\in W_h$, $\what{v}\in\what{\Lambda_h}$}
  \end{equation*}
and for all $c_1,c_2\in\mathbb{R}$, $(q_1,\what{q}_1),(q_2,\what{q}_2)\in W_h\times\what{\Lambda}_h$.
\section{Schur-complement multigrid solver}\label{sec:IMEXHDG}
  Constructing efficient solvers for Eq. \eqref{eqn:general_linear_system} is the topic of this paper. In the following section we give explicit expressions for $\what{\mathcal{A}}$ for the Lax-Friedrichs and upwind flux. We then describe a Schur complement approach which reduces the problem to solving an elliptic system in the flux variable $\what{q}$ on the facets. This flux system is solved with a new non-nested multigrid preconditioner based on ideas in \cite{Cockburn2014}.
\subsection{Linear equation}\label{sec:Schur_MG_linear_eqn}
Recall that $q=(\phi,\vec{u})\in W_h$ in the domain, $\what{q}=(\what{\phi},\what{\vec{u}})\in\what{\Lambda}_h$ on the facets, and write the corresponding test-functions as $v=(\psi,\vec{w})\in W_h$, $\what{v}=(\what{\psi},\what{\vec{w}})\in\what{\Lambda}_h$. Using the definitions in Eqs. \eqref{eqn:flux_linear_LaxFriedrichs} and \eqref{eqn:flux_linear_upwind}, the bilinear form $\what{\mathcal{A}}(q,\what{q},v,\what{v})$ on the left hand side of Eq. \eqref{eqn:general_linear_system} can be written down explicitly for the upwind- and Lax-Friedrichs flux. As shown in \cite{BuiThanh2016}, the resulting bilinear form $\mathcal{A}^{(\text{up})}$ for the upwind flux does not depend on $\what{\vec{u}}$. For the Lax-Friedrichs flux the form $\mathcal{A}^{(\text{LF})}$ does not depend on $\what{\phi}$. Explicit expressions for those forms are given as follows:
\begin{description}
\item{\textbf{Upwind flux:}}
\begin{equation}
  \begin{aligned}
    \what{\mathcal{A}}^{(\text{up})}(q,\what{q},v,\what{v})=  \what{\mathcal{A}}^{(\text{up})}(\phi,\vec{u},\what{\phi},\psi,\vec{w},\what{\psi})
    &:= \left(\phi\psi + \vec{u}\cdot\vec{w}\right)_{\Omega_h} - c_g\alpha\Delta t
  \Big[
    \left(\vec{u}\cdot\nabla\psi + \phi_B\phi\nabla\cdot\vec{w}\right)_{\Omega_h}\\
    &\quad-\;\;\left(\fjump{\vec{u}\psi}+2\sqrt{\phi_B}\left(\favg{\phi\psi}-\what{\phi}\favg{\psi}\right)+\phi_B\what{\phi}\fjump{\vec{w}}\right)_{\mathcal{E}_h}
    \Big]\\
  &\quad+\;\;\left(\what{\psi}\left[\fjump{\vec{u}}+2\sqrt{\phi_B}\left(\favg{\phi}-\what{\phi}\right)\right]\right)_{\mathcal{E}_h}
  \end{aligned}\label{eqn:bilinear_HDG_upwind}
\end{equation}
\item{\textbf{Lax-Friedrichs flux:}}
\begin{equation}
  \begin{aligned}
  \what{\mathcal{A}}^{(\text{LF})}(q,\what{q},v,\what{v})=  \what{\mathcal{A}}^{(\text{LF})}(\phi,\vec{u},\what{\vec{u}},\psi,\vec{w},\what{\vec{w}})
  &:=\left(\phi\psi + \vec{u}\cdot\vec{w}\right)_{\Omega_h} - c_g\alpha\Delta t
  \Big[
    \left(\vec{u}\cdot\nabla\psi + \phi_B\phi\nabla\cdot\vec{w}\right)_{\Omega_h}\\
    &\quad-\;\;\left(
    \what{\vec{u}}\fjump{\psi}+2\sqrt{\phi_B}\left(
    \favg{\vec{u}\cdot\vec{w}}-\what{\vec{u}}\cdot\favg{\vec{w}}
    \right)
    +\phi_B\fjump{\phi\vec{w}}
    \right)_{\mathcal{E}_h}
    \Big]\\
   &\quad+\;\;\left(\what{\vec{w}}\cdot\left[
  \phi_B\fjump{\phi}+2\sqrt{\phi_B}\left(
  \favg{\vec{u}}-\what{\vec{u}}
  \right)
  \right]\right)_{\mathcal{E}_h}
  \end{aligned}\label{eqn:bilinear_HDG_lax}
\end{equation}
\end{description}
The system in Eq. \eqref{eqn:general_linear_system} can be solved with a Schur-complement approach by eliminating the field $q$ defined on the elements and leaving a problem for $\what{q}$ defined on the facets. For this write
\begin{equation}
    \what{\mathcal{A}}(q,\what{q},v,\what{v}) = \left(\mathfrak{A}q,v\right)_{\Omega_h}+\left(\what{q}, \mathfrak{B} ^\top v\right)_{\mathcal{E}_h} + \left(\mathfrak{C}q,\what{v}\right)_{\mathcal{E}_h} - \left(\mathfrak{M}\what{q},\what{v}\right)_{\mathcal{E}_h}=\mathcal{R}(v) = \left(f,v\right)_{\Omega_h}\label{eqn:ABCMform}
\end{equation}
where $f\in W_h$ is the Riesz representer of $\mathcal{R}(v)$. The scalar products in Eq. \eqref{eqn:ABCMform} are defined as
\begin{equation*}
\begin{aligned}
  \left(q,v\right)_{\Omega_h}&=\left(\phi\psi+\vec{u}\cdot\vec{w}\right)_{\Omega_h}:=\int_{\Omega}\left(\phi(\vec{x})\psi(\vec{x})+\vec{u}(\vec{x})\cdot\vec{w}(\vec{x})\right)\;dx,\\
  \left(\what{q},\what{v}\right)_{\mathcal{E}_h}&=\left(\what{\phi}\what{\psi}+\what{\vec{u}}\cdot\what{\vec{w}}\right)_{\mathcal{E}_h}
:=\int_{\mathcal{E}_h}\left(\what{\phi}(\vec{x})\what{\psi}(\vec{x})+\what{\vec{u}}(\vec{x})\cdot\what{\vec{w}}(\vec{x})\right)\;ds
\end{aligned}
\end{equation*}
and the linear operators
\begin{xalignat}{4}
  \mathfrak{A}&: W_h\rightarrow W_h, &
  \mathfrak{B}&:\what{\Lambda}_h\rightarrow W_h, &
  \mathfrak{C}&: W_h\rightarrow\what{\Lambda}_h, &
  \mathfrak{M}&:\what{\Lambda}_h\rightarrow\what{\Lambda}_h.\label{eqn:ABCM_operators}
\end{xalignat}
Explicit expressions for the operators in Eq. \eqref{eqn:ABCM_operators} can be deduced from Eqs. \eqref{eqn:bilinear_HDG_lax} and \eqref{eqn:bilinear_HDG_upwind} and are given both for the Lax-Friedrichs- and for the upwind flux in \ref{sec:ABCM_operators}.
\subsection{Schur complement solve}\label{sec:schur_solve}
Let the inverse $\mathfrak{F}: W_h\rightarrow W_h$ of $\mathfrak{A}$ be defined in the obvious way as
\begin{equation}
  \left(\mathfrak{F}\mathfrak{A}q,v\right)_{\Omega_h}=(q,v)_{\Omega_h}\qquad\text{for all $q,v\in W_h$}.\label{eqn:AFinverse}
\end{equation}
It can now be shown that Eq. \eqref{eqn:general_linear_system} can be solved in three steps.
\begin{description}
\item[Step 1.] Given $f$ in Eq. \eqref{eqn:ABCMform}, calculate the modified right hand side
\begin{equation*}
\what{\mathcal{R}}(\what{v})=\left(\mathfrak{C}\mathfrak{F}f,\what{v}\right)_{\mathcal{E}_h}\qquad\text{for all $\what{v}\in\what{\Lambda}_h$}.
\end{equation*}
\item[Step 2.] Solve the Schur-complement system
  \begin{equation}
    \what{\mathcal{S}}(\what{q},\what{v})=
    \left(\mathfrak{M}\what{q},\what{v}\right)_{\mathcal{E}_h}+\left(\mathfrak{CFB}\what{q},\what{v}\right)_{\mathcal{E}_h} = \what{\mathcal{R}}(\what{v})\qquad\text{for all $\what{v}\in\what{\Lambda}_h$}
    \label{eqn:schur_system}
  \end{equation}
  for the flux unknown $\what{q}$.
\item[Step 3.] Reconstruct the solution $q$ as
  \begin{equation}
    q = \mathfrak{F}\left(f-\mathfrak{B}\what{q}\right).
    \label{eqn:reconstruct_solution}
  \end{equation}
\end{description}
Note that the operator $\widehat{\mathcal{A}}$ defined in Eq. \eqref{eqn:general_linear_system} contains the term $\Xi(q,\widehat{q},\widehat{v})$ and therefore the Schur-complement solution procedure implicitly enforces the continuity constraint in the second line of Eq. \eqref{eqn:theta_linear_system}. The crucial observation is that both $\mathfrak{A}$ and $\mathfrak{F}$ only couple unknowns within one grid cell and therefore can both be represented by a set of cell-local matrices. Since further $\mathfrak{B}$ and $\mathfrak{C}$ couple the unknowns in one grid cell to the flux unknowns on its boundary, the matrix representation of the operator $\mathfrak{CFB}:\what{\Lambda}_h\rightarrow\what{\Lambda}_h$ in Eq. \eqref{eqn:schur_system} is sparse. Recall further that for the Lax-Friedrichs flux only $\what{\vec{u}}$ and $\what{\vec{w}}$ appear in Eq. \eqref{eqn:general_linear_system} whereas for the upwind flux, the equation only contains $\what{\phi}$ and $\what{\psi}$. Therefore Eq. \eqref{eqn:schur_system} reduces to
\begin{subequations}
\begin{align}
\what{\mathcal{S}}^{(\text{up})}(\what{\phi},\what{\psi}) &= \what{\mathcal{R}}^{(\text{up})}(\what{\psi})&\text{for all $\what{\psi}\in \Lambda_h$}&\qquad\text{(upwind)}
\label{eqn:Schur_upwind}\\
    \what{\mathcal{S}}^{(\text{LF})}(\what{\vec{u}},\what{\vec{w}}) &= \what{\mathcal{R}}^{(\text{LF})}(\what{\vec{w}})&\text{for all $\what{\vec{w}}\in \Lambda_h\times\Lambda_h$} & \qquad\text{(Lax-Friedrichs)}
    \label{eqn:Schur_LaxFriedrichs}
\end{align}
\end{subequations}
in those cases. Before discussing the solution of Eq. \eqref{eqn:schur_system} in Section \ref{sec:non-nested-multigrid}, we briefly discuss the bilinear form in Eq. \eqref{eqn:bilinear_HDG_upwind} in the context of previous work.
\subsection{Relationship to previous work in \cite{Cockburn2014}}\label{sec:relationship_to_Cockburn}
For the HDG upwind flux, the Schur-complement system in Eq. \eqref{eqn:schur_system} is very similar to the elliptic problem written down in Theorem 2.1 of \cite{Cockburn2014}. To see this, observe that for constant $c$, $\tau_K$ Eq. (2.5) in \cite{Cockburn2014} can be written in our notation as
\begin{equation*}
\what{\mathcal{A}}^{(\text{CGDT})}(\phi,\vec{u},\what{\phi},\psi,\vec{w},\what{\psi}) = b(\vec{w},\psi)
\end{equation*}
for some right hand side $b$. The bilinear form $\what{\mathcal{A}}^{(\text{CGDT})}$ is given by
\begin{equation}
  \begin{aligned}
    \what{\mathcal{A}}^{(\text{CGDT})}(\phi,\vec{u},\what{\phi},\psi,\vec{w},\what{\psi})
    &:= \left(\vec{u}\cdot\vec{w}\right)_{\Omega_h} - c^{-1}
  \Big[
    \left(\kappa\vec{u}\cdot\nabla\psi + \phi\nabla\cdot\vec{w}\right)_{\Omega_h}\\
    &\quad-\;\;\left(\kappa\fjump{\vec{u}\psi}+2\kappa\tau_K\left(\favg{\phi\psi}-\what{\phi}\favg{\psi}\right)+\what{\phi}\fjump{\vec{w}}\right)_{\mathcal{E}_h}
    \Big]\\
  &\quad+\;\;\left(\what{\psi}\left[\fjump{\vec{u}}+2\tau_K\left(\favg{\phi}-\what{\phi}\right)\right]\right)_{\mathcal{E}_h}
  \end{aligned}\label{eqn:bilinear_HDG_Cockburn}
\end{equation}
where $\kappa$ is a constant. If we assume spatially non-varying bathymetry ($\phi_B=\text{constant}$) for the shallow water problem and set $\kappa=\phi_B^{-1}$, $c^{-1}=\phi_B c_g\alpha \Delta t$, $\tau_K=\sqrt{\phi_B}$ in Eq. \eqref{eqn:bilinear_HDG_Cockburn}, the resulting bilinear form differs from the one given in our Eq. \eqref{eqn:bilinear_HDG_upwind}  only by a zero order term, namely:
\begin{equation*}
  \what{\mathcal{A}}^{(\text{up})}(\phi,\vec{u},\what{\phi},\psi,\vec{w},\what{\psi}) = \what{\mathcal{A}}^{(\text{CGDT})}(\phi,\vec{u},\what{\phi},\psi,\vec{w},\what{\psi}) + \left(\phi,\psi\right)_{\Omega_h}.
\end{equation*}
It can then be shown \cite{Betteridge2019} that in addition to slightly modifying the lifting operators $\oldvec{\mathcal{Q}}$ and $\mathcal{U}$ defined on page 5 of \cite{Cockburn2014}, this gives rise to an additional, zero order term in the elliptic problem given in Theorem 2.1 of \cite{Cockburn2014}. This zero order term can be identified with the $\left(\mathfrak{M}\what{q},\what{v}\right)_{\mathcal{E}_h}$ contribution to $\what{S}(\what{q},\what{v})$ written down in Eq. \eqref{eqn:schur_system} above. The additional term has a positive sign and does not affect the positive definiteness of $\what{S}$. In fact, since it adds a positive shift to all eigenvalues, it reduces the condition number of the corresponding matrix and allows the use of a shallow multigrid hierarchy as discussed in \cite{Mueller2014,Sandbach2015}. The above observations motivated the construction of the non-nested multigrid algorithm described in Section \ref{sec:non-nested-multigrid}. Further mathematical analysis and an extension of the proof in \cite{Cockburn2014} to the HDG discretisation of the shallow water equations with the Lax-Friedrichs flux is beyond the scope of this paper and could be pursued in a future publication. 
\subsection{Non-nested multigrid algorithm}\label{sec:non-nested-multigrid}
\begin{figure}
\begin{center}
\includegraphics[width=0.75\linewidth]{\figdir/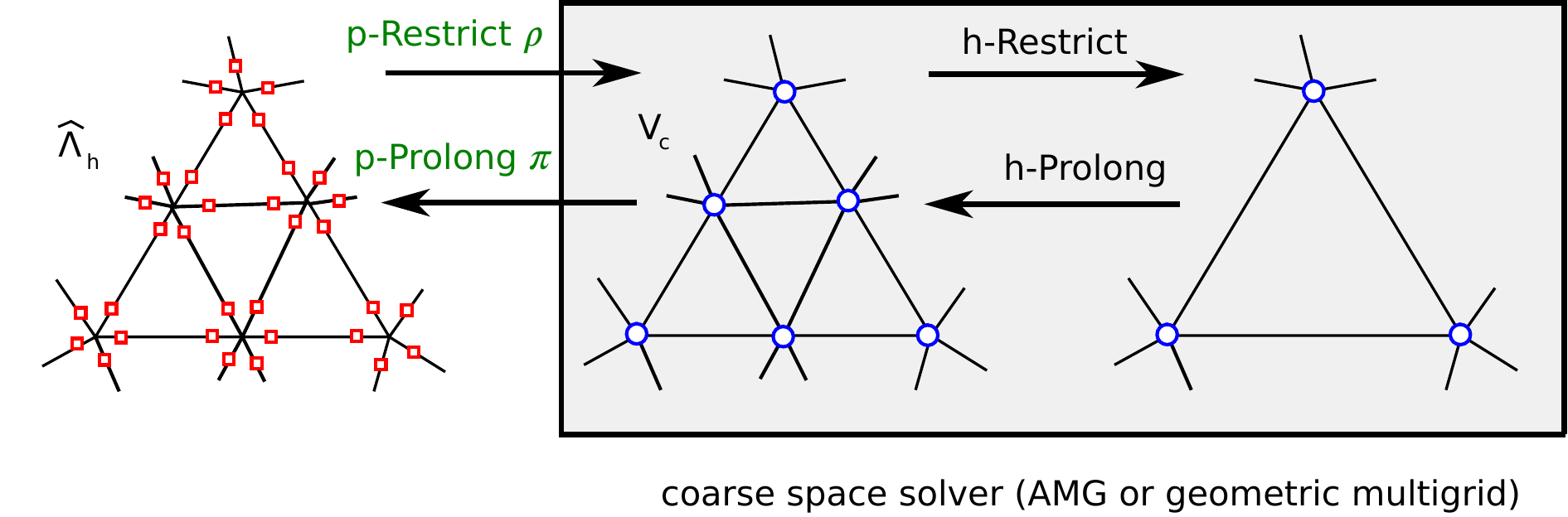}
\caption{Multigrid algorithm. The coarse space problem in the space $V_c$ is solved with AMG or geometric multigrid.}
\label{fig:multigrid_schematic}
\end{center}
\end{figure}

In \cite{Kang2020} a direct method is used to solve the linear system in Eq. \eqref{eqn:schur_system} for the flux unknowns. This is inefficient for large problems since the cost grows rapidly as the number of unknowns increases. To overcome this issue we now describe how Eq. \eqref{eqn:schur_system} can be solved with an iterative Krylov-subspace method, which is preconditioned with a bespoke multigrid algorithm. For this let $V_c$ be a coarse function space, which is not necessarily a subspace of $\what{\Lambda}_h$. Assume that there is prolongation operator $\pi: V_c\rightarrow \what{\Lambda}_h$, which naturally induces a corresponding restriction $\rho: \what{\Lambda}^*_h \rightarrow V^*_c$ on the dual spaces (marked by an asterisk $*$). For any $\mathcal{R}\in\what{\Lambda}^*_h$ the restriction is given as
\begin{equation*}
  \rho(\mathcal{R}) = \mathcal{R}_c=\mathcal{R}\circ\pi \in V_c^*, \quad\text{in other words $\mathcal{R}_c(q_c)=\mathcal{R}(\pi q_c)$ for all $q_c\in V_c$.}
\end{equation*}
Further assume the existence of a coarse space bilinear form\footnote{One possible way (not pursued here) of defining this would be to set $\mathcal{S}_c(q_c,v_c) = \what{\mathcal{S}}(\pi q_c,\pi v_c)$ for all $q_c, v_c\in V_c$.} $\mathcal{S}_c:V_c\times V_c \mapsto \mathbb{R}$. Based on this, a two-level V-cycle for approximately solving Eq. \eqref{eqn:schur_system} is written down in Alg. \ref{alg:twolevel_vcycle} and shown schematically in Fig. \ref{fig:multigrid_schematic}.
The left-hand side represents the hybridised DG space $\what{\Lambda}_h$ on the facets while the non-nested coarse space $V_c$ on the triangular mesh induced by the facets is shown in the centre of the figure; in this example we assume that $V_c$ is the lowest order piecewise linear conforming space (usually known as P1). Functions are transported between the two spaces $\what{\Lambda}_h$ and $V_c$ with the $p$-restriction $\rho$ and $p$-prolongation $\pi$ described above. The right-hand side of the figure represents one level of coarsening of the P1 space; the latter is applied recursively in a GMG iteration. More generally, the gray box in Fig. \ref{fig:multigrid_schematic} symbolises the solution of the coarse space problem with a geometric- or algebraic multigrid algorithm. For the upwind flux, the application of geometric multigrid is straightforward since the coarse level elliptic problem in Eq. \eqref{eqn:coarse_problem_upwind} below is discretised in a conforming piecewise linear function space.
\begin{algorithm}
  \caption{Non-nested two-level V-cycle for approximately solving $\what{\mathcal{S}}(\what{q},\what{v})=\mathcal{R}(\what{v})$ in Eq. \eqref{eqn:schur_system}}\label{alg:twolevel_vcycle}
  \begin{algorithmic}[1]
    \STATE{Presmooth $\what{q}\mapsto \textsf{Smooth}(\what{\mathcal{R}},\what{q};n_{\text{pre}})$}
    \STATE{Calculate residual $\what{\mathcal{B}}(v)=\what{\mathcal{R}}(\what{v})-\what{\mathcal{S}}(\what{q}, \what{v})$ for all $\what{v}\in \what{\Lambda}_h$}
    
    \STATE{Restrict residual $\mathcal{R}_c(v_c)=\what{\mathcal{B}}(\pi v_c)$ for all $v_c\in V_c$}
  \STATE{Solve $\mathcal{S}_c(q_c,v_c)=\mathcal{R}_c(v_c)$ for all $v_c\in V_c$}
  \STATE{Update $\what{q}\mapsto \what{q} + \pi q_c$}
  \STATE{Postsmooth $\what{q}\mapsto \textsf{Smooth}(\what{\mathcal{R}},\what{q};n_{\text{post}})$}
  \end{algorithmic}
\end{algorithm}

In Algorithm \ref{alg:twolevel_vcycle} ``$\textsf{Smooth}(\what{\mathcal{R}},\what{q};n)$'' stands for $n$ applications of a stationary smoother for the equation $\what{\mathcal{S}}(\what{q},\what{v})=\what{\mathcal{R}}(\what{v})$, such as a Jacobi iteration or a more sophisticated Chebyshev smoother. We use the following coarse level spaces, prolongation operators and bilinear forms for the upwind- and Lax-Friedrichs flux:
\begin{description}
\item[Upwind flux:]
  The coarse space is chosen to be the scalar valued piecewise linear conforming finite element space on $\Omega$ with respect to the mesh $\Omega_h$ (usually called P1), defined by
  \begin{equation*}
    V_c^{(\text{up})} = \left\{\phi_c\in H^1(\Omega): \phi_c|_K \in \mathcal{P}^1(K), \forall K\in \Omega_h\right\}.
  \end{equation*}
  In the following we assume that the polynomial degree of the DG space is $p\ge 1$. Since the restriction of $V_c^{(\text{up})}$ to the skeleton $\mathcal{E}_h$ is a subspace of $\Lambda_h$, the prolongation $\pi^{(\text{up})}:V_c^{(\text{up})}\rightarrow \Lambda_h$ is the natural injection onto $\mathcal{E}_h$:
  \begin{equation*}
    \pi^{(\text{up})} \phi_c = \what{\phi},\qquad\text{such that $\what{\phi}=\phi_c|_{\mathcal{E}_h}$ for all $\phi_c\in V_c^{(\text{up})}$}.
  \end{equation*}
  The coarse space bilinear form is the weak form of the second order positive definite div-grad-problem $\phi-(c_g\alpha\Delta t)^2\nabla\cdot\left(\nabla \phi\right)$, which would arise formally from the time-implicit discretisation of the continuum equations in Eq. \eqref{eqn:SWE_continuum_linear} after eliminating the field $\vec{u}$:
  \begin{equation}
      \mathcal{S}^{(\text{up})}_c(\phi_c,\psi_c) = \left(\phi_c,\psi_c\right)_{\Omega_h}+(c_g\alpha\Delta t)^2\left(\nabla\phi_c\cdot \nabla \psi_c\right)_{\Omega_h}\qquad\text{for all $\phi_c,\psi_c\in V_c^{(\text{up})}$}\label{eqn:coarse_problem_upwind}
  \end{equation}
Note that, as discussed in Section \ref{sec:relationship_to_Cockburn}, the coarse level problem in \cite{Cockburn2014} is very similar, but does not contain a zero order term.
\item[Lax-Friedrichs flux:]
  The coarse space is the lowest order Raviart-Thomas-space \cite{Raviart1977}
  \begin{equation}
    \vec{V}_c^{(\text{LF})} = \left\{\vec{u}_c\in H(\text{div},\Omega): \vec{u}_c|_K = \vec{\alpha}_K+\beta_K\vec{x}, \forall K\in \Omega_h\right\}.
  \end{equation}
  Here $H(\text{div},\Omega)$ is the space of all vector fields for which both the components and the weak divergence are square-integrable. $\vec{\alpha}_K$ and $\beta_K$ are constant on each cell $K$ and are chosen such that $\vec{u}_c$ has continuous normal component across facets. The prolongation $\pi^{(\text{LF})}:\vec{V}_c^{(LF)}\rightarrow\what{\Lambda}_h$ is given by the $L_2$ projection on the skeleton $\mathcal{E}_h$:
  \begin{equation*}
      \pi^{(\text{LF})} \vec{u}_c = \what{\vec{u}},\qquad\text{such that $\langle\what{\vec{u}}, \what{\vec{w}} \rangle_{\mathcal{E}_h}=\langle\favg{\vec{u}_c}, \what{\vec{w}} \rangle_{\mathcal{E}_h}$ for all $\what{\vec{w}}\in \what{\Lambda}_h$}.
  \end{equation*}
  In contrast to the upwind case, the coarse space bilinear-form $\mathcal{S}_c^{(\text{LF})}$ is now the weak form of the second order grad-div-problem $\vec{u}-(\alpha\Delta t)^2\nabla\left(\nabla\cdot \vec{u}\right)$, which would arise from the time-implicit discretisation of the continuum equations in Eq. \eqref{eqn:SWE_continuum_linear} after eliminating the field $\phi$:
  \begin{equation}
      \mathcal{S}^{(\text{LF})}_c(\vec{u}_c,\vec{w}_c) = \left(\vec{u}_c, \vec{w}_c\right)_{\Omega_h}+(c_g\alpha\Delta t)^2\left((\nabla\cdot\vec{u}_c),(\nabla\cdot \vec{w}_c)\right)_{\Omega_h}\qquad\text{for all $\vec{u}_c,\vec{w}_c\in \vec{V}_c^{(\text{LF})}$}\label{eqn:coarse_problem_LF}
  \end{equation}
\end{description}
\section{Timestepping methods and linear solver configuration}\label{sec:timesteppers}
To study the performance of our solvers in a wider context we use a range of different timestepping methods.
\subsection{IMEX methods}
The following semi-implicit time stepping methods are used:
\begin{description}[leftmargin=\parindent,labelindent=1ex]
\item[Theta($\theta$):] The method in Eq. \eqref{eqn:theta_method} with a particular value of the off-centering parameter $\theta$. For $\theta=\frac{1}{2}$ this corresponds to treating the linear part of the PDE with the Crank-Nicholson method \cite{Crank1947}, which is second order in time.
\item[ARS2(2,3,2):] The ARS2 three-stage IMEX method which is also used in \cite{Kang2020}. In general, ARS IMEX methods are developed in \cite{Ascher1995}; the matrices are written down explicitly in \cite{Weller2013} and are given for ARS2(2,3,2) as
  \begin{xalignat*}{3}
  a &= \begin{pmatrix}
    0 & 0 & 0 \\
    \gamma & 0 & 0 \\
    \delta & 1-\delta & 0
  \end{pmatrix},
  &
  \widetilde{a} &= \begin{pmatrix}
    0 & 0 & 0 \\
    0 & \gamma & 0 \\
    0 & 1-\gamma & \gamma
  \end{pmatrix},
  &
  b = \widetilde{b} &= \begin{pmatrix}
    0 \\ 1-\gamma \\ \gamma
  \end{pmatrix}
\end{xalignat*}
  with $\gamma = 1-\frac{1}{\sqrt{2}}$ and $\delta = -\frac{2}{3}\sqrt{2}$.
\item[SSP2(3,2,2):] The strong stability preserving method described in \cite{Pareschi2005} can be written as an IMEX method which is defined by the matrices
  \begin{xalignat*}{3}
  a &= \begin{pmatrix}
    0 & 0 & 0 \\
    0 & 0 & 0 \\
    0 & 1 & 0
  \end{pmatrix},
  &
  \widetilde{a} &= \begin{pmatrix}
    1/2 & 0 & 0 \\
    -1/2 & 1/2 & 0 \\
    0 & 1/2 & 1/2
  \end{pmatrix},
  &
  b = \widetilde{b} &= \begin{pmatrix}
    0 \\ 1/2 \\ 1/2
  \end{pmatrix}.
\end{xalignat*}
\item[ARS3(4,4,3):] The ARS3 five-stage IMEX method with
  \begin{xalignat*}{2}
    \begin{aligned}
  a &= \begin{pmatrix}
    0 & 0 & 0 & 0 & 0\\
    1/2 & 0 & 0 & 0 & 0\\
    11/18 & 1/18 & 0 & 0 & 0\\
    5/6 & -5/6 & 1/2 & 0 & 0\\
    1/4 & 7/4 & 3/4 & -7/4 & 0
  \end{pmatrix},
  &
  b &= \begin{pmatrix}
    1/4 \\ 7/4 \\ 3/4 \\ -7/4 \\ 0 
  \end{pmatrix},\\[1ex]
  \widetilde{a} &= \begin{pmatrix}
    0 & 0 & 0 & 0 & 0\\
    0 & 1/2 & 0 & 0 & 0\\
    0 & 1/6 & 1/2 & 0 & 0\\
    0 & -1/2 & 1/2 & 1/2 & 0\\
    0 & 3/2 & -3/2 & 1/2 & 1/2
  \end{pmatrix},
  &
  \widetilde{b} &= \begin{pmatrix}
    0 \\ 3/2 \\ -3/2 \\ 1/2 \\ 1/2 
  \end{pmatrix}.
  \end{aligned}
\end{xalignat*}
  \end{description}
As discussed in Section \ref{sec:IMEXmethods}, all semi-implicit methods require the solution of the ill-conditioned linear system in Eq. \eqref{eqn:general_linear_system}. In addition, mass-solvers are required to obtain the state $q^{(n+1)}$ at the next time step in Eq. \eqref{eqn:IMEX_HDG_II}. However, the corresponding mass matrices are block-diagonal. For each cell, a small, dense matrix has to be inverted, which is readily achieved with LU factorisation. Tab. \ref{tab:timestepper_comparison} lists the order as well as the number of linear solves and function evaluations for all discussed methods, including the explicit time integrators which will be discussed in Section \ref{sec:explicit_methods} below.

To assess the performance of the naive, approximate Schur-complement preconditioner, semi-implicit methods have also been implemented for the standard (non-hybridised) DG discretisation. The non-hybridised variant of the Theta($\theta$) method for Eq. \eqref{eqn:linear_non_hybridized} is
\begin{equation}
  \frac{\mathcal{M}(q^{(n+1)},v)-\mathcal{M}(q^{(n)},v)}{\Delta t}
  = \mathcal{N}(q^{(n)},v) + \theta \mathcal{L}(q^{(n+1)},v)
  + (1-\theta) \mathcal{L}(q^{(n)},v)\qquad\text{for all $v\in W_h$.}
\label{eqn:DG_theta_method}
\end{equation}
Generally, a non-hybridised $s$-stage IMEX method can be written as
\begin{equation}
  \begin{aligned}
    \mathcal{M}(Q^{(i)},v) &= \mathcal{M}(q^{(n)},v) + \Delta t\sum_{j=1}^{i-1}
    a_{ij} \mathcal{N}(Q^{(j)},v) + \Delta t \sum_{j=1}^{i}\widetilde{a}_{ij}\mathcal{L}(Q^{(j)},v),\qquad\text{for $i=1,\dots,s$}\\
      \mathcal{M}(q^{(n+1)},v) &= \mathcal{M}(q^{(n)},v) + \Delta t\sum_{i=1}^s b_i \mathcal{N}(Q^{(i)},v) + \Delta t\sum_{i=1}^s \widetilde{b}_i \mathcal{L}(Q^{(i)},v),\qquad\text{for all $v\in W_h$,}
  \end{aligned}
\label{eqn:DG_IMEX}
\end{equation}
which should be compared to Eqs. \eqref{eqn:IMEX_HDG_I} and \eqref{eqn:IMEX_HDG_II}.
\begin{table}
\begin{center}
\begin{tabular}{rlcccccc}
\hline
& & & \multicolumn{2}{c}{\# evaluations} & \multicolumn{2}{c}{\# solves} \\
& method & order & $\mathcal{L}$ or $\what{\mathcal{L}}$ & $\mathcal{N}$ or $\mathcal{N}_0$ & $\mathcal{M}$ & $\mathcal{M}-\alpha \what{\mathcal{L}}$\\
\hline\hline
\multirow{4}{8ex}{IMEX $\left\{\begin{matrix}\\[8ex]\end{matrix}\right.\hspace{-2ex}$}
& Theta($\theta$) & $1^*$ & $1$ & $1$ & --- & $1$\\
& ARS2(2,3,2) & $2$ & $2$ & $3$ & $2$ & $2$ \\
& SSP2(3,2,2) & $2$ & $3$ & $2$ & $1$ & $3$\\
& ARS3(4,4,3) & $3$ & $4$ & $4$ & $2$ & $4$\\\hline
\multirow{3}{9ex}{explicit $\left\{\begin{matrix}\\\\\end{matrix}\right.\hspace{-2ex}$}
& Explicit Euler & $1$ & $1$ & $1$ & $1$ & ---\\
& Heun & $2$ & $2$ & $2$ &$2$  & ---\\
& SSPRK3 & $3$ & $3$ & $3$ & $3$ & --- \\
\hline
\end{tabular}
\caption{Approximation order, number of function evaluations (listed separately for $\mathcal{L}$ or $\what{\mathcal{L}}$ and $\mathcal{N}$ or $\mathcal{N}_0$), linear solves with $\mathcal{M}-\alpha \what{\mathcal{L}}$
 and mass solves with $\mathcal{M}$ for all timestepping methods discussed in Section \ref{sec:timesteppers}. For the non-hybridised versions of the IMEX integrators, the linear operator $\what{\mathcal{L}}$ is replaced by $\mathcal{L}$ in $\mathcal{M}-\alpha\what{\mathcal{L}}$. Although for general values of $\theta$ the Theta($\theta$) method is first order, the linear part of the equations is integrated to second order accuracy for $\theta=0.5$; this is indicated by the asterisk in the table above.}
\label{tab:timestepper_comparison}
\end{center}
\end{table}

In the following section we provide further details on the configuration of the linear solvers which are used in our semi-implicit methods. 
\subsection{Linear solver}\label{sec:solver_setup}
To compare the performance of different time integrators and investigate speedups achieved by using the hybridised DG discretisation we work with two different setups:
\begin{description}
  \item[NL-LF:] the non-linear SWEs with a Lax-Friedrichs flux and spatially varying bathymetry
  \item[L-Up:] the linear SWEs with an upwind flux and constant bathymetry $\phi_B=1$
\end{description}
\subsubsection{Hybridised DG}\label{sec:solver_setup_HDG}
For the semi-implicit time integrators, the hybridised system in Eq. \eqref{eqn:general_linear_system} is solved with the (exact) Schur-complement reduction to flux-space described in Section \ref{sec:schur_solve}. We solve the Schur-complement system defined in Eq. \eqref{eqn:schur_system} with a Krylov-subspace method which is preconditioned with the non-nested multigrid method in Algorithm \ref{alg:twolevel_vcycle}. For a flat bathymetry ($\phi_B=1$), which is used in the linear setup (L-Up), the problem in Eq. \eqref{eqn:schur_system} is symmetric and a Conjugate Gradient (CG) solver \cite{Hestenes1952} can be used. For the non-linear setup (NL-LF) with spatially varying $\phi_B$ we use a GMRES \cite{Saad1986} iteration since the problem is no longer symmetric. In both cases the (preconditioned) residual is reduced by a relative factor of at least $\epsilon$, with $\epsilon$ to be specified. The bathymetry is always set to $\phi_B=1$ in the multigrid preconditioner. As argued in Section \ref{sec:relationship_to_Cockburn}, for the upwind flux the Schur-complement system \eqref{eqn:schur_system} only differs from the one in \cite{Cockburn2014} by a zero-order term.

It remains to specify the smoother $\textsf{Smooth}(\what{R},\what{q};n)$ in Algorithm \ref{alg:twolevel_vcycle} (lines 1 and 6) and the method for solving the coarse space problem $\mathcal{S}_c(q_c,v_c) = \mathcal{R}_c(v_c)$ (line 4). In our implementation we used a parallel Chebyshev iteration with SOR preconditioning and applied two pre- and post-smoothing steps, $n_{\text{pre}}=n_{\text{post}}=2$. The coarse level problem $\mathcal{S}_c(q_c,v_c) = \mathcal{R}_c(v_c)$ is solved with a geometric- or algebraic- multigrid V-cycle, using PETSc's GAMG implementation in the latter case. Two SOR-preconditioned Chebychev iterations are used as the smoother and coarse level solver in both AMG and the geometric multigrid solver.
\subsubsection{Native DG}\label{sec:solver_setup_DG}
The linear system in the non-hybridised variants of the time-integrators in Eqs. \eqref{eqn:DG_IMEX} and \eqref{eqn:DG_theta_method} can be written in the form
\begin{equation}
  \mathcal{A}(q,v)=\mathcal{M} (q,v) - \alpha\Delta t \mathcal{L}(q,v) = \mathcal{R}(v)\qquad\text{for all $v\in W_h$}.
  \label{eqn:native_dg_system}
\end{equation}
Eq. \eqref{eqn:native_dg_system} is solved with a preconditioned GMRES solver,
using the naive, approximate Schur-complement preconditioner which has been applied for conforming mixed finite elements in \cite{Mitchell2016,Melvin2019}. As for the hybridised case, the iteration is aborted once the (preconditioned) residual has been reduced by a relative factor of at least $\epsilon$. To construct the approximate Schur-complement preconditioner, consider the non-hybridised bilinear form $\mathcal{A}$ which is written down for the Lax-Friedrichs flux in Eq. \eqref{eqn:bilinear_standardDG_lax} and for the upwind flux in Eq. \eqref{eqn:bilinear_standardDG_upwind} in \secapp\ref{sec:standardDGlinear_forms}. Observe that in both cases $\mathcal{A}$ can be decomposed into four terms as
\begin{equation}
\mathcal{A}(q,v) = \mathcal{A}(\phi,\vec{u}) =
\mathcal{A}_{\phi\phi}(\phi,\psi) +
\mathcal{A}_{\phi u}(\vec{u},\psi) +
\mathcal{A}_{u\phi}(\phi,\vec{w}) +
\mathcal{A}_{uu}(\vec{u},\vec{w}).
\label{eqn:standard_DG_bilinear}
\end{equation}
Both $\mathcal{A}_{\phi\phi}$ and $\mathcal{A}_{uu}$ contain artificial diffusion terms which couple unknowns across adjacent cells. This is different from the hybridised case, where the term $\left(\mathfrak{A}q,v\right)_{\Omega_h}$ in Eq. \eqref{eqn:ABCMform} only contains couplings within one cell of the mesh.
To construct a preconditioner, consider the matrix-representation
\begin{equation}
  \begin{pmatrix}
    A_{\phi \phi} & A_{\phi u}\\
    A_{u \phi} & A_{u u}
  \end{pmatrix}
  \begin{pmatrix}
    \vec{\Phi}\\\vec{U}
  \end{pmatrix}
  =
  \begin{pmatrix}
    \vec{R}_\phi\\\vec{R}_u
  \end{pmatrix}\label{eqn:native_matrix_representation}
\end{equation}
of Eq. \eqref{eqn:native_dg_system} where $(\vec{\Phi},\vec{U})^T$ is the dof-vector of $(\phi,\vec{u})=q$. $A_{\phi\phi}$, $A_{\phi u}$, $A_{u \phi}$ and $A_{u u}$ are the (sparse) matrix representations of the four bilinear forms in Eq. \eqref{eqn:standard_DG_bilinear}. Given $(\vec{R}_\phi,\vec{R}_u)^T$, Eq. \eqref{eqn:native_matrix_representation} can be solved approximately as follows:
\begin{description}
\item[Step 1.] Approximately solve $A_{uu}\vec{V} = \vec{R}_u$ for $\vec{V}$.
\item[Step 2.] Approximately solve $S\vec{\Phi}=\vec{R}_\phi - A_{\phi u}\vec{V}$ for $\vec{\Phi}$ where
  \begin{equation}
    S=A_{\phi\phi}-A_{\phi u} (\operatorname{diag}(A_{uu}))^{-1} A_{u\phi}.
    \label{eqn:S_approx_diag}
    \end{equation}
\item[Step 3.] Approximately solve $A_{uu} \vec{U} = \vec{R}_u - A_{u\phi}\vec{\Phi}$
\end{description}
Note that the approximate Schur complement $S$ in step 2 is a sparse matrix since the diagonal matrix $\operatorname{diag}(A_{uu})$ can be inverted exactly and is also diagonal. An ILU algorithm is applied to approximately invert the matrix $A_{uu}$ in steps 1 and 3. One GAMG V-cycle is used to solve the Schur-complement system in step 2; two Chebyshev iterations (preconditioned with an approximate block-ILU-Jacobi method) are employed as the smoother and coarse-level solver. In our code, the PETSc ``fieldsplit'' preconditioner \cite{petsc-user-ref} was used to implement steps 1-3 of the approximate Schur-complement solver above.

The construction of the approximate Schur-complement $S$ in Eq. \eqref{eqn:S_approx_diag} uses the diagonal $\text{diag}(A_{uu})$ of $A_{uu}$. Intuitively, a better approach appears to be to use the \textit{block-diagonal} (i.e. the matrix $\text{blockdiag}(A_{uu})$ obtained from $A_{uu}$ by dropping all couplings of unknowns between different grid cells) instead, since this does not change the sparsity pattern of $S$ but uses more information from $A_{uu}$. However, most results in the paper were obtained by constructing the approximate Schur-complement based on the diagonal $\text{diag}(A_{uu})$, since this is the default option in the PETSc ``fieldsplit'' preconditioner. Potential improvements due to replacing $\text{diag}(A_{uu})\mapsto \text{blockdiag}(A_{uu})$ in the approximate Schur-complement in Eq. \eqref{eqn:S_approx_diag} are quantified in Section \ref{sec:blockdiag_schurapprox}.
\subsection{Explicit methods}\label{sec:explicit_methods}
For reference we also consider fully explicit time integrators. In contrast to semi-implicit methods they do not require the solution of an ill-conditioned system but suffer from tighter restrictions on the time step size. Results are reported for the following methods:
\begin{description}[leftmargin=\parindent,labelindent=1ex]
\item[Explicit Euler:] The fully explicit Euler method, i.e. Eq. \eqref{eqn:theta_method} with $\theta=0$.
  \item[Heun's method:] An improved second order variant of the explicit Euler method.
  \item[SSPRK3:] The third order strong stability preserving Runge Kutta method defined in \cite{Shu1988}.
\end{description}
In the context of DG discretisations, Heun's method and SSPRK3 also also known as RKDG2 and RKDG3 which are discussed in the context of a wider framework in \cite{Cockburn1989,Cockburn1998}. For reference, further details on the explicit time integrators introduced in this section can be found in \secapp \ref{sec:RK_details}.
\section{Implementation}\label{sec:Implementation}
All code was implemented in the open source Firedrake library \cite{Rathgeber2017}. This allows the expression of weak forms in Unified Form Language (UFL) \cite{Alnaes2014} in Python, from which highly efficient C-code is automatically generated. Firedrake uses PETSc \cite{petsc-efficient,petsc-user-ref} as a linear algebra backend, which provides a wide range of iterative solvers and preconditioners. PETSc is highly configurable, and supports the composition of user-defined operators to design bespoke preconditioners \cite{Kirby2018}. Combined with the multigrid support in Firedrake, this allows the implementation of the sophisticated non-nested multigrid preconditioners described in this paper.
\subsection{Local matrix algebra with Slate}\label{sec:slate}
The recently developed Slate package \cite{Gibson2019} for cell-wise linear algebra is crucial for the construction of the Schur-complement solver described in Section \ref{sec:schur_solve}. Slate converts the UFL expression of a weak form into a so-called \texttt{Tensor} object, which stores the \textit{locally} assembled matrix. \texttt{Tensor}s can be manipulated cell-wise. To illustrate this, consider the Schur-complement system in Eq. \eqref{eqn:schur_system} with $\mathfrak{F}$ as in Eq. \eqref{eqn:AFinverse}. Each of the operators $\mathfrak{M}$, $\mathfrak{C}$, $\mathfrak{A}$ and $\mathfrak{B}$ corresponds to a weak bilinear form, which is defined in Eqs. \eqref{eqn:bilinear_HDG_lax}, \eqref{eqn:bilinear_HDG_upwind} and \eqref{eqn:ABCMform}.

For example, consider the weak form $(\mathfrak{A}q,v)_{\Omega_h}$ defined in Eq. \eqref{eqn:ABCMform}. For the Lax-Friedrich flux $(\mathfrak{A}q,v)_{\Omega_h}$ is given by
\begin{equation}
  \begin{aligned}
  (\mathfrak{A}q,v)_{\Omega_h} &= \left(\phi\psi + \vec{u}\cdot\vec{w} - c_g\alpha\Delta t\left[\vec{u}\cdot\nabla\psi + \phi\nabla\cdot\vec{w}\right]\right)_{\Omega_h}\\
    &+ c_g\alpha\Delta t\sum_{e\in\mathcal{E}_h} \left(
    \sqrt{\phi_B}\left(\vec{u}^+\cdot\vec{w}^++\vec{u}^-\cdot\vec{w}^-\right)
  +\phi_B\left(\phi^+\vec{n}^+\cdot\vec{w}^++\phi^-\vec{n}^-\cdot\vec{w}^-\right)
  \right)_e
  \end{aligned}.\label{eqn:weak_form_example}
\end{equation}
where $q=(\phi,\vec{u}),v=(\psi,\vec{w})\in W_h$ are discontinuous fields. The bilinear form $(\mathfrak{A}q,v)_{\Omega_h}$ can be expressed in UFL and converted to a Slate tensor as follows:
\begin{lstlisting}[language={[firedrake]{python}}]
  a_form = (phi*psi+inner(u,w)-cg*alpha*dt*(inner(u,grad(psi))+phiB*phi*div(w)))*dx
         + cg*alpha*dt*(sqrt(phiB)*(inner(u('+'),w('+'))
                                   +inner(u('-'),w('-')))
                       +phiB*(phi('+')*inner(n('+'),w('+'))
                             +phi('-')*inner(n('-'),w('-'))))*dS
  A = Tensor(a_form)                       
\end{lstlisting}
Crucially, since the fields $q,v$ in Eq. \eqref{eqn:weak_form_example} are discontinuous, in this case the matrix representation of $\mathfrak{A}$ is block-diagonal, since it only couples unknowns in the same grid cell. Hence, the inverse $\mathfrak{F}$ can be obtained by inverting dense, cell-local matrices. The corresponding Slate tensor representation of $\mathfrak{F}$ is given by \texttt{A.inv}. By constructing the tensors \texttt{M}, \texttt{C} and \texttt{B} which represent $\mathfrak{M}$, $\mathfrak{C}$ and $\mathfrak{B}$, it is possible to construct the Slate tensor which corresponds to $\mathfrak{S} = \mathfrak{M} + \mathfrak{C}\mathfrak{F}\mathfrak{B}$ in Eq. \eqref{eqn:schur_system} as
\begin{lstlisting}[language={[firedrake]{python}}]
  S = M + C*A.inv*B
\end{lstlisting}
This tensor can now be passed to a Firedrake linear variational problem solver as usual. More specifically, to solve a problem with a given right hand side \verb|r_rhs| and return the solution in the variable \texttt{lambda}, write
\begin{lstlisting}[language={[firedrake]{python}}]
  trace_lvp = LinearVariationalProblem(a=S_expr,L=r_rhs,u=lmbda)
  trace_lvs = LinearVariationalSolver(trace_lvp,solver_parameters=param)
  trace_lvs.solve()
\end{lstlisting}
Note that the exact solver configuration is specified by passing the dictionary \verb|param| as the keyword argument \verb|solver_parameters|. This is used to build the sophisticated multigrid solvers described in the following section.

We stress that the generation of all Slate expressions which are required in the HDG Schur-complement construction is automated; this is achieved by using the static condensation preconditioner \textsf{firedrake.SCPC} described in the following section.
\subsection{Composing the multigrid solver}\label{sec:composing_multigrid}
A preconditioned iterative solver in Firedrake is characterised by the pair $(\textsf{KSP},\textsf{PC})$ where \textsf{KSP} is the Krylov-subspace solver to be used and $\textsf{PC}$ the preconditioner. For example, to solve the problem in the previous section with a GMRES solver, preconditioned with a Jacobi-iteration, set
\begin{lstlisting}[language={[firedrake]{python}}]
  solver_parameters={'ksp_type':'gmres','pc_type':'jacobi'}
\end{lstlisting}
when constructing the \verb|LinearVariationalSolver| object. PETSc provides a wide range of iterative solvers and preconditioners. Crucially, the solvers can be composed \cite{Kirby2018} to construct new, problem-specific preconditioners. This is achieved by setting \verb|'pc_type'='python'|, deriving a new python preconditioner object \verb|MyPC| from a suitable Firedrake base class and setting \verb|'pc_python_type'=MyPC|.

To solve the problem $\what{\mathcal{A}}(q,\what{q},v,\what{v})=\mathcal{R}(v)$
defined in Eq. \eqref{eqn:general_linear_system}, two bespoke Python preconditioners are constructed and composed
\begin{description}
\item[\textsf{firedrake.SCPC}:] Given Eq. \eqref{eqn:general_linear_system}, this preconditioner implements steps 1-3 of the Schur-complement solver described at the beginning of Section \ref{sec:schur_solve}. It uses Slate to reduce the problem to the Schur-complement system $\what{\mathcal{S}}(\what{q},\what{v})=\what{\mathcal{R}}(\what{v})$ defined in Eq. \eqref{eqn:schur_system} (steps 1 and 2). After solving this system with \textsf{firedrake.GTMGPC} (step 2), the DG field $q$ is reconstructed with Eq. \eqref{eqn:reconstruct_solution} (step 3).
  \item[\textsf{firedrake.GTMGPC}:] This solves the system $\what{\mathcal{S}}(\what{q},\what{v})=\what{\mathcal{R}}(\what{v})$ in the space $\what{\Lambda}_h$ by using the non-nested multigrid method described in Algorithm \ref{alg:twolevel_vcycle} in Section \ref{sec:non-nested-multigrid}. The user can specify the smoother in the DG space and the coarse level solver by passing options to the solver.
\end{description}
The \textsf{firedrake.GTMGPC} object requires additional, problem specific information to construct the bilinear form in the P1 or Raviart-Thomas space defined in Eqs. \eqref{eqn:coarse_problem_upwind} and \eqref{eqn:coarse_problem_LF}. This information is passed via an application-context (\verb|appctx|) keyword when constructing the \verb|LinearVariationalSolver|.

Schematically, given the bilinear form \verb!a_form! written down in the code listing in Section \ref{sec:slate}, the problem $\what{\mathcal{A}}(q,\what{q},v,\what{v})=\mathcal{R}(v)$ is solved in Firedrake as follows:
\begin{lstlisting}[language={[firedrake]{python}}]
  lvp = LinearVariationalProblem(a=a_form,L=r_rhs,u=q_qhat)
  param = {'mat_type': 'matfree',
           'ksp_type': 'preonly',
           'pc_type': 'python',
           'pc_python_type': 'firedrake.SCPC',
           'pc_sc_eliminate_fields': '0, 1',
           'condensed_field': {'ksp_type': 'cg',
                               'mat_type': 'matfree',
                               'pc_type': 'python',
                               'pc_python_type': 'firedrake.GTMGPC',
                               'gt': {...}
                               }
           }
  lvs = LinearVariationalSolver(lvp,solver_parameters=param,appctx=appctx)
  lvs.solve()
\end{lstlisting}
Note that since the reduction to the Schur-complement in Section \ref{sec:schur_solve} is exact, the outer problem is solved by applying the preconditioner once (this is achieved by setting \verb|'ksp_type'='preonly'|). The dictionary \verb|gt| is used to specify the smoother and coarse level solver of the \textsf{firedrake.GTMGPC} object. For reference, the full configuration is written down in \secapp \ref{sec:solver_configuration_HDG}.

As explained in Section \ref{sec:solver_setup_DG}, the approximate Schur complement preconditioner for the native DG method, is implemented with the PETSc fieldsplit preconditioner. The exact solver options for this are written down in \secapp \ref{sec:solver_configuration_nativeDG}. All code developed for this paper is openly available as a git repository at \url{https://bitbucket.org/em459/hybridizeddg}.
\section{Results}\label{sec:Results}
The numerical methods developed in the first part of the paper are now applied to different model problems. For this, we first consider an idealised, smooth stationary flow in a flat domain, which exhibits the typical separation between advective- and gravity-wave velocities observed in atmospheric flows. The problem setup is described in Section \ref{sec:problem_setup} and results on the accuracy, performance and parallel scalability are presented in sections \ref{sec:results_accuracy} to \ref{sec:results_scalability}. Results for an alternative approximate DG Schur-complement are shown in Section \ref{sec:blockdiag_schurapprox} and a comparison to explicit time integrators is presented in Section \ref{sec:results_explicit}.

Following this, we test our methods for several well-established test cases in spherical geometry from the Williamson et al. suite in \cite{Williamson1992} in Section \ref{sec:results_spherical}. After verifying expected spatial convergence rates for a smooth stationary test case (zonal flow with compact support), we run our code for two non-stationary examples: (i) flow over an isolated mountain and (ii) the Rossby-Haurwitz wave. In the latter two cases we compare the performance of our new multigrid algorithms to a direct solver, i.e. a setup which is similar to the one used in \cite{Kang2020}; we find that our preconditioners are competitive with a highly-optimised direct method. 
\subsection{Problem setup for flat geometry}\label{sec:problem_setup}
\subsubsection{Stationary vortex}\label{sec:stationary_vortex}
\begin{figure}
\begin{center}
\includegraphics[width=0.6\linewidth]{\figdir/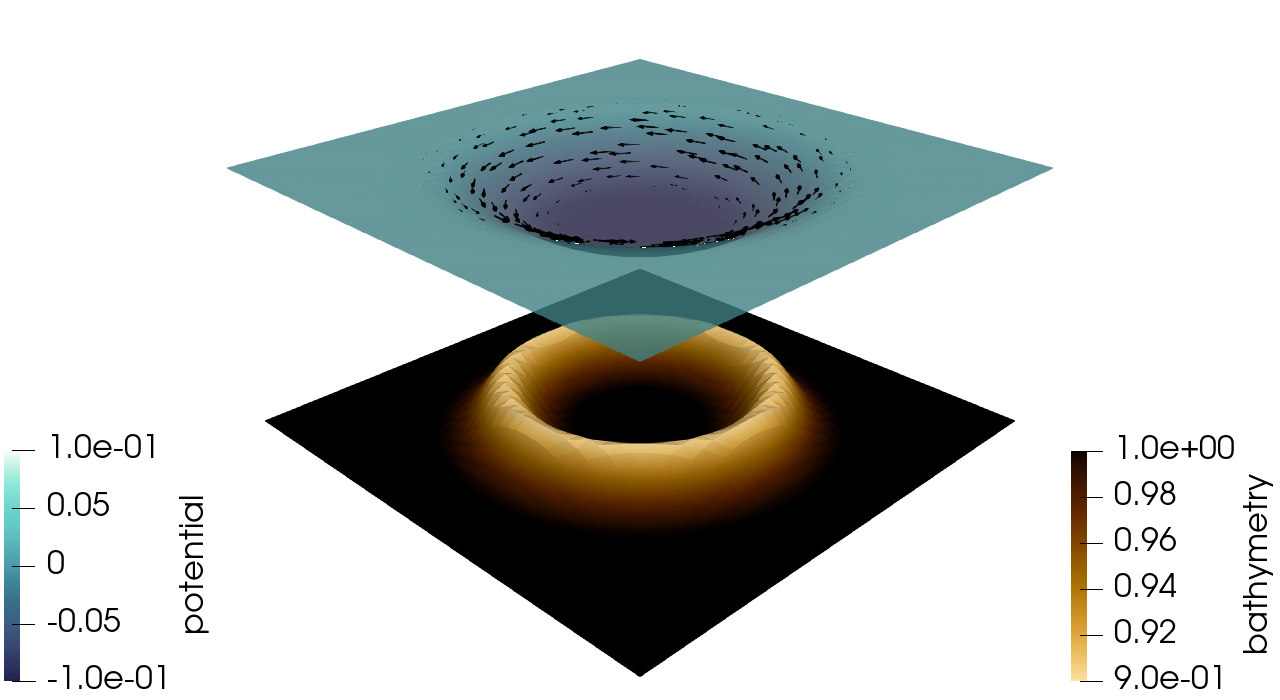}
\caption{Stationary vortex solution $\phi$, $\vec{u}$ as defined in Eqs. \eqref{eqn:stationary_vortex_phi} and \eqref{eqn:stationary_vortex_u}. The bathymetry given in Eq. \eqref{eqn:stationary_vortex_bathymetry} is shown at the bottom of the figure.}
\label{fig:stationary_vortex}
\end{center}
\end{figure}

\begin{figure}
\begin{center}
\includegraphics[width=0.6\linewidth]{\figdir/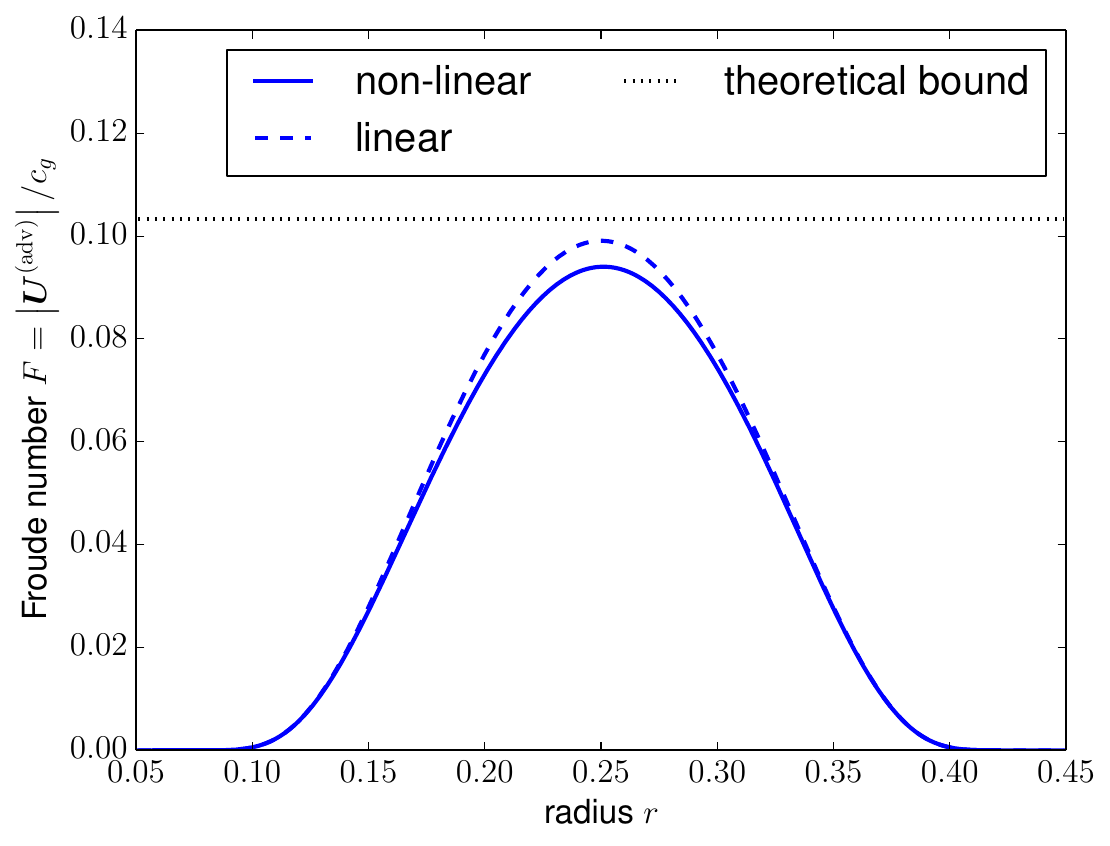}
\caption{Froude number defined in Eq. \eqref{eqn:froude_definition} for the linear and non-linear SWEs. The theoretical upper bound in Eq. \eqref{eqn:froude_bound} is also shown as a dotted line.}
\label{fig:froude}
\end{center}
\end{figure}

To assess the correctness of our code and quantify numerical accuracy we consider a smooth, time-independent rotationally invariant solution of the shallow water equations. As will be shown below, the numerical parameters can be adjusted such that the relative size of the gravity wave speed and advective velocities is similar to what is encountered in atmospheric models. To construct this stationary solution, all fields are assumed to be radially symmetric, and the bathymetry $\phi_B$ is described by
\begin{equation}
  \phi_B(r) = \begin{cases}
    1 - \delta_B\exp\left[\frac{1}{r - r_+} + \frac{4}{r_+ - r_-} - \frac{1}{r - r_-}\right] & \text{for $r_-\le r\le r_+$},\\
    1 & \text{otherwise}.
  \end{cases}\label{eqn:stationary_vortex_bathymetry}
\end{equation}
An exact solution of the SWEs is given by the height-perturbation
\begin{equation}
  \phi(r) = -\delta\cdot\begin{cases}
      1 & \text{for $r<r_-$},\\
    \frac{1}{2}\left(1+\tanh\left[\frac{\sigma}{r-r_-}+\frac{\sigma}{r-r_+}\right]\right) & \text{for $r_-\le r\le r_+$},\\
    0 & \text{for $r_+<r$},
  \end{cases}\label{eqn:stationary_vortex_phi}
\end{equation}
and purely tangential velocity $\vec{u}(r)=u(r)\vec{e}_{\theta}$ in the tangential direction. For the non-linear SWEs in Eq. \eqref{eqn:SWE_continuum} the function $u(r)$ is given by
\begin{equation}
  u(r) = \frac{r}{2L_R}(\phi + \phi_B)
  \left(-1+\sqrt{1+4L_R^2\frac{\phi'}{r}}
  \right).\label{eqn:stationary_vortex_u}
\end{equation}
where $L_R = c_g/f$ is the Rossby radius of deformation, measured in units of the reference length scale $\refr{L}$.
For the linear SWEs in Eq. \eqref{eqn:SWE_continuum_linear}, the tangential velocity simplifies to
\begin{equation}
  u(r) = L_R (\phi + \phi_B) \phi'.\label{eqn:stationary_vortex_u_linear}
\end{equation}
The solution is smooth ($\phi,u\in C^\infty$) and has compact support, i.e. $\phi$ and $\vec{u}$ are non-zero only on a disk of radius $r_+$ around the origin. In particular, if we set $r_+<\frac{1}{2}$, the computational domain can be set to the unit square $\Omega=[-\frac{1}{2},\frac{1}{2}]\times[-\frac{1}{2},\frac{1}{2}]$. Both fields are visualised together with the bathymetry $\phi_B$ in Fig. \ref{fig:stationary_vortex}. An important parameter is the relative size of the advection velocity $\vec{U}^{(\text{adv})}=\vec{u}/(\phi_B+\phi)$ and the gravity wave speed $\phi_B^{1/2}$. This ratio is given by the spatially varying non-dimensional Froude number
\begin{equation}
  F := \frac{\left|\vec{U}^{(\text{adv})}\right|}{\phi_B^{1/2}} = 
\begin{cases}
  \frac{r}{2L_R} \phi_B^{-1/2}
  \left(-1+\sqrt{1+4L_R^2\frac{\phi'}{r}}
  \right) & \text{(non-linear)},\\[2ex]
L_R\phi_B^{-1/2}\phi' & \text{(linear)}.
\end{cases}\label{eqn:froude_definition}
\end{equation}
It can be shown that for sufficiently large values of $(r_+-r_-)/\sigma$ the function $\phi'$ has a single maximum at $r=\frac{1}{2}(r_++r_-)$ and in both cases the Froude number can be bounded as
\begin{equation}
F\le L_R \phi_B^{-1/2}\phi' \le \frac{4\sigma L_R}{(r_+-r_-)^2}\left(1+\delta_B\right)\delta.
\label{eqn:froude_bound}
\end{equation}
To obtain the results in the rest of this paper we use the following numerical values: The reference depth is set to $\refr{H}_B=2\cdot 10^{3}\unit{m}$, the reference length is chosen to be the radius of the earth $\refr{L}=R_{\text{earth}}=6.37\cdot 10^{6}\unit{m}$ and the time scale is $\refr{T}=1\unit{day}=8.64\cdot 10^{4}\unit{s}$. The Coriolis parameter is set to $2\cos(\vartheta)\Omega$, where $\Omega=2\pi\;\unit{day}^{-1}$ is the angular speed of rotation of the earth and $\vartheta$ is the latitude (measured from the pole). For simplicity we further use the constant value of $f$ at the pole, i.e. set $\vartheta=0$. This results in the following non-dimensional values
\begin{xalignat*}{3}
  c_g &= 1.89, &
  f &= 4\pi \approx 12.567, &
  L_R &= 0.15.
\end{xalignat*}
We further set
\begin{xalignat*}{4}
  \delta &= \delta_B = 0.1, &
  r_- &= 0.05, &
  r_+ &= 0.45, &
  \sigma &= 0.25.
\end{xalignat*}
The Froude number is plotted for this choice of parameters in the range $r_-\le r \le r_+$ in Fig. \ref{fig:froude}. In all numerical experiments below the initial condition $q_0=(\phi_0,\vec{u}_0)$ was chosen such that $\phi_0$ is given by Eq. \eqref{eqn:stationary_vortex_phi} and $\vec{u}_0=u(r)\vec{e}_{\theta}$ with $u(r)$ given by either Eq. \eqref{eqn:stationary_vortex_u} for the non-linear SWEs or Eq. \eqref{eqn:stationary_vortex_u_linear} for the linearised equations. With this choice the exact solution $q_{\text{exact}} = (\phi_{\text{exact}},\vec{u}_{\text{exact}})=(\phi_0,\vec{u}_0)$ is stationary, which allows the easy quantification of numerical errors.
\subsubsection{Time step size}\label{sec:timestep_size}
For a given grid resolution $h$ and polynomial degree $p$ of the DG spaces defined in Eqs. \eqref{eqn:DG_space_def} and \eqref{eqn:DG_trace_def} the time step size is chosen as follows. First, a marginally stable explicit timestep size $\Delta t_{\text{expl}}$ is found by requiring that the Courant number is expressed as
\begin{equation}
  \frac{c_g\Delta t_{\text{expl}}}{h} = \frac{\rho}{2p+1}.
  \label{eqn:Deltat_stability}
\end{equation}
Here the dependency on $p$ is chosen as proposed by \cite{Cockburn2001}. Numerical experiments confirm that explicit time integrators are marginally stable, i.e. the timestep $\Delta t_{\text{expl}}$ is close to the stability limit, if $\rho=0.2$ in Eq. \eqref{eqn:Deltat_stability}. We then set the timestep for the semi-implicit IMEX methods to $\Delta t=\gamma \Delta t_{\text{expl}}$ where $\gamma>1$ is chosen such that the \textit{advective} Courant number $|\vec{U}^{(\text{adv})}|\Delta t/h$ is sufficiently small, but again close to the stability limit for advective transport. This can be achieved by requiring that $F\gamma \le 1$ (where $F$ is the Froude number defined in Eq. \eqref{eqn:froude_definition}). In atmospheric models the gravity wave speed is typically one order of magnitude larger than advective velocities, and setting $\gamma=10$ in the numerical experiments therefore replicates this scale separation in an idealised scenario (as can be seen from Fig. \ref{fig:froude}, the Froude number is bound by approximately $0.1$, so this choice of $\gamma$ guarantees that $F\gamma\lesssim 1$). The final time was always set to $T=\frac{1}{2}$.
\subsubsection{Grid layout}
The grid was constructed by first partitioning the domain $\Omega$ into $n^2$ squares of width $h=1/n$ and then subdividing each cell into two triangles, resulting in $N=2n^2$ triangular cells overall. The number $n$ is chosen such that $n=2^{r}$ where $r\in\mathbb{N}$ is referred to as the \textit{refinement level} in the following. Tab. \ref{tab:grid_setup} shows the grid sizes used. To highlight the relative size of the original DG system and the hybridised system, we also show the total number of unknowns for varying polynomial degrees $p$. For a fully periodic triangular grid with $N_{\text{cell}}$ cells the number of unknowns for storing the fields $\phi$ and $\vec{u}$ on cells is given by $N_{\text{dof}}^{(\text{cell})}$; storing the flux unknowns on the facets requires $N_{\text{dof}}^{(\text{facet})}$ variables with
\begin{xalignat}{2}
N_{\text{dof}}^{(\text{cell})} &= \frac{3}{2}(p+1)(p+2)N_{\text{cell}}, &
N_{\text{dof}}^{(\text{facet})} &= \nu^{(\text{flux})}\frac{3}{2}(p+1)N_{\text{cell}}.\label{eqn:ndofs}
\end{xalignat}
$\nu^{(\text{flux})}$ depends on the numerical flux. Since only scalar valued unknowns are stored on the facets for the upwind flux, whereas vectors are used in the Lax-Friedrichs flux, we have $\nu^{(\text{LF})}=2$ and $\nu^{(\text{up})}=1$. The Schur-complement flux-system in Eq. \eqref{eqn:schur_system} is a factor $\nu^{(\text{flux})}/(p+2)$ smaller than the original DG system in the variables $\phi$, $\vec{u}$.
\begin{table}
  \begin{tabular}{|r|r|r|rp{0.5ex}r|rp{0.5ex}r|rp{0.5ex}r|r|}
    \hline
    \multicolumn{1}{|c|}{$r$} & 
    \multicolumn{1}{|c|}{$N_{\text{cell}}$} & 
    \multicolumn{1}{|c|}{$h=2^{-r}$} &
    \multicolumn{3}{c|}{$p=1$} &
    \multicolumn{3}{c|}{$p=3$} &
    \multicolumn{3}{c|}{$p=5$}\\
    \hline\hline
    4 & 512 & 0.0625 & 4 608 & / & 1 536 & 15 360 & / & 3 072& 32 256 & / & 4 608 \\
    5 & 2 048 & 0.0313 & 18 432 & / & 6 144 & 61 440 & / &  12 288 & 129 024 & / & 18 432\\
    6 & 8 192 & 0.0156 & 73 728 & / & 24 576 & 245 760 & / & 49 152 & 516 096 & / & 73 728 \\
    7 & 32 768 & 0.0078 & 294 912 & / & 98 304 & 983 040 & / & 196 608 & 2 064 384 & / & 294 912\\
    8 & 131 072 & 0.0039 & 1 179 648 & / & 393 216 & 3 932 160 & / & 786 432 & & & \\
    \hline
  \end{tabular}
  \caption{Grid layout and numbers of unknowns for different polynomial degrees $p$. Each of the final three columns shows the total number of unknowns associated with cells and facets as calculated by Eq. \eqref{eqn:ndofs} for the scalar-valued flux problem ($\nu^{(\text{flux})}=\nu^{(\text{up})}=1$) in the form $N_{\text{dof}}^{(\text{cell})} / N_{\text{dof}}^{(\text{facet})}$.}
  \label{tab:grid_setup}
\end{table}
\subsubsection{Linear solver configurations}\label{sec:solver_configs_results}
The following solver configurations are used in our numerical experiments:
\begin{description}
\item[HDG+GTMG+GMG]: As described in Section \ref{sec:IMEXHDG}, the implicit system for the hybridised DG formulation in Eq. \eqref{eqn:general_linear_system} is solved with a Schur-complement approach, which requires the solution of a linear system in the flux-space $\Lambda_h$. The resulting problem (which is given in Eq. \eqref{eqn:schur_system}) is solved with GMRES or CG and preconditioned with the non-nested multigrid method described in Section \ref{sec:non-nested-multigrid}; this multigrid algorithm is written down as a two-level method in Alg. \ref{alg:twolevel_vcycle}. The coarse level system $\mathcal{S}_c(q_c,v_c) = \mathcal{R}_c(v_c)$ is solved with geometric multigrid (see Sections \ref{sec:solver_setup_DG} and \ref{sec:composing_multigrid} for details on the implementation and \secapp\ref{sec:solver_configuration_HDG} for the PETSc solver options).
\item[HDG+GTMG+AMG]: This configuration is very similar to HDG+GTMG+GMG, but the coarse level problem in line 4 of Alg. \ref{alg:twolevel_vcycle} is solved with AMG instead of a geometric multigrid method. 
\item[HDG+AMG]: For reference, the flux system in Eq. \eqref{eqn:schur_system} is also solved directly with AMG instead of the non-nested two-level method in Alg. \ref{alg:twolevel_vcycle} (again, see \secapp\ref{sec:solver_configuration_HDG} for details on the PETSc solver options).
\item[DG+ApproxSchur]: The non-hybridised $(\phi,\vec{u})$ system in Eq. \eqref{eqn:native_dg_system} is solved with GMRES. The preconditioner is the approximate Schur complement approach with AMG for the $\phi$-system described in Section \ref{sec:solver_setup_DG} (see \secapp\ref{sec:solver_configuration_nativeDG} for PETSc solver options).\end{description}
\subsection{Computational resources and code version}
All runs were carried out on the ``Balena'' supercomputer at the University of Bath. Each node of the machine contains two Intel Xeon E5-2650v2 (IvyBridge) processors with eight cores each, resulting in 16 cores per node. The results in the paper can be reproduced with the code released as \cite{hdg_release}. All runtimes reported in Sections \ref{sec:results_solver_comparison} and \ref{sec:results_other_timesteppers} were obtained by running on a single fully occupied node (16 cores).
\subsection{Accuracy}\label{sec:results_accuracy}
\begin{figure}
\begin{center}
\includegraphics[width=0.9\linewidth]{\figdir/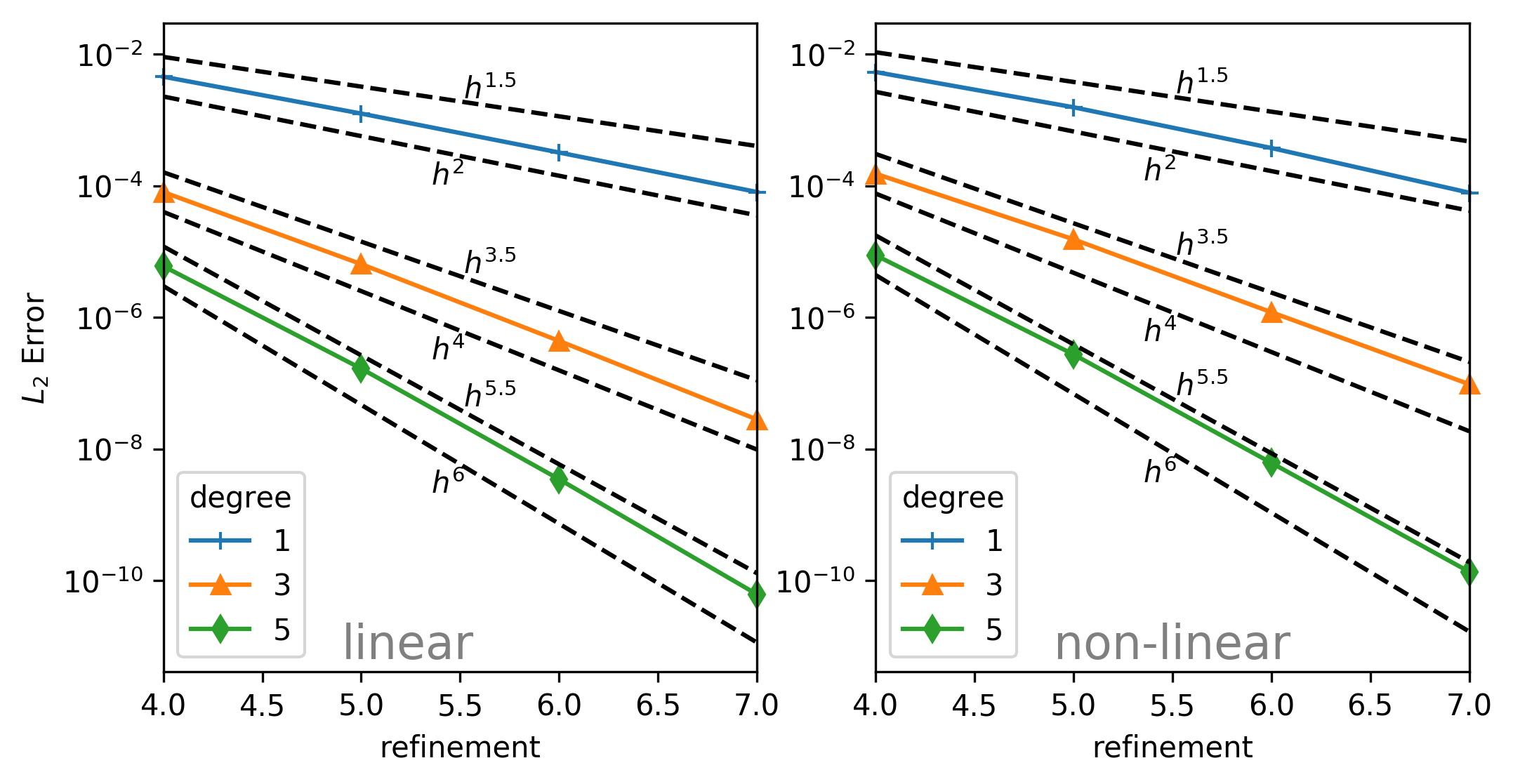}
\caption{$L_2$ error $\Delta_2$ as defined in EQ. \eqref{eqn:L2_error} for the linearised equations with flat bathymetry (left) and the fully non-linear SWEs with spatially varying bathymetry (right) as a function of the grid refinement $r$ for different polynomial degrees $p$. All results are obtained with the ARS2(2,3,3) timestepper. The Lax-Friedrichs flux was used in the fully non-linear case and the upwind flux for the linear problem. The grid spacing is $h=2^{-r}$, where $r$ is the refinement shown on the horizontal axis.}
\label{fig:DG_error}
\end{center}
\end{figure}

We begin by verifying the accuracy of the DG discretisation, which serves as an important test of the correctness of the implementation. Theoretically \cite{Kang2020} we expect the spatial discretisation error (measured in a suitable norm defined in \cite{Kang2020}) to be $\mathcal{O}(h^{p+\frac{1}{2}})$, where $p$ is the polynomial degree of the DG discretisation. To quantify the error, both the fully non-linear SWEs with bathymetry (the NL-Up configuration) and the linearised equations with flat bathymetry (L-Up) are solved with an ARS(2,3,2) IMEX timestepper. We verified that the DG+ApproxSchur and the HDG+GTMG+AMG solver give the same solution in both cases. For a given grid spacing $h$ and polynomial degree $p$ the $L_2$ error is defined as
\begin{equation}
\Delta_2 := ||q - q_{\text{exact}}||_2 = \left(\int_{\Omega} \left((\phi-\phi_{\text{exact}})^2+(\vec{u}-\vec{u}_{\text{exact}})^2\right)\;dx\right)^{1/2}.
\label{eqn:L2_error}
\end{equation}
Here $q_{\text{exact}}=(\phi_{\text{exact}},\vec{u}_{\text{exact}})$ is the known exact solution of the stationary vortex problem described in Section \ref{sec:stationary_vortex}, and this is used as the initial condition. $q=(\phi,\vec{u})$ denotes the numerical solution at the final time $T=\frac{1}{2}$. The implicit time step size is chosen as in Section \ref{sec:timestep_size}. Fig. \ref{fig:DG_error} shows the error $\Delta_2$ for varying grid refinement $r$ and different polynomial degrees $p$. The error decreases at a rate of at least $\mathcal{O}(h^{p+\frac{1}{2}})$ (and empirically no more than $\mathcal{O}(h^{p+1})$).
\subsection{Comparison of linear solvers}\label{sec:results_solver_comparison}
\begin{figure}
\begin{center}
\includegraphics[width=0.9\linewidth]{\figdir/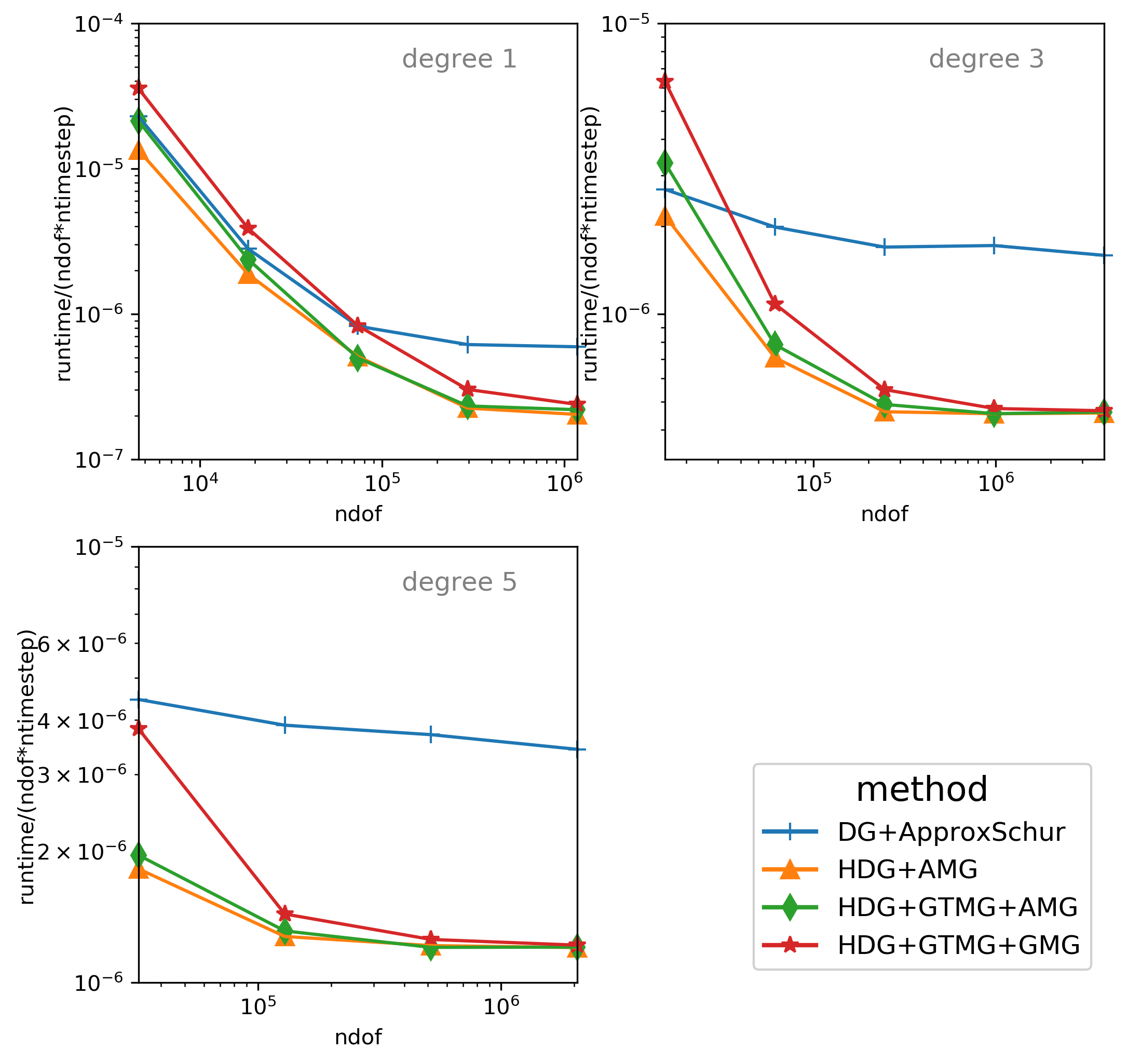}
\caption{Comparison of the performance of different linear solvers for the linearised SWEs without bathymetry (L-Up testcase). The runtime per timestep and per unknown is shown as a function of the number of degrees of freedom for different polynomial degrees $p$. The ARS2(2,3,2) timestepper is used in all cases and results are given in seconds.}
\label{fig:SolverComparisonAll}
\end{center}
\end{figure}

Next, we verify that the non-nested HDG solver described in Sections \ref{sec:schur_solve} and \ref{sec:non-nested-multigrid} is indeed superior to the naive DG solver in Section \ref{sec:solver_setup_DG} and demonstrate that the non-nested multigrid method developed in this paper (see Alg. \ref{alg:twolevel_vcycle}) is competitive with a black-box AMG solver for the hybridised system in Eq. \eqref{eqn:schur_system}. Fig. \ref{fig:SolverComparisonAll} compares the performance of the four solver setups listed in Section \ref{sec:solver_configs_results} for different polynomial degrees $p$ and increasing spatial resolution. The linear SWEs with flat bathymetry (L-Up) are solved with a Theta($\theta=0.5$) timestepper and again the timestep size is chosen as in Section \ref{sec:timestep_size}.
The plot demonstrates that for sufficiently large problems there is very little difference between the performance of the three hybridised solvers, namely HDG+GTMG+AMG, HDG+GTMG+GMG and HDG+AMG. To understand this, the total runtime for the simulations with refinement $r=7$ in Fig. \ref{fig:SolverComparisonAll} is broken down into the time spent in the solver for the Schur-complement system in Eq. \eqref{eqn:schur_system} and the combined time spent both in the fine-level smoother and residual calculation in Tab. \ref{tab:runtime_breakdown}. The results show that for high polynomial degrees the performance of the three solvers (GTMG+GMG, GTMG+AMG and AMG) differs by around $10\%$. This small variation in performance is readily explained by the fact that the solve time is dominated by the fine space smoother and residual calculation, which is the same for all methods.
\begin{table}
\begin{center}
\begin{tabular}{rlrrr}
\hline
& solver & $t_{\text{total}}$ & $t_{\text{solve}}$ & $t_{\text{fine}}$ \\
\hline\hline
\multirow{3}{9ex}{$p=1$\hspace{2ex} $\left\{\begin{matrix}\\[4ex]\end{matrix}\right.\hspace{-2ex}$} &GTMG+GMG & 21.0 & 6.9 & 1.6 \\
& GTMG+AMG & 17.7 & 3.5 & 1.5 \\
& AMG & 16.9 & 2.7 & 1.5\\
\hline
\multirow{3}{9ex}{$p=3$\hspace{2ex} $\left\{\begin{matrix}\\[4ex]\end{matrix}\right.\hspace{-2ex}$} & GTMG+GMG & 207.3 & 38.3 & 23.0 \\
& GTMG+AMG & 198.9 & 31.1 & 23.4 \\
& AMG & 200.7 & 32.3 & 26.4 \\
\hline
\multirow{3}{9ex}{$p=5$\hspace{2ex} $\left\{\begin{matrix}\\[4ex]\end{matrix}\right.\hspace{-2ex}$} & GTMG+GMG & 1691.7 & 121.1 & 90.3 \\
& GTMG+AMG & 1680.5 & 112.2 & 91.2 \\
& AMG & 1675.8 & 108.2 & 93.0 \\
\hline
\end{tabular}
\caption{Breakdown of runtime for the HDG-based solvers for refinement $r=7$ and different polynomial degrees $p$. The tolerance in the iterative solver is set to $\epsilon=10^{-8}$. In addition to the total runtime $t_{\text{total}}$, the time $t_{\text{solve}}$ spent in the solver for the Schur-complement system \eqref{eqn:schur_system} and the combined time $t_{\text{fine}}$ spent in the fine level smoother and residual calculation are shown. All results are given in seconds.}
\label{tab:runtime_breakdown}
\end{center}
\end{table}

\begin{figure}
\begin{center}
\includegraphics[width=0.9\linewidth]{\figdir/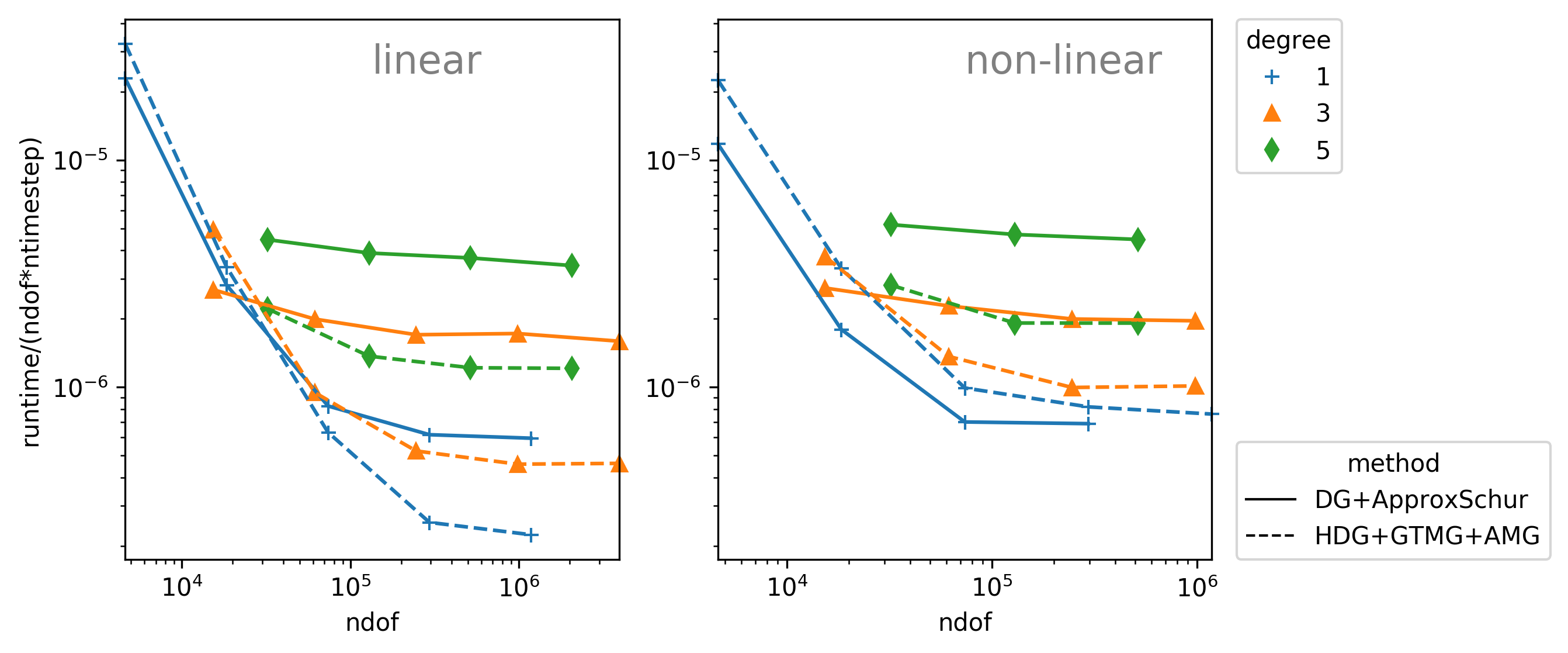}
\caption{Runtime per unknown per timestep as a function of the total number of DG unknowns $N_{\text{dof}}^{(\text{cell})}$ for different solvers. Results for the non-linear setup with spatially varying bathymetry (NL-LF) are shown on the right and times for the linear SWEs with constant bathymetry (L-Up) on the left. All results are given in seconds.}
\label{fig:results_theta_runtime}
\end{center}
\end{figure}

\begin{figure}
\begin{center}
\includegraphics[width=0.9\linewidth]{\figdir/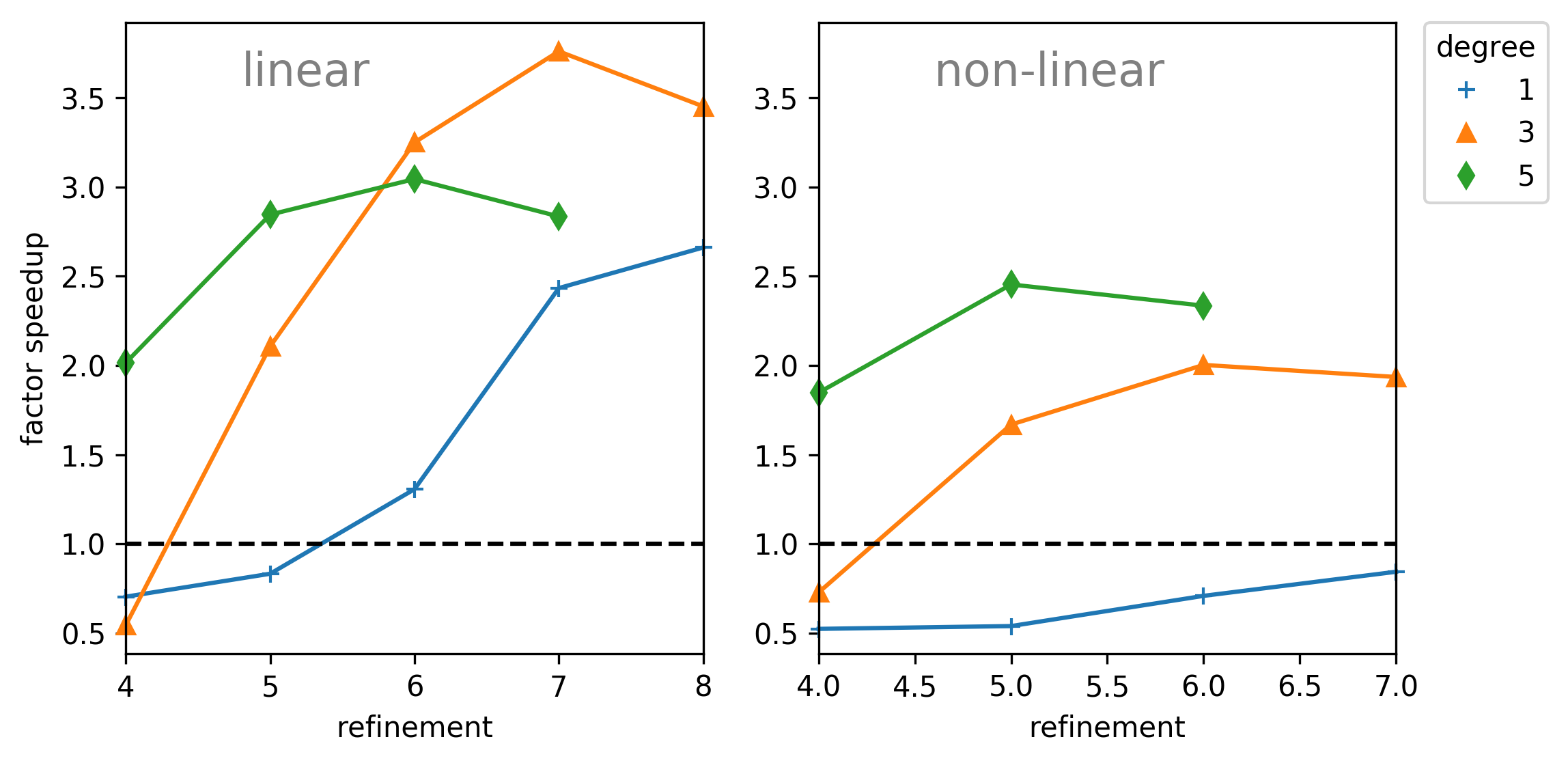}
\caption{Speedup of the hybridised HDG+GTMG+AMG solver relative to the DG+ApproxSchur method as a function of the total number of DG unknowns $N_{\text{dof}}^{(\text{cell})}$ given in Eq. \eqref{eqn:ndofs}. The Theta($\theta=0.5$) timestepper is used in both cases. Results for the non-linear setup with spatially varying bathymetry (NL-LF) are shown on the left and speedups for the linear SWEs with constant bathymetry (L-Up) on the right.}
\label{fig:results_theta_speedup}
\end{center}
\end{figure}

All solvers based on the hybridised DG discretisation perform significantly better the native DG method with an approximate Schur complement preconditioner (DG+ApproxSchur). At the highest considered polynomial degree ($p=5$) the speedup is approximately a factor of $3$. Fig. \ref{fig:results_theta_runtime} shows the runtimes achieved with the DG+ApproxSchur and HDG+GTMG+AMG method both for the linear system (L-Up, left) and the non-linear problem with spatially varying bathymetry (NL-LF, right). The relative speedup of the HDG+GTMG+AMG method is quantified in both cases in Fig. \ref{fig:results_theta_speedup}. As can be seen from Fig. \ref{fig:results_theta_runtime} the drop in speedup for higher refinements can be explained by the fact that the runtime of the DG+ApproxSchur decreases slightly more than expected as $r$ is increased from 7 to 8 for $p=3$ and from 6 to 7 for $p=5$.

This speedup can be partly attributed to the larger number of GMRES iterations in the ``outer'' solve of the $(\phi,\vec{u})$ system for the DG+ApproxSchur solver and the relatively smaller number of Krylov-solver iterations required to solve the flux-system with the non-nested multigrid preconditioner for the HDG+GTMG+AMG method; in both cases the residual was reduced by a factor of $\epsilon=10^{-8}$. This number of iterations is shown for the linear L-Up problem (which uses a CG-solver for the symmetric positive definite flux system) in Tab. \ref{tab:nIterLinear} and for  the non-linear NL-LF problem in Tab. \ref{tab:nIterNonlinear}. Compared to the native DG method, the hybridised DG solver requires significantly fewer iterations and is robust with respect to the polynomial degree $p$.
\begin{table}
\begin{center}
\begin{tabular}{|cc|cc|cc|cc|}
\hline
& & \multicolumn{2}{|c|}{$p=1$} 
& \multicolumn{2}{|c|}{$p=3$} 
& \multicolumn{2}{|c|}{$p=5$} \\
$r$ & $h=2^{-r}$ & DG & HDG & DG & HDG & DG & HDG\\
\hline\hline
4 & 0.0625 & 14.1 & 8.1 & 15.5 & 8.0 & 17.3 & 8.0 \\
5 & 0.0313 & 13.8 & 8.1 & 14.9 & 7.0 & 16.5 & 8.0 \\
6 & 0.0078 & 13.0 & 8.0 & 14.0 & 7.0 & 15.9 & 8.0 \\
7 & 0.0078 & 12.9 & 8.0 & 14.0 & 7.0 & 15.0 & 8.0 \\
8 & 0.0039 & 12.0 & 8.0 & 13.0 & 7.0 & --- & ---\\
\hline
\end{tabular}
\end{center}
\caption{Average number of GMRES/CG iterations for solving the linear problem arising in the linearised SWEs with flat bathymetry (L-Up) to a tolerance of $\epsilon=10^{-8}$. For the DG+SchurApprox solver the number of GMRES iterations over the outer $(\phi,\vec{u})$ system in Eq. \eqref{eqn:native_dg_system} are given in the ``DG'' column. The number of CG iterations required for solving the flux problem in Eq. \eqref{eqn:schur_system} is reported for the HDG+GTMG+AMG method in the ``HDG'' column.}
\label{tab:nIterLinear}
\end{table}
\begin{table}
\begin{center}
\begin{tabular}{|cc|cc|cc|cc|}
\hline
& & \multicolumn{2}{|c|}{$p=1$} 
& \multicolumn{2}{|c|}{$p=3$} 
& \multicolumn{2}{|c|}{$p=5$} \\
$r$ & $h=2^{-r}$ & DG & HDG & DG & HDG & DG & HDG\\
\hline\hline
4 & 0.0625 & 15.9 & 10.1 & 18.0 & 8.9 & 21.0 & 8.8 \\
5 & 0.0313 & 15.0 & 10.2 & 17.3 & 9.0 & 20.4 & 8.8 \\
6 & 0.0078 & 14.1 & 10.1 & 16.3 & 9.0 & 19.4 & 9.0 \\
7 & 0.0078 & 14.0 & 10.1 & 16.0 & 8.9 & --- & --- \\
8 & 0.0039 & ---  & 10.0 & ---  & --- & --- & --- \\
\hline
\end{tabular}
\caption{Average number of GMRES iterations for the linear problem arising in the fully non-linear SWEs with bathymetry (NL-LF) to a tolerance of $\epsilon=10^{-8}$. For the DG+SchurApprox solver the number of iterations over the outer $(\phi,\vec{u})$ system in Eq. \eqref{eqn:native_dg_system} are given in the ``DG'' column. The number of iterations required for solving the flux problem in Eq. \eqref{eqn:schur_system} is reported for the HDG+GTMG+AMG method in the ``HDG'' column.}
\label{tab:nIterNonlinear}
\end{center}
\end{table}
Another advantage of the HDG+GTMG+AMG solver, in particular for higher polynomial degree $p$, is that the flux system, whose solution dominates the runtime, has significantly fewer unknowns than the $(\phi,\vec{u})$-system which is solved by the DG+ApproxSchur solver. As discussed at the end of section \ref{sec:solver_configs_results}, the $(\phi,\vec{u})$ system is $p+2$ times larger than the flux-system for the linearised SWEs (L-Up). In the fully non-linear case (NL-LF) the relative size is only $\frac{1}{2}(p+2)$ since the Lax-Friedrich flux will result in two flux unknowns per facet. This partly explains the smaller speedup in this case.
\subsection{Comparison of IMEX timesteppers}\label{sec:results_other_timesteppers}
To assess the overall performance of our methods when solving the time-dependent shallow water equations, we compare different IMEX timesteppers introduced in Section \ref{sec:timesteppers}. Tab. \ref{tab:timesteppercomparison} gives the runtime and $L_2$ error $\Delta_2$ (defined in Eq. \eqref{eqn:L2_error}) for the fully non-linear setup with bathymetry (NL-LF). For all IMEX methods the average number of iterations is reported both for DG and hybridised DG discretisations. The polynomial degree is $p=3$ and the refinement $r=6$. Results are given for linear solver tolerances of $\epsilon=10^{-8}$, $10^{-6}$ and $10^{-4}$. For the NL-LF setup the HDG+GTMG+AMG method is used to solve the Schur-complement problem for the flux variables.
For the timesteppers considered here the HDG solver is typically twice as fast as the corresponding DG method for $\epsilon=10^{-8}$, which confirms the findings in Section \ref{sec:results_solver_comparison}. For looser tolerances the advantage of the HDG method is less pronounced. This is readily explained by the following observation: for the DG method the cost is roughly proportional to the number of outer solver iterations, which becomes smaller as $\epsilon$ increases. For the HDG method, however, a significant amount of time is spent in the construction of the right hand side of the Schur-complement flux system and the reconstruction of $\phi$ and $\vec{u}$ (it is well known that the current Slate implementation of those parts of the algorithm is sub-optimal and will be improved in a future version of Firedrake). Since the solution of the flux system is relatively cheap, any reduction in the number of solver iterations with looser tolerances has a smaller impact on the overall runtime. Off-centering the Theta($\theta$)-method slightly by setting $\theta=0.55$ instead of $\theta=0.5$ reduces the error by two orders of magnitude at the cost of a small increase in the number of linear solver iterations and overall runtime (this increase is not observed for the hybridised solver). Looking back at Tab. \ref{tab:timestepper_comparison}, the significantly higher cost of the ARS2(2,3,2), SSP2(3,2,2) and ARS4(4,4,3) integrators (which are not really more accurate than the Theta$(\theta=0.55)$ method for this particular testcase) is largely explained by the fact that they require two, three and four linear solves per timestep respectively. In virtually all cases the HDG based solver with our non-nested multigrid algorithm performs significantly better than the naive DG based method.
\begin{table}
\begin{center}
\begin{tabular}{lr|rr|rr|c}
\hline
& tolerance & \multicolumn{2}{|c|}{runtime [s]} 
& \multicolumn{2}{|c|}{$n_{\text{iter}}$} 
& error $\Delta_2$ \\
timestepper & $\epsilon$ & DG & HDG & DG & HDG & \\
\hline\hline
\multirow{3}{*}{Theta($\theta=0.5$)}
  & $10^{-8}$ & 107.4 & 56.1 & 16.3 & 9.0 & $5.29\cdot 10^{-6}$\\
  & $10^{-6}$ & 78.5 & 49.7 & 12.0 & 7.0 & $5.14\cdot 10^{-6}$\\
  & $10^{-4}$ & 53.0 & 49.3 & 7.7 & 5.0 & $4.77\cdot 10^{-5}$\\\hline
\multirow{3}{*}{Theta($\theta=0.55$)}
  & $10^{-8}$ & 117.0 & 56.9 & 17.9 & 8.9 & $5.38\cdot 10^{-8}$ \\
  & $10^{-6}$ & 84.5 & 49.7 & 13.0 & 7.0 & $5.40\cdot 10^{-8}$ \\
  & $10^{-4}$ & 59.3 & 47.1 & 8.8 & 5.0 & $2.79\cdot 10^{-7}$ \\\hline
\multirow{3}{*}{ARS2(2,3,2)}
  & $10^{-8}$ & 184.4 & 112.0 & 14.0 & 8.6 & $5.39\cdot 10^{-8}$ \\
  & $10^{-6}$ & 137.7 & 102.7 & 10.0 & 7.0 & $5.39\cdot 10^{-8}$ \\
  & $10^{-4}$ & 95.2 & 97.0 & 6.4 & 5.0 & $2.43\cdot 10^{-7}$ \\\hline
\multirow{3}{*}{SSP2(3,2,2)}
  & $10^{-8}$ & 327.0 & 158.7 & 17.0 & 8.0 & $7.36\cdot 10^{-6}$ \\
  & $10^{-6}$ & 235.8 & 146.1 & 12.0 & 6.0 & $7.36\cdot 10^{-6}$ \\
  & $10^{-4}$ & 169.6 & 140.6 & 8.0 & 5.0 & $7.36\cdot 10^{-6}$ \\\hline
\multirow{3}{*}{ARS3(4,4,3)}
  & $10^{-8}$ & 456.0 & 212.7 & 17.0 & 8.0 & $5.38\cdot 10^{-8}$ \\
  & $10^{-6}$ & 357.4 & 205.9 & 13.0 & 7.0 & $5.38\cdot 10^{-8}$ \\
  & $10^{-4}$ & 230.6 & 195.0 &  8.0 & 5.0 & $1.10\cdot 10^{-7}$ \\\hline
\hline
\end{tabular}
\caption{Comparison of runtime, number of solver iterations and $L_2$ error for different IMEX timesteppers. Results are reported for solver tolerances of $\epsilon=10^{-8}$, $10^{-6}$ and $10^{-4}$. The polynomial degree is $p=3$ and the refinement $r=6$.}
\label{tab:timesteppercomparison}
\end{center}
\end{table}
\subsection{Parallel scalability}\label{sec:results_scalability}
\begin{figure}
\begin{center}
\includegraphics[width=0.9\linewidth]{\figdir/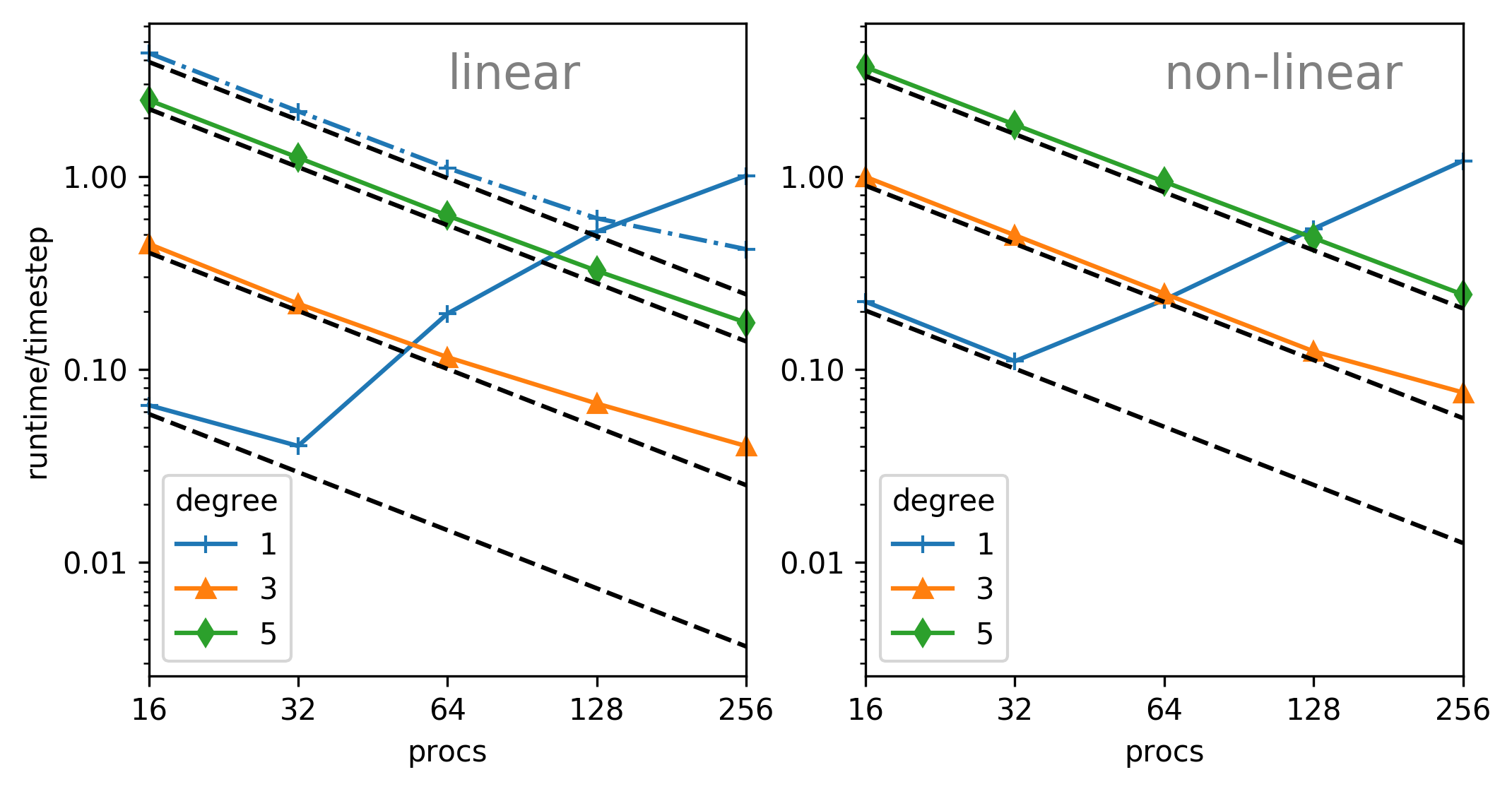}
\caption{Strong scaling of the time per timestep for the HDG+GTMG+AMG solver on 16 to 256 processors. Results are shown for both for the linearised problem with constant bathymetry (L-Up, left) and the fully non-linear SWEs with spatially varying bathymetry (NL-LF, right). The dot-dashed curve in the left plot shows the results for degree $p=1$ with refinement $r=10$, the solid curves use a refinement of $r=7$. The black dashed lines correspond to optimal speedup for which the runtime per timestep would be inversely proportional to the number of processors.}
\label{fig:results_strongscaling}
\end{center}
\end{figure}

Finally, the parallel scalability of the code is quantified for different polynomial degrees $p$. In a strong scaling experiment the global problem size is kept fixed at $2.95\cdot 10^{5}$ ($p=1$), $9.83\cdot 10^{5}$ ($p=3$) and $2.06\cdot 10^{6}$ ($p=5$) unknowns while increasing the number of processor cores. The chosen global problem sizes correspond to a refinement of $r=7$ for all polynomial degrees. As will be discussed below, a higher refinement is required for the lowest polynomial degree to achieve good parallel scalability. To illustrate this, results with $r=10$ are also reported for $p=1$ and the linearised SWEs (L-Up setup). The local problem size for all considered processor counts are shown in Tab. \ref{tab:strong_scaling_problemsizes}.
\begin{table}
\begin{center}
\begin{tabular}{c|ccccc}
\hline
$p$ $\backslash$ \#procs& 16 & 32 & 64 & 128 & 256\\
\hline\hline
$1$  & $ 1.84\cdot 10^{4}$ & $ 9.22\cdot 10^{3}$ & $ 4.61\cdot 10^{3}$ & $ 2.30\cdot 10^{3}$ & $ 1.15\cdot 10^{3}$\\
$3$  & $ 6.14\cdot 10^{4}$ & $ 3.07\cdot 10^{4}$ & $ 1.54\cdot 10^{4}$ & $ 7.68\cdot 10^{3}$ & $ 3.84\cdot 10^{3}$\\
$5$  & $ 1.29\cdot 10^{5}$ & $ 6.45\cdot 10^{4}$ & $ 3.23\cdot 10^{4}$ & $ 1.61\cdot 10^{4}$ & $ 8.06\cdot 10^{3}$\\\hdashline
$1^*$  & $ 1.18\cdot 10^{6}$ & $ 5.90\cdot 10^{5}$ & $ 2.95\cdot 10^{5}$ & $ 1.47\cdot 10^{5}$ & $ 7.37\cdot 10^{4}$\\
\hline
\end{tabular}
\caption{Local problem sizes for strong scaling experiments for different polynomial degrees $p$ and numbers of processor cores. The refinement level is $r=7$ for the first three rows, the final row (marked by an asterisk $*$) shows the local problem sizes for $p=1$ with refinement $r=10$.}
\label{tab:strong_scaling_problemsizes}
\end{center}
\end{table}
Fig. \ref{fig:results_strongscaling} shows the strong scaling of the time per timestep. As can be seen from this plot, scalability is reasonably good for $p=3,5$ until the local problem size reaches a few thousand unknows on 256 cores. For the lowest polynomial degree ($p=1$), however, the code antiscales if more than two nodes are used. This issue can only be overcome by significantly increasing the global problem size by using a refinement level of $r=10$ instead.

In summary, the results clearly demonstrate that increasing the polynomial degree improves scalability. This can be explained by the fact that, compared to the cost of computations, the time spent communicating data decreases as the polynomial degree increases. 

\begin{figure}
\begin{center}
\includegraphics[width=0.9\linewidth]{\figdir/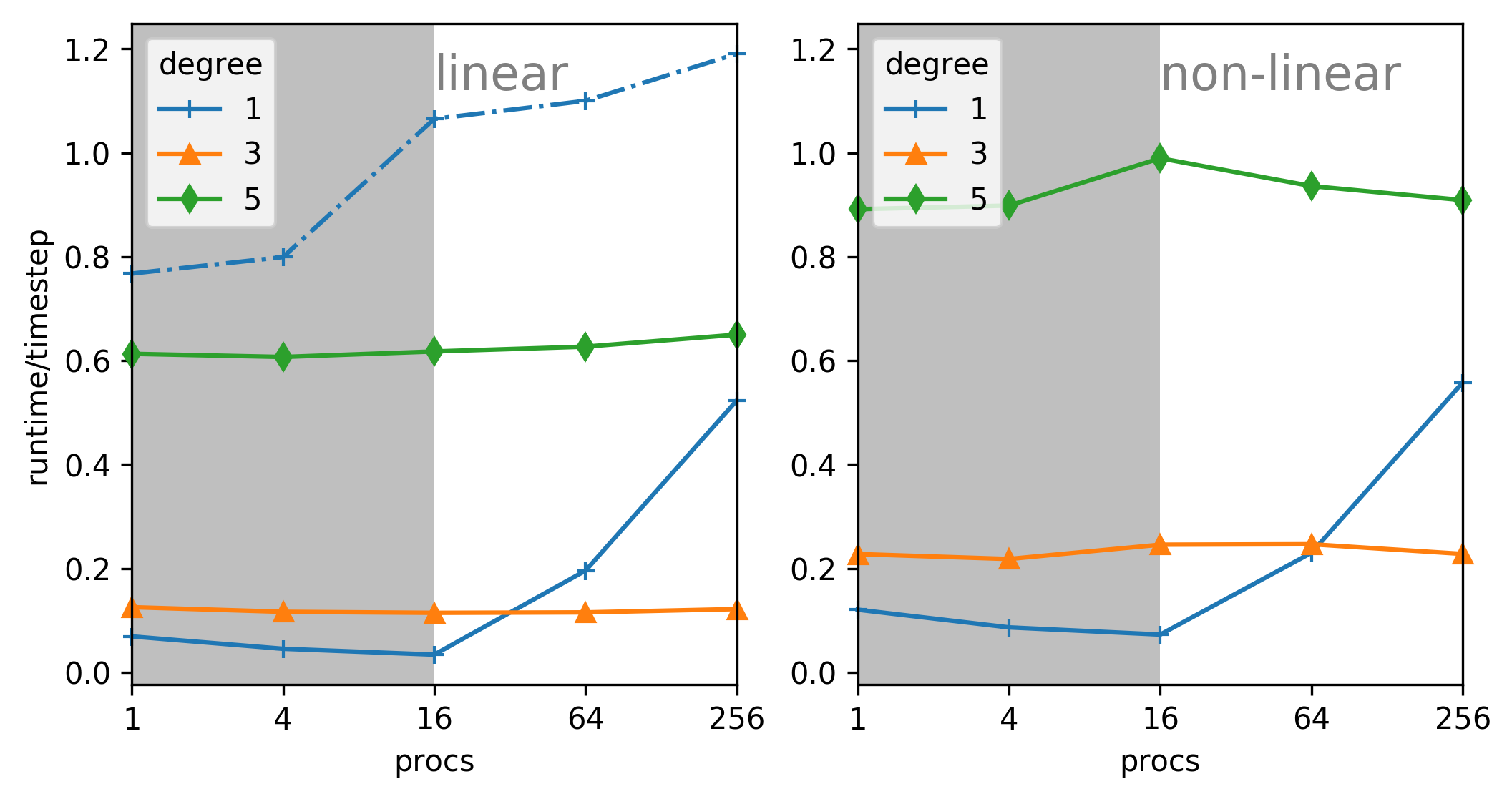}
\caption{Weak scaling of the time per timestep for the HDG+GTMG+AMG solver on 16 to 256 processors. Results are shown for both the fully non-linear SWEs with spatially varying bathymetry (NL-LF, left) and for the linearised problem with constant bathymetry (L-Up, right). The gray region shows scaling within a node.}
\label{fig:results_weakscaling}
\end{center}
\end{figure}

For the weak scaling experiment the total problem size is increased in proportion with the number of processors while keeping the local problem size fixed. On one processor the number of unknowns is $4.61\cdot 10^{3}$ ($p=1$), $1.54\cdot 10^{4}$ ($p=3$) and $3.23\cdot 10^{4}$ ($p=5$); this corresponds to a refinement of $r=4$ for all polynomial degrees. Tab. \ref{tab:weak_scaling_problemsizes} lists the global problem sizes. As this table shows, problems with more than a million unknowns are solved on 256 cores. Analogously to the strong scaling experiment, for degree $p=1$ we also report results for a higher refinement ($r=7$ in this case).
\begin{table}
\begin{center}
\begin{tabular}{c|ccccc}
\hline
$p$ $\backslash$ \#procs& 1 & 4 & 16 & 64 & 256\\
\hline\hline
$1$  & $ 4.61\cdot 10^{3}$ & $ 1.84\cdot 10^{4}$ & $ 7.37\cdot 10^{4}$ & $ 2.95\cdot 10^{5}$ & $ 1.18\cdot 10^{6}$\\
$3$  & $ 1.54\cdot 10^{4}$ & $ 6.14\cdot 10^{4}$ & $ 2.46\cdot 10^{5}$ & $ 9.83\cdot 10^{5}$ & $ 3.93\cdot 10^{6}$\\
$5$  & $ 3.23\cdot 10^{4}$ & $ 1.29\cdot 10^{5}$ & $ 5.16\cdot 10^{5}$ & $ 2.06\cdot 10^{6}$ & $ 8.26\cdot 10^{6}$\\\hdashline
$1^*$  & $ 2.95\cdot 10^{5}$ & $ 1.18\cdot 10^{6}$ & $ 4.72\cdot 10^{6}$ & $ 1.89\cdot 10^{7}$ & $ 7.55\cdot 10^{7}$\\
\hline
\end{tabular}
\caption{Global problem sizes for weak scaling experiments for different polynomial degrees $p$ and numbers of processor cores. The refinement level is $r=4$ for the first three rows, the final row (marked by an asterisk $*$) shows the local problem sizes for $p=1$ with refinement $r=7$.}
\label{tab:weak_scaling_problemsizes}
\end{center}
\end{table}
The time per timestep for increasing problem sizes is shown in Fig. \ref{fig:results_weakscaling}. 
Again, we find that scalability is very good for high polynomial degrees ($p=3,5$). Scaling breaks down for $p=1$ once inter-node communication is required. As for the strong scaling experiment, this is caused by the poorer calculation-to-communication ratio in this case. Increasing the global problem size for $p=1$ by setting $r=7$ improves this, but the code still does not scale as well as for higher polynomial degrees. As usual, interpreting results within a single node (i.e. for 16 or less cores) is very challenging since the variation of the runtime can be influenced by saturation of the memory bandwidth and the complexities of the cache hierarchy.
\subsection{Block-diagonal approximation in Schur complement}\label{sec:blockdiag_schurapprox}
As discussed at the end of Section \ref{sec:solver_setup_DG}, the approximate Schur-complement preconditioner for the native DG method (the DG+ApproxSchur solver) can be potentially improved by using the block-diagonal instead of the diagonal of the matrix $A_{uu}$.
\begin{figure}
\begin{center}
\includegraphics[width=0.7\linewidth]{\figdir/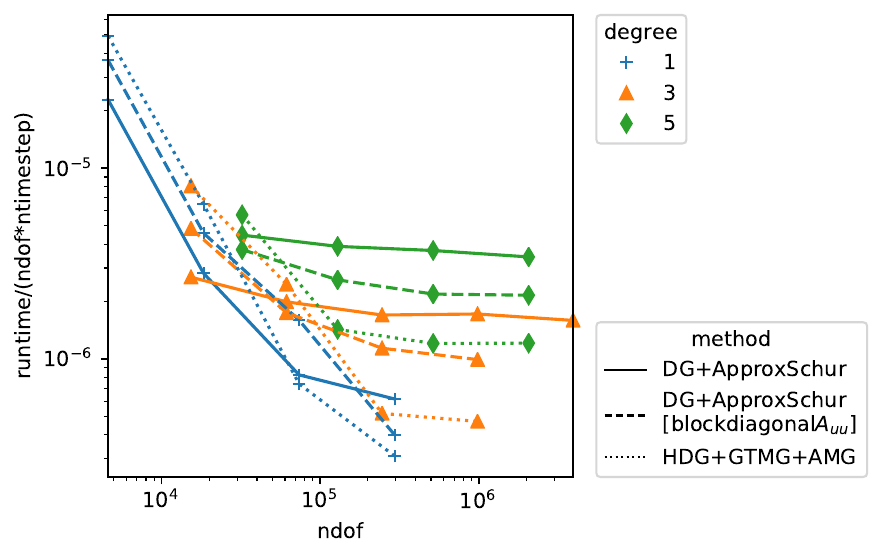}
\caption{Runtime per unknown per timestep as a function of the total number of DG unknowns $N_{\text{dof}}^{(\text{cell})}$ for the HDG+GTMG+AMG solver and different approximate Schur complements, using either the diagonal (DG+ApproxSchur) or the block-diagonal (DG+ApproxSchur [blockdiagonal $A_{uu}$]) of the matrix $A_{uu}$. All results are given in seconds and were obtained for the linear SWEs with constant bathymetry (L-Up). The tolerance in the linear solver is set to $\epsilon=10^{-8}$.}
\label{fig:results_blockdiag_schurapprox}
\end{center}
\end{figure}
To quantify any improvements due to this modification, the runs for the linearised SWE testcase (L-Up) from Section \ref{sec:results_solver_comparison} were repeated for a variant of the DG+ApproxSchur preconditioner, in which $\text{diag}(A_{uu})$ has been replaced by $\text{blockdiag}(A_{uu})$ in Eq. \eqref{eqn:S_approx_diag}. Fig. \ref{fig:results_blockdiag_schurapprox} shows those results, together with the ones for DG+ApproxSchur and HDG+GTMG+AMG which are already plotted in Fig. \ref{fig:results_theta_runtime} (left). As those results demonstrate, the approximate Schur-complement preconditioner can be improved by this modification, but is still not competitive with solvers based on the hybridised DG method.
\subsection{Comparison to explicit time integrators}\label{sec:results_explicit}
For completeness, we also present a tentative comparison of our IMEX-HDG solvers with fully explicit time integrators. We stress, however, that this is not the main focus of this paper, and that the comparison of absolute runtimes should be interpreted with caution: the simplified considered setup here does not replicate the conditions encountered for realistic flows and a more reliable comparison should be carried out for example for the spherical testcase in \cite{Williamson1992} to assess the relative merits of explicit and semi-implicit methods.
\begin{table}
\begin{center}
\begin{tabular}{lcc}
\hline
timestepper & runtime [s] & error $\Delta_2$ \\
\hline\hline
Explicit Euler & 32.0 & $6.25\cdot 10^{-2}$\\
Heun & 35.3 & $5.41\cdot 10^{-8}$\\
SSPRK3 & 59.1 & $5.41\cdot 10^{-8}$ \\
\hline
\end{tabular}
\caption{Comparison of runtime and $L_2$ error for different explicit timesteppers. The polynomial degree is $p=3$ and the refinement $r=6$.}
\label{tab:timesteppercomparison_explicit}
\end{center}
\end{table}

We consider the three explicit time integrators introduced in Section \ref{sec:explicit_methods}: the lowest order explicit (forward) Euler method, Heun's method and the third order accurate SSPRK3 integrator. The absolute runtime and errors for the same setup as used in Tab. \ref{tab:timesteppercomparison} are shown in Tab. \ref{tab:timesteppercomparison_explicit}. Since the Froude number is around $F=10$, those results were obtained with an explicit timestep which is a factor ten shorter than the implicit timestep size used for the IMEX integrators in Section \ref{sec:results_other_timesteppers} to guarantee stability.

While the forward Euler integrator is extremely simple and leads to the shortest runtime, not very surprisingly the error is very large in this case; potentially the method is also affected by a time-stepping instability, which is seeded by the imperfect balance of the initial state. The Heun method, which requires an additional function evaluation and two instead of one mass solves, reduces the error significantly, while only increasing the runtime by around $10\%$. Comparing the most efficient IMEX method (Theta($\theta=0.55$)) to the fastest explicit method with the same $L_2$ error (Heun), we conclude that for this particular setup using hybridisation allows the design of IMEX timesteppers which (for a linear solver tolerance of $10^{-6}$) are only $1.4\times$ slower than the significantly simpler explicit integrators. The difference in performance can be understand by looking at the number of function evaluations and linear solves given in Tab. \ref{tab:timestepper_comparison}. The Theta($\theta$) method requires one expensive linear solve and one function evaluation per timestep while two function evaluations and two mass-solves are needed for the Heun-method.

Although here we restrict ourselves to the case where the implicit time step size $\Delta t$ is not more than ten times larger than the explicit step size $\Delta t_{\text{expl}}$, the authors of \cite{Kang2020} also consider scenarios where this ratio is larger. They find that for $\Delta t/\Delta t_{\text{expl}}=20$ their IMEX HDG-ARS2 code is nearly as fast as an RKDG2 (SSPRK3) integrator, whereas for $\Delta t/\Delta t_{\text{expl}}=200$ their HDG-based IMEX solver is ten times faster than the explicit integrator, albeit with a larger error.
\subsection{Spherical geometry}\label{sec:results_spherical}
We now apply our methods to well-established test cases in spherical geometry \cite{Williamson1992}. Care has to be taken when extending the DG discretisation to non-Euclidean geometries since in general the cell-normals of two neighbouring cells (see Fig. \ref{fig:cells}) do not necessarily satisfy $\vec{n}^-=-\vec{n}^+$. To address this issue, we followed the approach described in \cite{Bernard2009}. As can be seen for example in Eq. \eqref{eqn:bilinear_standardDG_lax}, this introduces additional ``curvature'' terms in the linear- and non-linear standard DG-fluxes. It is not necessary to modify the HDG flux since in this case unknowns in neighbouring cells only couple indirectly through the hybridised variables on the facets.

The main aim of the following numerical experiments is to compare the performance of our non-nested multigrid methods to direct solvers which were also used in \cite{Kang2020}.
\subsubsection{Problem setup}\label{sec:spherical_setup}
We consider three configurations from \cite{Williamson1992} which are labelled W3, W5 and W6 for future reference:
\begin{description}
\item[W3]: Steady state zonal geostrophic flow with compact support described in \cite[Sec. 3.3]{Williamson1992}. Since this test case has an analytic solution, we use it to verify the correctness of our code and study convergence under $h$- and $p$-refinement. Note also that the setup is conceptually very similar to the steady state vortex in flat geometry described in Section \ref{sec:stationary_vortex}.
\item[W5]: Zonal flow over an isolated mountain, as described in \cite[Sec. 3.5]{Williamson1992}. Although the initial condition is the same as for W3, in this case the bathymetry is not constant and describes an isolated conical mountain in the Northern hemisphere.
\item[W6]: Rossby-Haurwitz waves described in \cite[Sec 3.6]{Williamson1992} are analytical solutions of the nonlinear barotropic vorticity equation (but now of the SWEs) on the sphere.
\end{description}
The parameters used in our numerical experiments are the same as in \cite{Williamson1992}; for reference they are collected in Tab. \ref{tab:williamson_parameters} in \secapp~\ref{sec:williamson_parameters}; in the following all quantities are non-dimensionalised as described in Section \ref{sec:SWEs}. For this choice of parameters the maximal value of the Froude number $F := \left|\vec{U}^{(\text{adv})}\right|\phi_B^{-1/2}$ at the initial time is approximately $0.23$ for W3, $0.17$ for W5 and $0.34$ for W6, and hence the scale separation is slightly less pronounced than for the steady-state vortex in Section \ref{sec:stationary_vortex}. Since the test cases considered are non-linear, the Lax-Friedrichs flux formulation of the discretised SWEs is used in all cases. The computational grid is obtained by recursively sub-dividing the cells of an icosahedral coarse mesh consisting of 20 triangles, taking into account the curvature of the sphere. If $r$ denotes the number of refinement steps, the average side length of a triangular grid cell is (in units of the radius of the Earth)
\begin{equation}
  h = 2^{-r}h_0\qquad\text{with $h_0 = \sqrt{\frac{4\pi}{5\sqrt{3}}}\approx 1.20459$}.
\end{equation}
Tab. \ref{tab:grid_setup_spherical} shows the grid layout and number of unknowns for a range of refinement levels and polynomial degrees.
\begin{table}
\begin{center}
\begin{tabular}{|r|c|r|rrr|rrr|}
\hline
\multicolumn{1}{|c|}{$r$} & \multicolumn{1}{|c|}{$N_{\text{cell}}$} & $h$& \multicolumn{3}{|c|}{$p=3$}& \multicolumn{3}{|c|}{$p=5$}
\\\hline\hline
2 & 320  & 0.3011 & 9 600  & /  & 1 920  & 20 160  & /  & 2 880 \\
3 & 1 280  & 0.1506 & 38 400  & /  & 7 680  & 80 640  & /  & 11 520 \\
4 & 5 120  & 0.0753 & 153 600  & /  & 30 720  & 322 560  & /  & 46 080 \\
5 & 20 480  & 0.0376 & 614 400  & /  & 122 880  & 1 290 240  & /  & 184 320 \\
6 & 81 920  & 0.0188 & 2 457 600  & /  & 491 520  & 5 160 960  & /  & 737 280 \\
7 & 327 680  & 0.0094 & 9 830 400  & /  & 1 966 080  & 20 643 840  & /  & 2 949 120 \\
\hline
\end{tabular}
\caption{Grid layout and numbers of unknowns for different polynomial degrees $p$ for the numerical experiments in Section \ref{sec:results_spherical}. Each of the final two columns shows the total number of unknowns associated with cells and facets as calculated by Eq. \eqref{eqn:ndofs} for the scalar-valued flux problem ($\nu^{(\text{flux})}=\nu^{(\text{up})}=1$) in the form $N_{\text{dof}}^{(\text{cell})} / N_{\text{dof}}^{(\text{facet})}$.}
\label{tab:grid_setup_spherical}
\end{center}
\end{table}
Integrals over cells and facets in the weak form require evaluation of geometric quantities to capture the local curvature of the sphere, such as the Jacobian of the pullback to a flat reference cell. In the following the coordinate field is approximated by a polynomial of degree 3 in each grid cell. For all numerical experiments we use the ARS2 time integrator and choose the timestep size $\Delta t$ such that
\begin{equation}
  \frac{\max{\{|\vec{U}^{(\text{adv})}_0|\}} \Delta t}{h} = \frac{\rho}{2p+1}
  \label{eqn:spherical_timestep_constraint}
\end{equation}
where $\vec{U}^{(\text{adv})}_0$ is the initial advection velocity and $\rho$ is a parameter that is adjusted such that the semi-implicit timestepper (which treats advective processes explicitly) is stable. We find that the choice $\rho=0.1/\sqrt{2}\approx 0.0707$ works well for the majority of considered test cases, larger values lead to instabilities. The only exception are the numerical experiments for polynomial degree $p=5$ in the (non-linear) W5- and W6- setups; in this case we had to reduce the timestep size by an additional factor of two to ensure stability over the entire simulation window of 15 days.
\subsubsection{Linear solvers}\label{sec:spherical_solver_configs}
We use three different methods to solve the linear problem that arises in the semi-implicit time integrator:
\begin{description}
\item[HDG+GTMG+AMG]: The non-nested multigrid approach with AMG solver for the system in the Raviart-Thomas subspace, as described in more detail in Section \ref{sec:solver_configs_results}.
\item[HDG+LU]: In this configuration the hybridised linear system in Eq. \eqref{eqn:general_linear_system} is solved with a direct method from the MUMPS suite \cite{MUMPS1,MUMPS2}; this is similar to the setup in \cite{Kang2020}.
\item[DG+ApproxSchur]: As described in Section \ref{sec:solver_configs_results}, the non-hybridised $(\phi,\vec{u})$ system in Eq. \eqref{eqn:native_dg_system} is solved with an iterative solver preconditioned with an approximate Schur complement.
\end{description}
Recall that the nature of the hybridised system in Eq. \eqref{eqn:general_linear_system} is very different if the upwind flux is used: in contrast to the Lax-Friedrichs flux, the hybridised unknowns on each facet are scalar-valued and the coarse-level problem is formulated in the P1-subspace in this case. To also explore the performance of our solvers in this setup, we constructed linearised variants $\text{W3}^{(\text{lin})}$, $\text{W5}^{(\text{lin})}$ and $\text{W6}^{(\text{lin})}$ of the test cases in \cite{Williamson1992}. Those are obtained by dropping all terms in the initial conditions which are quadratic in the velocity and using the linear shallow water equations with upwind flux. Note that with those modifications the solution of $\text{W3}^{(\text{lin})}$ is still stationary and known analytically.
\subsubsection{Convergence}
\begin{figure}
  \begin{minipage}{0.3\linewidth}
    \begin{center}
      \includegraphics[width=1.0\linewidth]{\figdir/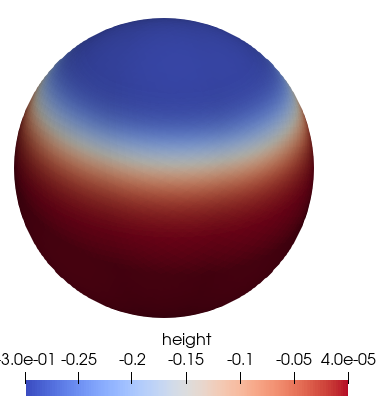}
  \end{center}
\end{minipage}
\hfill
  \begin{minipage}{0.3\linewidth}
    \begin{center}
            \includegraphics[width=1.0\linewidth]{\figdir/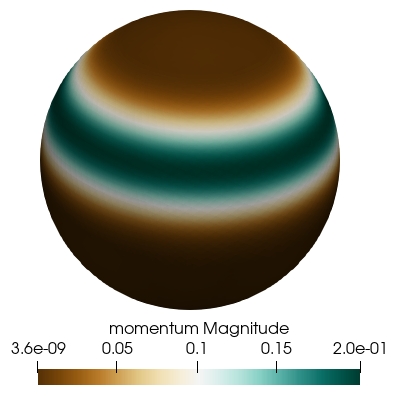}
  \end{center}
  \end{minipage}
\hfill
  \begin{minipage}{0.3\linewidth}
    \begin{center}
            \includegraphics[width=1.0\linewidth]{\figdir/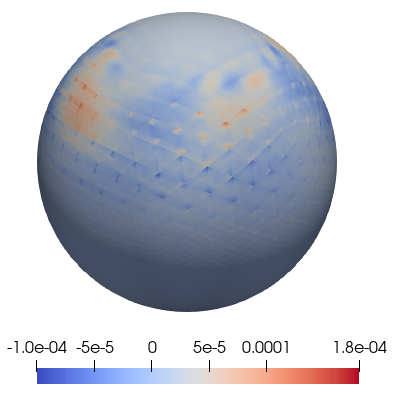}
  \end{center}
  \end{minipage}
  \caption{Potential $\phi$ (left), absolute value of the momentum $|\vec{u}|$ (center) and relative error in the potential $\phi+\phi_B$ (right) for the non-linear W3 (steady-state zonal flow) testcase at $t=12\unit{days}$. The polynomial degree is $p=3$ and a refinement level of $r=3$ was used. The timestep size in physical units is $650.6\unit{s}$. The linear flux system is solved with the HDG+GTMG+AMG solver.}
  \label{fig:W3_field_and_error}
\end{figure}

\begin{figure}
  \centering
  \includegraphics[width=0.8\linewidth]{\figdir/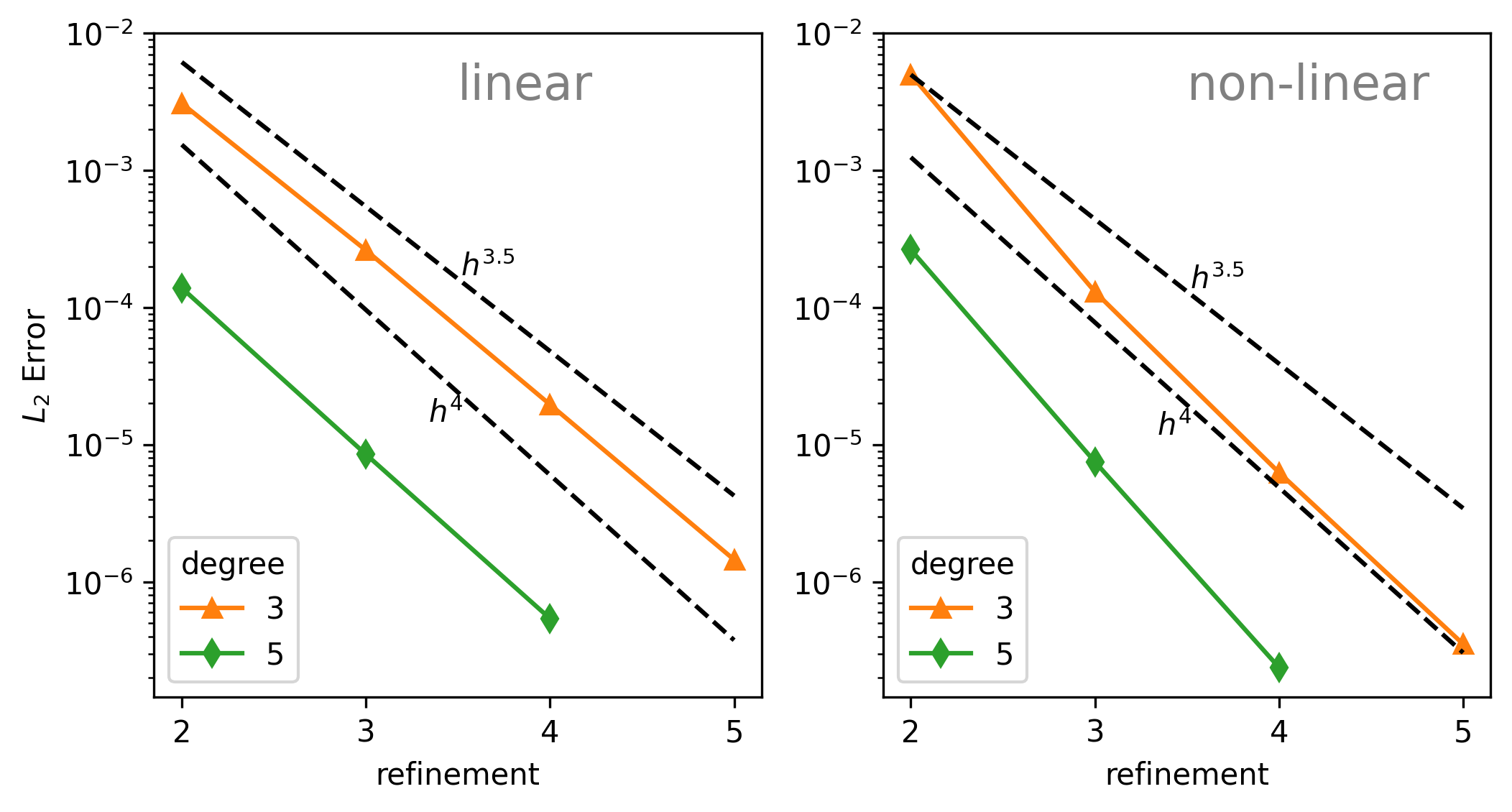}
  \caption{Total $L_2$ error defined in Eq. \eqref{eqn:L2_error} as a function of the refinement level for $p=3$ and $p=5$. The code was run to the final time $T=5$, using both W3 (right) and $\text{W3}^{(\text{lin})}$ (left).}
  \label{fig:W3_error}
\end{figure}

To verify the accuracy and expected convergence rates of the error as the grid is refined we solved the stationary W3 problem and compared the numerical result to the exact solution. Fig. \ref{fig:W3_field_and_error} shows the potential $\phi$, absolute momentum $|\vec{u}|$ and the relative error \mbox{$(\phi-\phi^{(\text{exact})})/(\phi_B+\phi^{(\text{exact})})$} of the total potential $\phi_B+\phi$ after 12 days for polynomial degree $p=3$ and refinement level $r=3$ for the HDG+GTMG+AMG solver. As the figure shows, the absolute value of the relative error of the total potential is of the order $10^{-4}$.
Fig. \ref{fig:W3_error} shows the $L_2$ norm of the total error as defined in Eq. \eqref{eqn:L2_error} for increasing refinement levels and two different polynomial degrees $p$. For each value of $h$ and $p$ the timestep size $\Delta t$ was adjusted according to Eq. \eqref{eqn:spherical_timestep_constraint}. For $p=3$ the error decreases with a rate between $h^{3.5}$ and $h^4$, which is consistent with the results in Section \ref{sec:results_accuracy}. Although the error is over an order of magnitude smaller for $p=5$, asymptotically it does not appear to be decrease with the naively expected rate between $h^{p+1/2}$ and $h^{p+1}$. This, however, is readily explained by the fact that in our implementation the coordinate field is approximated by a cubic polynomial in each grid cell, which introduces an additional error that limits the asymptotic rate of convergence. Nevertheless, even in this context, the benefit of using $p=5$ instead of $p=3$ is evident.
\subsubsection{Computational performance}
\begin{figure}
  \begin{minipage}{0.45\linewidth}
    \begin{center}
      \includegraphics[width=0.8\linewidth]{\figdir/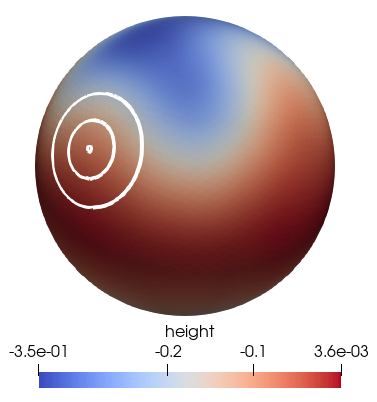}
  \end{center}
\end{minipage}
\hfill
  \begin{minipage}{0.45\linewidth}
    \begin{center}
            \includegraphics[width=0.8\linewidth]{\figdir/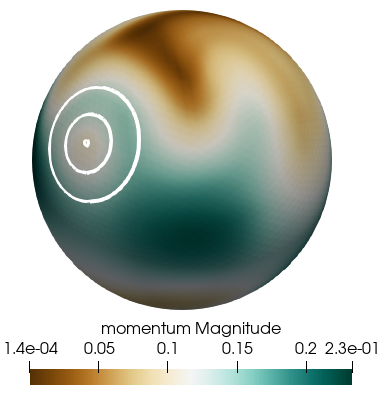}
  \end{center}
  \end{minipage}
  \caption{Potential $\phi$ (left) and absolute value of the momentum $|\vec{u}|$ (right) for the non-linear W5 (isolated mountain) testcase at $t=15\unit{days}$, with the location of the mountain marked by concentric circles. The polynomial degree is $p=3$ and a refinement level of $r=3$ was used. The timestep size in physical units is $783.2\unit{s}$. The linear flux system is solved with the HDG+GTMG+AMG solver.}
  \label{fig:W5_field}
\end{figure}

\begin{figure}
  \begin{minipage}{0.45\linewidth}
    \begin{center}
      \includegraphics[width=0.8\linewidth]{\figdir/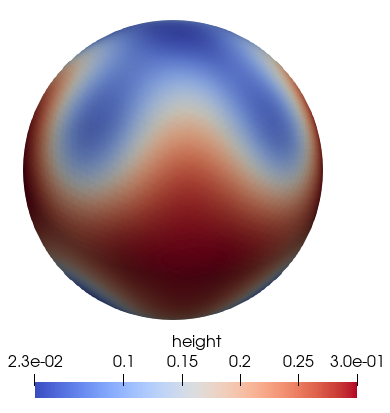}
  \end{center}
\end{minipage}
\hfill
  \begin{minipage}{0.45\linewidth}
    \begin{center}
            \includegraphics[width=0.8\linewidth]{\figdir/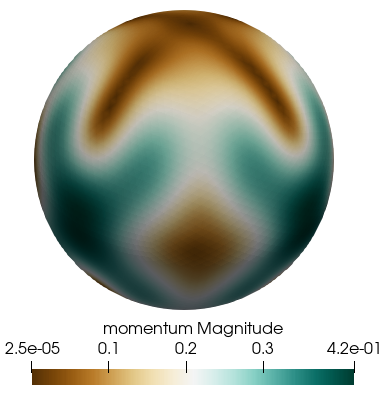}
  \end{center}
  \end{minipage}
  \caption{Potential $\phi$ (left) and absolute value of the momentum $|\vec{u}|$ (right) for the non-linear W6 (Rossby-Haurwitz) testcase at $t=14\unit{days}$. The polynomial degree is $p=3$ and a refinement level of $r=3$ was used. The timestep size in physical units is $306.8\unit{s}$. The linear flux system is solved with the HDG+GTMG+AMG solver.}
    \label{fig:W6_field}
  \end{figure}

The performance of the different solver configurations discussed in Section \ref{sec:spherical_solver_configs} is compared by running the isolated mountain (W5) and Rossby-Haurwitz (W6) test. Fig. \ref{fig:W5_field} shows the potential $\phi$ and momentum $\vec{u}$ for the W5 test after 15 days for polynomial degree $p=3$ and refinement $r=3$; the corresponding results for the W6 test at the same resolution and after 14 days are shown in Fig. \ref{fig:W6_field}. The figures should be compared for example to Figs. 9 and 11 in \cite{Kang2020}.
  For each solver configuration the time $t_{\text{dof,iter}}$ per timestep and per (DG) unknown is measured over the first 15 days of the run for both W5 and W6. More specifically, $t_{\text{dof,iter}}$ is defined by
  \begin{equation}
    t_{\text{dof,iter}} = \frac{\text{total runtime}}{\text{(\# unknowns)}\times\text{(\# timesteps)}}.
  \end{equation}
  For the DG+ApproxSchur and HDG+GTMG+AMG solvers the average number $n_{\text{iter}}$ of solver iterations required to reduce the relative residual by a factor of at least $10^{-8}$ is also recorded.
  Again, the timestep size is adjusted following Eq. \eqref{eqn:spherical_timestep_constraint} in all cases. Tab. \ref{tab:performance_W5} shows $n_{\text{iter}}$ and $t_{\text{dof,iter}}$ for the non-linear test-cases W5 and W6, the corresponding results for the linear setups $\text{W5}^{(\text{lin})}$ and $\text{W6}^{(\text{lin})}$ are shown in Tab. \ref{tab:performance_W6}.
  \begin{table}
    \begin{center}
      \begin{tabular}{|l|rr|rr|rr|rr|}
        \hline
        & \multicolumn{4}{|c|}{W5 (mountain)}
        & \multicolumn{4}{|c|}{W6 (Rossby-Haurwitz)}\\ 
        & \multicolumn{2}{|c}{$p=3$}
        & \multicolumn{2}{c|}{$p=5$}
        & \multicolumn{2}{|c}{$p=3$}
        & \multicolumn{2}{c|}{$p=5$}\\
        & \multicolumn{2}{|c}{($\Delta t = 391.5\unit{s}$)}
        & \multicolumn{2}{c|}{($\Delta t = 124.6\unit{s}$)}
        & \multicolumn{2}{|c}{($\Delta t = 153.4\unit{s}$)}
        & \multicolumn{2}{c|}{($\Delta t = 48.8\unit{s}$)}
        \\\hline\hline
        solver & $n_{\text{iter}}$ & $t_{\text{dof,iter}}$ & $n_{\text{iter}}$ & $t_{\text{dof,iter}}$
        & $n_{\text{iter}}$ & $t_{\text{dof,iter}}$ & $n_{\text{iter}}$ & $t_{\text{dof,iter}}$\\
        \hline
		DG+ApproxSchur &  $11.0$ & $3.99$ &  $9.8$ & $5.61$ & $9.0$ & $3.48$ &    $8.9$ & $5.23$ \\
		HDG+LU         &   --- &  $5.45$ &      --- & $10.86$ & --- & $5.24$ &        --- &     $10.73$ \\
		HDG+GTMG+AMG   &   $8.0$ & $6.29$ &      $8.0$ & $11.78$ & $8.0$ & $6.12$ &   $8.0$ &     $11.52$ \\
		\hline
      \end{tabular}
      \caption{Average number $n_{\text{iter}}$ of solver iterations and time $t_{\text{dof,iter}}$ (measured in $\mu\unit{s}$) per time step and per unknown for the non-linear W5 (isolated mountain) and W6 (Rossby-Haurwitz) testcases. In all cases the refinement level is $r=4$ and the SWEs are integrated to a final time of $T=15\unit{days}$ with the timestep size $\Delta t$ given in the relevant column. The iterative solver tolerance is $10^{-8}$.}
      \label{tab:performance_W5}
    \end{center}
  \end{table}
  \begin{table}
    \begin{center}
      \begin{tabular}{|l|rr|rr|rr|rr|}
        \hline
        & \multicolumn{4}{|c|}{$\text{W5}^{(\text{lin})}$ (mountain)}
        & \multicolumn{4}{|c|}{$\text{W6}^{(\text{lin})}$ (Rossby-Haurwitz)}\\ 
        & \multicolumn{2}{|c}{$p=3$}
        & \multicolumn{2}{c|}{$p=5$}
        & \multicolumn{2}{|c}{$p=3$}
        & \multicolumn{2}{c|}{$p=5$}\\
        & \multicolumn{2}{|c}{($\Delta t = 391.5\unit{s}$)}
        & \multicolumn{2}{c|}{($\Delta t = 249.2\unit{s}$)}
        & \multicolumn{2}{|c}{($\Delta t = 153.4\unit{s}$)}
        & \multicolumn{2}{c|}{($\Delta t = 97.6\unit{s}$)}
        \\\hline\hline
        solver & $n_{\text{iter}}$ & $t_{\text{dof,iter}}$ & $n_{\text{iter}}$ & $t_{\text{dof,iter}}$
        & $n_{\text{iter}}$ & $t_{\text{dof,iter}}$ & $n_{\text{iter}}$ & $t_{\text{dof,iter}}$\\
        \hline
		DG+ApproxSchur &    $10.0$ & $3.49$ &    $11.0$ & $6.17$ &        $9.0$ & $3.27$ &        $9.0$ & $5.21$ \\
		HDG+LU         &        - & *$4.42$ &        - & $9.13$ &        - & $4.23$ &        - & $8.95$ \\
		HDG+GTMG+AMG   &        *$8.0$ & *$4.42$ &    $8.0$ & $9.13$ &        $8.0$ & $4.20$ &    $7.3$ & $8.95$ \\
		\hline
      \end{tabular}
      \caption{Average number $n_{\text{iter}}$ of solver iterations and time $t_{\text{dof,iter}}$ (measured in $\mu\unit{s}$) per time step and per unknown for the linear $\text{W5}^{(\text{lin})}$ (isolated mountain) and $\text{W6}^{(\text{lin})}$ (Rossby-Haurwitz) testcases. In all cases the refinement level is $r=5$ and the SWEs are integrated to a final time of $T=15\unit{days}$ with the timestep size $\Delta t$ given in the relevant column. The iterative solver tolerance is $10^{-8}$. Results marked with an asterisk (*) were compiled with slightly different optimisation flags to avoid a known Intel compiler bug.}
      \label{tab:performance_W6}
    \end{center}
  \end{table}
  As the results show, the performance of the different solvers (as measured by $t_{\text{dof,iter}}$) differs by no more than a factor of around two.
In contrast to what we found in Section \ref{sec:results_solver_comparison} the non-hybridised DG+ApproxSchur solver gives the best overall performance. This might be due inefficiencies in the Slate implementation in the spherical case and further investigation is required to understand this. Note, however, that the hybridised HDG+GTMG+AMG solver consistently requires less iterations to reduce the residual by the same factor. While for the parameters chosen here the iterative GTMG+AMG solver is not faster than the direct solver, several points should be taken into account when comparing those two methods: firstly, MUMPS \cite{MUMPS1,MUMPS2} is a highly optmised sparse direct solver, provided as a monolithic compiled library, whereas there are likely to be inefficiencies due to Python overheads in our Firedrake implementation of GTMG+AMG. Secondly, reducing the relative residual by at least eight orders of magnitude is probably overly cautious. In fact, we find that for the chosen tolerance the linear problem is solved to almost machine precision. Since the performance of the GTMG+AMG solver is directly proportional on the number of iterations, we expect it to perform better for tolerances of, say, $10^{-4}$ which are more relevant to atmospheric simulations. Finally, the runs presented in this section were carried out on a single compute node. While there is no reason why iterative solvers and multigrid preconditioners would not scale to very large problem sizes on massively parallel machines (see the review in \cite{Mueller2014}), this is less clear for direct solvers. Overall, we believe that our results show that our new methods provide a competitive alternative to existing methods.
\section{Conclusion}\label{sec:Conclusion}
Since IMEX methods are able to remove the tight timestep size restrictions due to fast modes, they are of particular interest when simulating large-scale geophysical flow. Combining IMEX time integrators with a hybridised DG discretisation in space leads to an elliptic system of equations for the flux-unknowns on the facets of the spatial grid. In this paper we described an efficient non-nested multigrid preconditioner for this elliptic problem by extending the algorithm in \cite{Cockburn2014}. In contract to the direct solver described in \cite{Kang2020}, we expect our approach to allow the efficient simulation of much larger problems. The first coarsening step projects the problem onto a conforming P1- or lowest-order Raviart-Thomas- space, where it can be solved with established geometric- or algebraic- multigrid methods. Compared to the standard DG discretisation with a naive Schur-complement preconditioner, we show numerically that our method is algorithmically robust under both $p$- and $h$-refinement. Since the problem in pressure space has fewer unknowns and converges in a smaller number of iterations, it leads to better performance overall for an idealised stationary test case in a flat domain. Compared to the standard DG approach, the hybridised method with our non-nested multigrid algorithm approximately halves the overall runtime for a linear solver tolerance of $\epsilon=10^{-8}$ in this idealised setup.

For more complex flow problems in spherical geometry \cite{Williamson1992} the performance of our bespoke multigrid solver is comparable to that of the highly-optimised state-of-the-art direct MUMPS solver \cite{MUMPS1,MUMPS2} on a single compute node. However, we expect that further improvements to the implementation will eventually lead to superior performance of the non-nested multigrid when simulating at high resolution for large core counts, especially if the solver tolerance is reduced to levels that are more common in atmospheric simulations. In contrast to the simplified Euclidean setup, we find that for the considered test cases from the Williamson et al. suite the non-hybridised DG solver with an approximate Schur-complement preconditioner is faster than the HDG based method.

All code was implemented in the Firedrake library. In particular we used the recently developed Slate package \cite{Gibson2019} to realise the Schur-complement reduction to an elliptic system in flux-space. The PETSc library allowed the easy composition of our complex solver hierarchy. For an idealised flow problem good parallel scalability was demonstrated on up to 256 cores, in particular for higher-order DG discretisations. Strong scalability was good until the size of the local problem was reduced to a few thousand unknowns per core. The largest problems solved in the weak scaling runs had several million degrees of freedom.

While the numerical results imply that solving the elliptic flux system with a black-box AMG method gives approximately the same performance as our sophisticated non-nested multigrid preconditioner, the latter can have several  advantages:
\begin{enumerate}
  \item It does not suffer from the high setup costs, in particular for the coarse level operator construction via the Galerkin triple matrix-product.
  \item If matrix-free smoothers for the flux-space are combined with a geometric multigrid solver for the coarse space problem, the memory requirements are significantly reduced, allowing the solution of much larger problems.
\end{enumerate}

There are several ways of extending the work presented here. First of all, the performance of the non-nested multigrid preconditioner can be improved further by tailoring the individual components for the problem at hand. For example, recall that the bilinear form $\what{\mathcal{S}}$ in Eq. \eqref{eqn:schur_system} couples the unknowns on each facet and unknowns on facets which lie on the boundary of the same cell. Hence, the corresponding matrix has a block-structure. To exploit this structure, which is particularly important at higher polynomial degree $p$, one could use a block-Jacobi smoother in which the diagonal blocks are inverted exactly via LU-factorisation. In the future this could be replaced by the approximate matrix-free iterative inversion described in \cite{Bastian2019} which can lead to further speedups for high polynomial orders. The coarse level grad-div problem for the Lax-Friedrichs flux (the corresponding bilinear form is given in Eq. \eqref{eqn:coarse_problem_LF}) is currently solved with AMG. In the future the bespoke patch-based smoothers \cite{Farrell2019} could be used for this instead, which would result in a fully geometric method that can again potentially be optimised with a matrix-free implementation. While the numerical results presented in this work imply that our method is efficient and robust under both $h$- and $p$-refinement, this should be confirmed theoretically by adapting the proof in \cite{Cockburn2014}, as suggested in Section \ref{sec:relationship_to_Cockburn}. Comparing our approach to time-explicit integrators, we find that IMEX methods are only slightly slower if the methods are chosen such that they have the same accuracy for our simplified setup. Furthermore, it is well known that the construction of the HDG Schur-complement system and the reconstruction of the momentum- and height-perturbation field in each time step is currently not implemented very efficiently in Slate; improvements in future versions of Firedrake will likely remove this bottleneck and make the IMEX-HDG methods more competitive. Finally, the non-nested multigrid approach should also be applied to more complicated physical problems, such as the compressible Euler- and Navier-Stokes equations, for which a hybridised DG discretisation has been developed in \cite{Peraire2010}. Ultimately it is of course desirable to use the computationally most efficient discretisation. For this, it is necessary to compare the performance of the IMEX-HDG methods with the multigrid preconditioners described in this paper to other approaches such as spectral- and finite-volume discretisations. While such a comparison if of course extremely important, it is a major undertaking which is well beyond the scope of this paper and which should be pursued in a future publication.
\section*{Acknowledgments}
This research made use of the Balena High Performance Computing (HPC) Service at the University of Bath. The PhD of Jack Betteridge was funded by EPSRC as part of the SAMBa CDT in Bath, grant number EP/L015684/1. We would like to thank all Firedrake developers for their continuous support, in particular Lawrence Mitchell for help with the multigrid implementation. We are grateful to Patrick Farrell (Oxford) for helpful comments. Part of this work was carried out during a short research visit (SRV) to Imperial College in May 2018, funded by the UK-Fluids network (EPSRC grant EP/N032861/1).
\appendix
\section{Bilinear forms for standard DG}\label{sec:standardDGlinear_forms}
The standard DG discretisation of the bilinear form $\what{\mathcal{A}}(q,v)=\mathcal{M}(q,v) - c_g\alpha\Delta t \mathcal{L}(q,v)$ is given for the Lax-Friedrichs flux in Eq. \eqref{eqn:bilinear_standardDG_lax} and for the upwind flux in Eq. \eqref{eqn:bilinear_standardDG_upwind}. Those two expressions, which only differ by one term ($\fdiff{\vec{u}}\cdot\fdiff{\vec{w}}$ vs. $\fjump{\vec{u}}\fjump{\vec{w}}$), should be compared to the corresponding HDG versions in Eqs. \eqref{eqn:bilinear_HDG_lax} and \eqref{eqn:bilinear_HDG_upwind}.
\begin{description}
\item{\textbf{Lax-Friedrichs flux:}}
\begin{equation}
\begin{aligned}
\mathcal{A}^{(\text{LF})}(q,v)=  \mathcal{A}^{(\text{LF})}(\phi,\vec{u},\psi,\vec{w})
  &=\left(\phi\psi + \vec{u}\cdot\vec{w}\right)_{\Omega_h} - c_g\alpha\Delta t
  \Big[
    \left(\vec{u}\cdot\nabla\psi + \phi\nabla\cdot\vec{w}\right)_{\Omega_h}+\mathcal{K}(\vec{u},\vec{w})\\
    &-\left(
      \frac{1}{2}\fdiff{\vec{u}\cdot\vec{n}}\fdiff{\psi} 
+\frac{1}{2}\sqrt{\phi_B}\left(\fdiff{\phi}\fdiff{\psi} + \fdiff{\vec{u}}\cdot\fdiff{\vec{w}}\right)+\phi_B\favg{\phi}\fjump{\vec{w}}
\right)_{\mathcal{E}_h}
    \Big]
\end{aligned}\label{eqn:bilinear_standardDG_lax}
\end{equation}
with the ``curvature'' term 
\begin{equation*}
\mathcal{K}(\vec{u},\vec{w})=-\left(\frac{1}{2}\sqrt{\phi_B}\left((\vec{u}_{+}\cdot\vec{n}_{+})(\vec{n}_{+}+\vec{n}_{-})\cdot \vec{w}_{-}+(\vec{u}_{-}\cdot\vec{n}_{-})(\vec{n}_{+}+\vec{n}_{-})\cdot \vec{w}_{+}\right)\right)_{\mathcal{E}_h}
\end{equation*}
which can be derived following \cite{Bernard2009}. Note that $\mathcal{K}(\vec{u},\vec{w})=0$ for flat geometries where $\vec{n}_{-}=-\vec{n}_{+}$.
\item{\textbf{Upwind flux:}}
\begin{equation}
\begin{aligned}
\mathcal{A}^{(\text{up})}(q,v)=  \mathcal{A}^{(\text{up})}(\phi,\vec{u},\psi,\vec{w})
  &=\left(\phi\psi + \vec{u}\cdot\vec{w}\right)_{\Omega_h} - c_g\alpha\Delta t
  \Big[
    \left(\vec{u}\cdot\nabla\psi + \phi\nabla\cdot\vec{w}\right)_{\Omega_h}\\
    &-\left(
      \frac{1}{2}\fdiff{\vec{u}\cdot\vec{n}}\fdiff{\psi} 
+\frac{1}{2}\sqrt{\phi_B}\left(\fdiff{\phi}\fdiff{\psi} + \fjump{\vec{u}}\fjump{\vec{w}}\right)+\phi_B\favg{\phi}\fjump{\vec{w}}
    \right)_{\mathcal{E}_h}
    \Big]
\end{aligned}\label{eqn:bilinear_standardDG_upwind}
\end{equation}
\end{description}
\section{Linear operators}\label{sec:ABCM_operators}
Let
\begin{equation*}
  \mathcal{A}_0(\phi,\vec{u};\psi,\vec{w}) =
  \left(\phi\psi+\vec{u}\cdot\vec{w}\right)_{\Omega_h}-c_g\alpha\Delta t\left[
    \left(\vec{u}\cdot\nabla\psi+\phi\nabla\cdot\vec{w}\right)_{\Omega_h}
    \right].
\end{equation*}
The linear operators defined in Eq. \eqref{eqn:ABCMform} are given the following expressions for the two fluxes considered in this paper: 
\begin{description}
  \item[Lax-Friedrichs flux]
\begin{equation*}
  \begin{aligned}
  (\mathfrak{A}^{(\text{LF})}q,v)_{\Omega_h} &=
  \mathcal{A}_0(\phi,\vec{u};\psi,\vec{w}) + c_g\alpha \Delta t\left(
  2\sqrt{\phi_B}\favg{\vec{u}\cdot\vec{w}}+\phi_B\fjump{\phi\vec{w}}
  \right)_{\Omega_h}\\
  (\what{q}, (\mathfrak{B}^{(\text{LF})})^\top v)_{\mathcal{E}_h} &= c_g\alpha \Delta t
  \left(\what{\vec{u}}\cdot\left(\fjump{\psi}-2\sqrt{\phi_B}\favg{\vec{w}}\right)\right)_{\mathcal{E}_h}\\
  (\mathfrak{C}^{(\text{LF})}q,\what{v})_{\mathcal{E}_h} &=
  \left(\what{\vec{w}}\cdot\left(\phi_B\fjump{\phi}+2\sqrt{\phi_B}\favg{\vec{u}}\right)\right)_{\mathcal{E}_h}\\
  (\mathfrak{M}^{(\text{LF})}\what{q},\what{v})_{\mathcal{E}_h} &= \left(\what{\vec{w}}\cdot\what{\vec{u}}\right)_{\mathcal{E}_h}
  \end{aligned}
\end{equation*}
  \item[upwind flux]
\begin{equation*}
  \begin{aligned}
  (\mathfrak{A}^{(\text{up})}q,v)_{\Omega_h} &=
  \mathcal{A}_0(\phi,\vec{u};\psi,\vec{w}) + c_g\alpha \Delta t\left(
  2\sqrt{\phi_B}\favg{\phi\psi}+\fjump{\vec{u}\psi}
  \right)_{\Omega_h}\\
  (\what{q},(\mathfrak{B}^{(\text{up})})^\top v)_{\mathcal{E}_h} &= c_g\alpha \Delta t
  \left(\what{\phi}\left(\phi_B\fjump{\vec{w}}-2\sqrt{\phi_B}\favg{\psi}\right)\right)_{\mathcal{E}_h}\\
  (\mathfrak{C}^{(\text{up})}q,\what{v})_{\mathcal{E}_h} &=
  \left(\what{\psi}\cdot\left(\fjump{\vec{u}}+2\sqrt{\phi_B}\favg{\phi}\right)\right)_{\mathcal{E}_h}\\
  (\mathfrak{M}^{(\text{up})}\what{q},\what{v})_{\mathcal{E}_h} &= \left(\what{\psi}\what{\phi}\right)_{\mathcal{E}_h}
  \end{aligned}
\end{equation*}
\end{description}
\section{Runge Kutta methods}\label{sec:RK_details}
Generally, an explicit $s$-stage Runge Kutta (RK) method can be written as
\begin{equation}
  \begin{aligned}
    \mathcal{M}(Q^{(i)},v) &= \mathcal{M}(q^{(n)},v) + \Delta t\sum_{j=1}^{i-1}
    \overline{a}_{ij} \left(\mathcal{N}(Q^{(j)},v) + \mathcal{L}(Q^{(j)},v)\right),\qquad\text{for $i=1,\dots,s$ and all $v\in W_h$,}\\
      \mathcal{M}(q^{(n+1)},v) &= \mathcal{M}(q^{(n)},v) + \Delta t\sum_{i=1}^s \overline{b}_i \left(\mathcal{N}(Q^{(i)},v) + \mathcal{L}(Q^{(i)},v)\right),\qquad\text{for all $v\in W_h$.}
  \end{aligned}
\label{eqn:RungeKutta}
\end{equation}
Since the $\what{\mathcal{L}}$ in \eqref{eqn:theta_method} does not
make a contribution in the case of explicit integrators, it is not necessary to use hybridisation, i.e. only fields in the standard DG space $W_h$ are required. In general, $s$ evaluations of $\mathcal{N}+\mathcal{L}$ and $s+1$ mass-solves on each element are required per timestep. For reference, the matrices $\overline{a}$ and $\overline{b}$ for the explicit RKDG methods used in this paper are given in the following equation:
\begin{equation*}
\begin{aligned}
\text{Heun's method:}\qquad
\overline{a} & = 
\begin{pmatrix}
0 & 0 \\
1 & 0
\end{pmatrix}, 
& \overline{b} &=
\begin{pmatrix}
1/2 \\
1/2
\end{pmatrix}\\[2ex]
\text{SSPRK3:}\qquad
\overline{a} & = 
\begin{pmatrix}
0 & 0 & 0 \\
1 & 0 & 0 \\
1/4 & 1/4 & 0\\
\end{pmatrix}, 
& \overline{b} &=
\begin{pmatrix}
1/6 \\
1/6 \\
2/3
\end{pmatrix}
\end{aligned}
\end{equation*}
\section{Solver configuration}
\subsection{Hybridised DG}\label{sec:solver_configuration_HDG}
To solve the problem in Eq. \eqref{eqn:general_linear_system}, with the HDG+GTMG+AMG solver configuration described in Section \ref{sec:solver_setup_HDG}, the following parameter dictionary is passed to the Firedrake \verb|LinearVariationalSolver| object:
\begin{lstlisting}[language={[firedrake]{python}}]  
  param = {'mat_type': 'matfree',
           'ksp_type': 'preonly',
           'pc_type': 'python',
           'pc_python_type': 'firedrake.SCPC',
           'pc_sc_eliminate_fields': '0, 1',
           'condensed_field': {'ksp_type': inner_ksp_type,
                               'ksp_rtol':rtol,
                               'mat_type': 'aij',
                               'pc_type': 'python',                          
                               'pc_python_type': 'firedrake.GTMGPC',
                               'gt': {'mat_type': 'aij',
                                      'mg_levels': {'ksp_type': 'chebysheb',
                                                    'pc_type': 'bjacobi',
                                                    'sub_pc_type':'sor',
                                                    'ksp_max_it': 2},
                                      'mg_coarse': {'ksp_type': 'preonly',
                                                    'pc_type': "gamg",
                                                    'ksp_rtol': rtol,
                                                    'pc_mg_cycles': 'v',
                                                    'mg_levels':
                                                       {'ksp_type': 'chebyshev',
                                                        'ksp_max_it': 2,
                                                        'pc_type': 'sor'},
                                                    'mg_coarse':
                                                       {'ksp_type':'chebyshev',
                                                        'ksp_max_it':2,
                                                        'pc_type':'sor'}
                                                    }
                                      }
                                      
                               }
                                
         }
\end{lstlisting}
Here \verb|rtol| is the relative tolerance (on the preconditioned residual) to which the linear system is solved. The solver for the Schur-complement system \eqref{eqn:schur_system} is either GMRES for the non-linear shallow-water equations or CG in the linearised case. This is controlled by setting the variable \verb|inner_ksp_type| to \verb|gmres| or \verb|cg|. To use the HDG+AMG solver instead, the \verb|'condensed_field'| dictionary in the code above is replaced by:
\begin{lstlisting}[language={[firedrake]{python}}]
  'condensed_field': {'ksp_type': inner_ksp_type,
                      'ksp_rtol': rtol,
                      'mat_type': 'aij',
                      'pc_type': 'gamg',
                      'mg_levels': {'ksp_type': 'chebyshev',
                                    'ksp_max_it': 2,
                                    'pc_type': 'bjacobi',
                                    'sub_pc_type': 'sor'}
                      }
\end{lstlisting}
To use the HDG+GMG solver (for the upwind flux) this becomes
\begin{lstlisting}[language={[firedrake]{python}}]
  'condensed_field': {'ksp_type': inner_ksp_type,
                      'ksp_rtol': rtol,
                      'mat_type': 'aij',
                      'pc_type': 'python',
                      'pc_python_type': 'firedrake.GTMGPC'
                      'gt': {'mat_type': 'aij',
                             'mg_levels': {'ksp_type': 'chebyshev',
                                           'ksp_max_it': 2,
                                           'pc_type': 'bjacobi',
                                           'sub_pc_type': 'sor'}
                             }
                      }
\end{lstlisting}
\subsection{Native DG}\label{sec:solver_configuration_nativeDG}
To solve the native DG system $\mathcal{A}(q,v)=\mathcal{R}(v)$ in Eq. \eqref{eqn:native_dg_system} with the DG+ApproxSchur solver, the following parameters are used:
\begin{lstlisting}[language={[firedrake]{python}}]
  
  solver_param = {'ksp_type': 'gmres',
                  'ksp_rtol': rtol,
                  'pc_type': 'fieldsplit',
                  'pc_fieldsplit_type': 'schur',
                  'pc_fieldsplit_schur_fact_type': 'FULL',
                  'pc_fieldsplit_schur_precondition': 'selfp',
                  'fieldsplit_0': {'ksp_type': 'preonly',
                                   'pc_type': 'gamg',
                                   'mg_levels': {'ksp_type': 'chebyshev',
                                                 'ksp_max_it': 2,
                                                 'pc_type': 'bjacobi',
                                                 'sub_pc_type': 'sor'}
                  'fieldsplit_1': {'ksp_type': 'preonly',
                                   'pc_type': 'bjacobi',
                                   'sub_pc_type': 'ilu'},
                                  }
                  }
\end{lstlisting}
As above, the system is solved to a relative tolerance \verb!rtol!.
\subsection{Numerical parameters for Williamson et al. testcases}\label{sec:williamson_parameters}
Tab. \ref{tab:williamson_parameters} show the parameters used for the numerical experiments in Section \ref{sec:results_spherical}.
\begin{table}[h]
  \begin{center}
    \begingroup
    \renewcommand{\arraystretch}{1.5}
    \setlength{\tabcolsep}{20pt}
    \begin{tabular}{ll}
      \hline
      parameter & value\\
      \hline\hline
      earth radius & $R_{\text{earth}}=6.37122\cdot 10^{6}\unit{m}$\\
      angular rotation frequency & $\Omega=7.292\cdot{10}^{-5}\unit{s}^{-1}$\\
      reference height (see Section \ref{sec:SWEs}) & $\overline{H}_B=\begin{cases}
        3000\unit{m} & \text{for W3}\\
        5960\unit{m} & \text{for W5}\\
        8000\unit{m} & \text{for W6}
      \end{cases}$\\
      velocity & $u_0=\begin{cases}2\pi R_{\text{earth}}/(12\unit{days})&\text{for W3}\\ 20\unit{ms}^{-1}&\text{for W5}\end{cases}$\\
      zonal flow parameters for W3 & $\theta_b=\pi/6$, $\theta_e=\pi/2$, $x_e=0.3$\\
      mountain height and radius for W5 & $h_{s_0}=2000\unit{m}$, $R=\pi/9$\\
      latitude and longitude of mountain for W5 & $\lambda_c=3\pi/2$, $\theta_c=\pi/6$\\
      parameters for W6 & $\omega=K=7.848\cdot 10^{-6}\unit{s}^{-1}$
    \end{tabular}
    \endgroup
    \caption{Numerical parameters used for numerical experiments in Section \ref{sec:results_spherical}.}
    \label{tab:williamson_parameters}
  \end{center}
\end{table}
\clearpage
\bibliographystyle{unsrt}

\begin{thebibliography}{10}
\expandafter\ifx\csname url\endcsname\relax
  \def\url#1{\texttt{#1}}\fi
\expandafter\ifx\csname urlprefix\endcsname\relax\def\urlprefix{URL }\fi
\expandafter\ifx\csname href\endcsname\relax
  \def\href#1#2{#2} \def\path#1{#1}\fi

\bibitem{Wood2014}
N.~Wood, A.~Staniforth, A.~White, T.~Allen, M.~Diamantakis, M.~Gross,
  T.~Melvin, C.~Smith, S.~Vosper, M.~Zerroukat, et~al., {An inherently
  mass-conserving semi-implicit semi-Lagrangian discretization of the
  deep-atmosphere global non-hydrostatic equations}, Quarterly Journal of the
  Royal Meteorological Society 140~(682) (2014) 1505--1520.

\bibitem{Ascher1995}
U.~M. Ascher, S.~J. Ruuth, B.~T. Wetton, {Implicit-explicit methods for
  time-dependent partial differential equations}, SIAM Journal on Numerical
  Analysis 32~(3) (1995) 797--823.

\bibitem{Ascher1997}
U.~M. Ascher, S.~J. Ruuth, R.~J. Spiteri, {Implicit-explicit Runge-Kutta
  methods for time-dependent partial differential equations}, Applied Numerical
  Mathematics 25~(2-3) (1997) 151--167.

\bibitem{Pareschi2000}
L.~Pareschi, G.~Russo, {Implicit-explicit Runge-Kutta schemes for stiff systems
  of differential equations}, Recent trends in numerical analysis 3 (2000)
  269--289.

\bibitem{Kennedy2003}
C.~A. Kennedy, M.~H. Carpenter, {Additive Runge--Kutta schemes for
  convection--diffusion--reaction equations}, Applied Numerical Mathematics
  44~(1-2) (2003) 139--181.

\bibitem{Weller2013}
H.~Weller, S.-J. Lock, N.~Wood, {Runge--Kutta IMEX schemes for the horizontally
  explicit/vertically implicit (HEVI) solution of wave equations}, Journal of
  Computational Physics 252 (2013) 365--381.

\bibitem{Staniforth2012}
A.~Staniforth, J.~Thuburn, {Horizontal grids for global weather and climate
  prediction models: a review}, Quarterly Journal of the Royal Meteorological
  Society 138~(662) (2012) 1--26.

\bibitem{Mitchell2016}
L.~Mitchell, E.~H. M{\"u}ller, {High level implementation of geometric
  multigrid solvers for finite element problems: Applications in atmospheric
  modelling}, Journal of Computational Physics 327 (2016) 1--18.

\bibitem{Marras2016}
S.~Marras, J.~F. Kelly, M.~Moragues, A.~M{\"u}ller, M.~A. Kopera,
  M.~V{\'a}zquez, F.~X. Giraldo, G.~Houzeaux, O.~Jorba, {A review of
  element-based Galerkin methods for numerical weather prediction: Finite
  elements, spectral elements, and discontinuous Galerkin}, Archives of
  Computational Methods in Engineering 23~(4) (2016) 673--722.

\bibitem{Muething2017}
S.~M{\"u}thing, M.~Piatkowski, P.~Bastian, High-performance implementation of
  matrix-free high-order discontinuous galerkin methods, arXiv preprint
  arXiv:1711.10885 (2017).

\bibitem{Kang2020}
S.~Kang, F.~X. Giraldo, T.~Bui-Thanh, {IMEX HDG-DG: A coupled implicit
  hybridized discontinuous Galerkin and explicit discontinuous Galerkin
  approach for shallow water systems}, Journal of Computational Physics 401
  (2020) 109010.

\bibitem{Cockburn2009a}
B.~Cockburn, B.~Dong, J.~Guzm{\'a}n, M.~Restelli, R.~Sacco, {A hybridizable
  discontinuous Galerkin method for steady-state convection-diffusion-reaction
  problems}, SIAM Journal on Scientific Computing 31~(5) (2009) 3827--3846.

\bibitem{Cockburn2009b}
B.~Cockburn, J.~Gopalakrishnan, R.~Lazarov, {Unified hybridization of
  discontinuous Galerkin, mixed, and continuous Galerkin methods for second
  order elliptic problems}, SIAM Journal on Numerical Analysis 47~(2) (2009)
  1319--1365.

\bibitem{Egger2009}
H.~Egger, J.~Sch{\"o}berl, {A hybrid mixed discontinuous Galerkin
  finite-element method for convection--diffusion problems}, IMA Journal of
  Numerical Analysis 30~(4) (2009) 1206--1234.

\bibitem{Nguyen2009}
N.~C. Nguyen, J.~Peraire, B.~Cockburn, {An implicit high-order hybridizable
  discontinuous Galerkin method for linear convection--diffusion equations},
  Journal of Computational Physics 228~(9) (2009) 3232--3254.

\bibitem{BuiThanh2015}
T.~Bui-Thanh, {From Godunov to a unified hybridized discontinuous Galerkin
  framework for partial differential equations}, Journal of Computational
  Physics 295 (2015) 114--146.

\bibitem{BuiThanh2016}
T.~Bui-Thanh, {Construction and analysis of HDG methods for linearized shallow
  water equations}, SIAM Journal on Scientific Computing 38~(6) (2016)
  A3696--A3719.

\bibitem{Duff2017}
I.~S. Duff, A.~M. Erisman, J.~K. Reid, {Direct methods for sparse matrices},
  Oxford University Press, 2017.

\bibitem{Gopalakrishnan2009}
J.~Gopalakrishnan, S.~Tan, {A convergent multigrid cycle for the hybridized
  mixed method}, Numerical Linear Algebra with Applications 16~(9) (2009)
  689--714.

\bibitem{Cockburn2014}
B.~Cockburn, O.~Dubois, J.~Gopalakrishnan, S.~Tan, {Multigrid for an HDG
  method}, IMA Journal of Numerical Analysis 34~(4) (2014) 1386--1425.

\bibitem{Gibson2019}
T.~H. Gibson, L.~Mitchell, D.~A. Ham, C.~J. Cotter,
  \href{https://arxiv.org/abs/1802.00303}{{Slate: extending Firedrake's
  domain-specific abstraction to hybridized solvers for geoscience and
  beyond.}} (2020).
\newblock \href {http://arxiv.org/abs/1802.00303} {\path{arXiv:1802.00303}},
  \href {https://doi.org/10.5194/gmd-13-735-2020}
  {\path{doi:10.5194/gmd-13-735-2020}}.
\newline\urlprefix\url{https://arxiv.org/abs/1802.00303}

\bibitem{Rathgeber2017}
F.~Rathgeber, D.~A. Ham, L.~Mitchell, M.~Lange, F.~Luporini, A.~T. McRae, G.-T.
  Bercea, G.~R. Markall, P.~H. Kelly, {Firedrake: automating the finite element
  method by composing abstractions}, ACM Transactions on Mathematical Software
  (TOMS) 43~(3) (2017) 24.

\bibitem{Williamson1992}
D.~L. Williamson, J.~B. Drake, J.~J. Hack, R.~Jakob, P.~N. Swarztrauber, {A
  standard test set for numerical approximations to the shallow water equations
  in spherical geometry}, Journal of Computational Physics 102~(1) (1992)
  211--224.

\bibitem{MUMPS1}
P.~Amestoy, I.~S. Duff, J.~Koster, J.-Y. L'Excellent, {A Fully Asynchronous
  Multifrontal Solver Using Distributed Dynamic Scheduling}, SIAM Journal on
  Matrix Analysis and Applications 23~(1) (2001) 15--41.

\bibitem{MUMPS2}
P.~Amestoy, A.~Buttari, J.-Y. L'Excellent, T.~Mary, {Performance and
  Scalability of the Block Low-Rank Multifrontal Factorization on Multicore
  Architectures}, ACM Transactions on Mathematical Software 45 (2019)
  2:1--2:26.

\bibitem{Wildey2018}
T.~Wildey, S.~Muralikrishnan, T.~Bui-Thanh, {Unified geometric multigrid
  algorithm for hybridized high-order finite element methods}, arXiv preprint
  arXiv:1811.09909 (2018).

\bibitem{Muralikrishnan2019}
S.~Muralikrishnan, T.~Bui-Thanh, J.~N. Shadid, {A Multilevel Approach for Trace
  System in HDG Discretizations}, arXiv preprint arXiv:1903.11045 (2019).

\bibitem{Rusanov1961}
V.~V. Rusanov, {The calculation of the interaction of non-stationary shock
  waves with barriers}, Zhurnal Vychislitel'noi Matematiki i Matematicheskoi
  Fiziki 1~(2) (1961) 267--279.

\bibitem{Leveque2002}
R.~J. LeVeque, Finite volume methods for hyperbolic problems, Vol.~31,
  Cambridge university press, 2002.

\bibitem{Melvin2019}
T.~Melvin, T.~Benacchio, B.~Shipway, N.~Wood, J.~Thuburn, C.~Cotter, A mixed
  finite-element, finite-volume, semi-implicit discretization for atmospheric
  dynamics: Cartesian geometry, Quarterly Journal of the Royal Meteorological
  Society 145~(724) (2019) 2835--2853.

\bibitem{Hesthaven2007}
J.~S. Hesthaven, T.~Warburton, {Nodal discontinuous Galerkin methods:
  algorithms, analysis, and applications}, Springer Science \& Business Media,
  2007.

\bibitem{Betteridge2019}
J.~Betteridge, {Efficient elliptic solvers for higher order DG discretisations
  on modern architectures and applications in atmospheric modelling}, Ph.D.
  thesis, University of Bath (2019).

\bibitem{Mueller2014}
E.~H. M{\"u}ller, R.~Scheichl, {Massively parallel solvers for elliptic partial
  differential equations in numerical weather and climate prediction},
  Quarterly Journal of the Royal Meteorological Society 140~(685) (2014)
  2608--2624.

\bibitem{Sandbach2015}
S.~Sandbach, J.~Thuburn, D.~Vassilev, M.~G. Duda, {A semi-implicit version of
  the MPAS-Atmosphere dynamical core}, Monthly Weather Review 143~(9) (2015)
  3838--3855.

\bibitem{Raviart1977}
P.-A. Raviart, J.-M. Thomas, {A mixed finite element method for 2-nd order
  elliptic problems}, in: Mathematical aspects of finite element methods,
  Springer, 1977, pp. 292--315.

\bibitem{Crank1947}
J.~Crank, P.~Nicolson, {A practical method for numerical evaluation of
  solutions of partial differential equations of the heat-conduction type}, in:
  Mathematical Proceedings of the Cambridge Philosophical Society, Vol.~43,
  Cambridge University Press, 1947, pp. 50--67.

\bibitem{Pareschi2005}
L.~Pareschi, G.~Russo, {Implicit--explicit Runge--Kutta schemes and
  applications to hyperbolic systems with relaxation}, Journal of Scientific
  computing 25~(1) (2005) 129--155.

\bibitem{Hestenes1952}
M.~R. Hestenes, E.~Stiefel, {Methods of conjugate gradients for solving linear
  systems}, Vol.~49, NBS Washington, DC, 1952.

\bibitem{Saad1986}
Y.~Saad, M.~H. Schultz, {GMRES: A generalized minimal residual algorithm for
  solving nonsymmetric linear systems}, SIAM Journal on scientific and
  statistical computing 7~(3) (1986) 856--869.

\bibitem{petsc-user-ref}
S.~Balay, S.~Abhyankar, M.~F. Adams, J.~Brown, P.~Brune, K.~Buschelman,
  L.~Dalcin, V.~Eijkhout, W.~D. Gropp, D.~Kaushik, M.~G. Knepley, L.~C.
  McInnes, K.~Rupp, B.~F. Smith, S.~Zampini, H.~Zhang, {PETS}c users manual,
  Tech. Rep. ANL-95/11 - Revision 3.6, Argonne National Laboratory (2015).

\bibitem{Shu1988}
C.-W. Shu, S.~Osher, {Efficient implementation of essentially non-oscillatory
  shock-capturing schemes}, Journal of computational physics 77~(2) (1988)
  439--471.

\bibitem{Cockburn1989}
B.~Cockburn, C.-W. Shu, {TVB Runge-Kutta local projection discontinuous
  Galerkin finite element method for conservation laws. II. General framework},
  Mathematics of computation 52~(186) (1989) 411--435.

\bibitem{Cockburn1998}
B.~Cockburn, C.-W. Shu, {The Runge--Kutta discontinuous Galerkin method for
  conservation laws V: multidimensional systems}, Journal of Computational
  Physics 141~(2) (1998) 199--224.

\bibitem{Alnaes2014}
M.~S. Aln{\ae}s, A.~Logg, K.~B. {\O}lgaard, M.~E. Rognes, G.~N. Wells, {Unified
  Form Language: A Domain-specific Language for Weak Formulations of Partial
  Differential Equations}, ACM Transactions on Mathematical Software 40~(2)
  (2014) 9:1--9:37.
\newblock \href {http://arxiv.org/abs/1211.4047} {\path{arXiv:1211.4047}}.

\bibitem{petsc-efficient}
S.~Balay, W.~D. Gropp, L.~C. McInnes, B.~F. Smith, Efficient management of
  parallelism in object oriented numerical software libraries, in: E.~Arge,
  A.~M. Bruaset, H.~P. Langtangen (Eds.), Modern Software Tools in Scientific
  Computing, Birkh\"{a}user Press, 1997, pp. 163--202.

\bibitem{Kirby2018}
R.~C. Kirby, L.~Mitchell, {Solver composition across the PDE/linear algebra
  barrier}, SIAM Journal on Scientific Computing 40~(1) (2018) C76--C98.

\bibitem{Cockburn2001}
B.~Cockburn, C.-W. Shu, {Runge--Kutta discontinuous Galerkin methods for
  convection-dominated problems}, Journal of scientific computing 16~(3) (2001)
  173--261.

\bibitem{hdg_release}
J.~Betteridge, I.~Graham, T.~H. Gibson, E.~H. Mueller,
  \href{https://doi.org/10.5281/zenodo.4081310}{{Code Release: Multigrid
  preconditioners for the hybridized Discontinuous Galerkin discretisation of
  the shallow water equations}} (Oct. 2020).
\newblock \href {https://doi.org/10.5281/zenodo.4081310}
  {\path{doi:10.5281/zenodo.4081310}}.
\newline\urlprefix\url{https://doi.org/10.5281/zenodo.4081310}

\bibitem{Bernard2009}
P.-E. Bernard, J.-F. Remacle, R.~Comblen, V.~Legat, K.~Hillewaert, {High-order
  discontinuous Galerkin schemes on general 2D manifolds applied to the shallow
  water equations}, Journal of Computational Physics 228~(17) (2009)
  6514--6535.

\bibitem{Bastian2019}
P.~Bastian, E.~H. M{\"u}ller, S.~M{\"u}thing, M.~Piatkowski, {Matrix-free
  multigrid block-preconditioners for higher order Discontinuous Galerkin
  discretisations}, Journal of Computational Physics (2019).

\bibitem{Farrell2019}
P.~E. Farrell, L.~Mitchell, F.~Wechsung, {An Augmented Lagrangian
  Preconditioner for the 3D Stationary Incompressible Navier--Stokes Equations
  at High Reynolds Number}, SIAM Journal on Scientific Computing 41~(5) (2019)
  A3073--A3096.

\bibitem{Peraire2010}
J.~Peraire, N.~Nguyen, B.~Cockburn, {A hybridizable discontinuous Galerkin
  method for the compressible Euler and Navier-Stokes equations}, in: 48th AIAA
  Aerospace Sciences Meeting Including the New Horizons Forum and Aerospace
  Exposition, 2010, p. 363.

\end{thebibliography}

\end{document}